\documentclass[modern]{aastex62}

\usepackage{graphicx, color}
\usepackage{dcolumn}
\usepackage{bm}
\usepackage{booktabs}
\usepackage{natbib}

\newcommand {\vect}[1]{\mbox{\boldmath $#1$}}

\newcommand {\dif}[3][]{\frac{d^{#1}#2}{d#3^{#1}}}
\newcommand {\pdif}[3][]{\frac{\partial^{#1}#2}{\partial#3^{#1}}}

\makeatletter
\def\mart{\@ifnextchar[{\mart@@}{\mart@}}
\def\mart@@[#1]#2{\sqrt[#1]{\mathstrut{#2}}}
\def\mart@#1{\sqrt{\mathstrut{#1}}}
\makeatother
\newcommand {\Alfven}{Alfv\'{e}n}

\newcommand{\myemail}{minoshim@jamstec.go.jp}

\newcommand{\divb}{$\nabla \cdot \vect{B}$ }
\newcommand{\divbzero}{$\nabla \cdot \vect{B} = 0$ }

\long\def\symbolfootnote[#1]#2{\begingroup%
\def\thefootnote{\fnsymbol{footnote}}\footnote[#1]{#2}\endgroup}

\begin{document}

\title{A high-order weighted finite difference scheme with a multi-state approximate Riemann solver for divergence-free magnetohydrodynamic simulations}
\shorttitle{A high-order shock-capturing scheme for divergence-free MHD}


\author{Takashi Minoshima}
\affiliation{Center for Mathematical Science and Advanced Technology, Japan Agency for Marine-Earth Science and Technology, 3173-25, Syowa-machi, Kanazawaku, Yokohama 236-0001, Japan}

\author{Takahiro Miyoshi}
\affiliation{Graduate School of Science, Hiroshima University, 1-3-1, Kagamiyama, Higashi-hiroshima, 739-8526, Japan}

\author{Yosuke Matsumoto}
\affiliation{Department of Physics, Chiba University, 1-33, Yayoi-cho, Inage-ku, Chiba, 263-8522, Japan}

\correspondingauthor{Takashi Minoshima}
\shortauthors{Minoshima et al.}
\email{\myemail}

\begin{abstract}
We design a conservative finite difference scheme for ideal magnetohydrodynamic simulations that attains high-order accuracy, shock-capturing, and divergence-free condition of the magnetic field.
The scheme interpolates pointwise physical variables from computational nodes to midpoints through a high-order nonlinear weighted average.
The numerical flux is evaluated at the midpoint by a multi-state approximate Riemann solver for correct upwinding, and its spatial derivative is approximated by a high-order linear central difference to update the variables with designed order of accuracy and conservation.
The magnetic and electric fields are defined at staggered grid points employed in the Constrained Transport (CT) method by \citeauthor{1988ApJ...332..659E} (1988).
We propose a new CT variant, in which the staggered electric field is evaluated so as to be consistent with the base one-dimensional Riemann solver and the staggered magnetic field is updated to be divergence-free as designed high-order finite difference representation.
We demonstrate various benchmark tests to measure the performance of the present scheme.
We discuss the effect of the choice of interpolation methods, Riemann solvers, and the treatment for the divergence-free condition on the quality of numerical solutions in detail.
\end{abstract}


\section{Introduction}\label{sec:introduction}
The magnetohydrodynamic (MHD) modeling has been extensively applied to various macroscopic dynamics in space and astrophysical plasmas such as flares and coronal mass ejections in the solar corona, evolution of the solar wind through the interplanetary space, auroral substorms and magnetic storms in the planetary magnetosphere, formation of the heliosphere through the interaction between the solar wind plasma and the interstellar medium, turbulence and dynamo in magnetized accretion disks, and so on.
The MHD simulation is an indispensable tool to study such plasmas since the system of equations is highly nonlinear and a variety of phenomena are coupled with each other.
Rapid development in  computational technology and science enables us to conduct high-resolution and long-term numerical simulations for global dynamics as well as local physics.

A common strategy to build an MHD simulation code for space and astrophysical plasmas would be based on an upwind-type shock-capturing method, the so-called Godunov's method.
The Godunov-type scheme is based on a finite volume scheme that updates volume-averaged variables defined in computational cells through an integral form of hyperbolic conservation laws in space and time \citep{1959GODUNOV}.
Godunov's method evaluates a numerical flux at a cell interface through an exact or approximate solution of the Riemann problem for left- and right-side variables as an initial state.
The solution of the Riemann problem retains upwind property for eigenmodes in a hyperbolic system, thus implicitly introduces numerical diffusion enough to stabilize a simulation.
The method is suited for the plasma that frequently contains supersonic flows, shocks and discontinuities.
Godunov's method has been extensively succeeded in hydrodynamic simulations, and utilized for modern MHD simulation codes as well \citep{2005JCoPh.205..509G,2007ApJS..170..228M,2008JCoPh.227.4123G,2008ApJS..178..137S,2009JCoPh.228..952L,2010JCoPh.229.2117M,2013JCoPh.243..269L,2016arXiv161101775M,2017JCoPh.341..230L}.
The choice of Riemann solvers impacts the quality of numerical solutions.

Since the original Godunov's method assumes variables as piecewise constant in a cell, a numerical error of the solution is reduced only as first order.
Improving the accuracy of the method is necessary for practical simulations within a reasonable computational cost.
A straightforward way is increasing the order of accuracy of a scheme.
For Godunov-type schemes, it is achieved by reconstructing Riemann states with a desired order of accuracy.
Piecewise linear interpolation is one of conventional methods to obtain left and right Riemann states in second-order accuracy, and the intermediate state through the Riemann solver.
However, high-order linear interpolation methods generally suffer from numerical oscillation and then tend to crash the simulation of nonlinear hyperbolic equations.
The robustness as well as the accuracy of the simulation rely on interpolation methods.
A nonlinear interpolation method is designed to preserve high order of accuracy in smooth regions but degrade to first order at discontinuous regions so as to avoid the oscillation.
The second-order MUSCL scheme \citep{1979JCoPh..32..101V} is one of standard nonlinear interpolation methods, and has been widely adopted for practical MHD simulations. 
To obtain a second-order accurate solution, one may combine a spatially second-order scheme with a multi-step time integration method.
{Many nonlinear second-order schemes are built to be total variation diminishing (TVD) that inevitably degrade the accuracy at a profile extremum.}
To preserve the profile extremum, some third- and higher-order schemes have been implemented to MHD simulation codes \citep{1999JCoPh.150..561J,2000JCoPh.160..405B,2008CoPhC.179..289M,2010JCoPh.229.7893L,2010JCoPh.229.5896M,2016arXiv161101775M,2017JCoPh.341..230L}.
Another approach is to statically or adaptively refine mesh at regions to be resolved finely.
The Adaptive Mesh Refinement (AMR) technique has been implemented to the MHD simulation code to tackle problems including extreme dynamic range in space and time, for example, star formation and heliosphere \citep{1999JCoPh.154..284P,2001JCoPh.174..614B,2006A&A...457..371F,2007PASJ...59..905M,2008CoPhC.179..227Z,2008ApJS..178..137S,2012ApJS..198....7M}.
The finite volume scheme is usually combined with the AMR since the scheme is flexible about the mesh structure.
In AMR, single step time integration methods are better suited for the reduction of a computational cost of communications between different mesh levels \citep{2005JCoPh.205..509G,2008JCoPh.227.4123G,2009JCoPh.228..952L,2010JCoPh.229.2117M,2013JCoPh.243..269L}.

The Godunov-type scheme can be extended to multi-dimension by a dimension-by-dimension reconstruction.
However, a multidimensional finite volume Godunov-type scheme is computationally expensive to ensure an exact order of accuracy better than two.
This is because one needs to reconstruct the numerical flux along orthogonal directions for face average: this is not identical to a solution of the Riemann problem for face-averaged variables \citep{2011CCoPh...9..807Z,2014JSC...61..343,2017JCoPh.341..230L}.
{Alternatively, a finite difference scheme is designed for hyperbolic conservation laws to retain high-order of accuracy, conservation, and upwind property \citep{1994JCoPh.115..200L,1996JCoPh.126..202J,2000JCoPh.165...22D}.}
Extension of the high-order finite difference scheme to multi-dimension is straightforward because it updates pointwise variables at computational nodes rather than volume-averaged ones.
The calculation of source terms is also straightforward for the finite difference scheme, while the finite volume scheme needs their high-order reconstruction in cells to retain the spatial accuracy better than two.
Note that the two schemes are essentially identical at the second-order level.

Multidimensional MHD simulations should take special care of the divergence-free condition for the magnetic field.
The Godunov-type scheme discretizes hyperbolic conservation laws into a divergence form.
Generally in the baseline Godunov scheme without any special divergence control, the magnetic field is not necessarily divergence-free and a finite divergence error may lead to an unphysical solution, even though the field is updated in the divergence form \citep{1980JCoPh..35..426B}.
Recipes against this problem are classified into two types, divergence-cleaning and divergence-free methods.
The divergence-cleaning method keeps a divergence error within a tolerable level by correcting the magnetic field through the Poisson equation \citep{1980JCoPh..35..426B,1995ApJ...452..785R,2005JCoPh.203..422C}, additional source terms \citep{1999JCoPh.154..284P}, or additional equations \citep{2002JCoPh.175..645D}.
The divergence-free method utilizes staggered grid spacing so as to discretize the induction equation to be consistent with the Faraday's law, termed as the Constrained Transport (CT) method \citep{1988ApJ...332..659E}.
{The CT method defines the in-plane magnetic field at cell interfaces and the out-of-plane electric field at cell corners to preserve the divergence-free condition in a discretized manner within a machine precision \citep[e.g.,][]{2000JCoPh.161..605T}.
Since the staggered electric field is not directly obtained from a one-dimensional Riemann solver, a variety of methods have been proposed to calculate it with retaining the upwind property \citep{1995CoPhC..89..127H,1998ApJ...494..317D,1998ApJ...509..244R,1999JCoPh.149..270B,2000ApJ...530..508L}.
\cite{1999JCoPh.149..270B} calculated the electric field through an arithmetic average of the neighboring numerical flux for the in-plane magnetic field obtained from a one-dimensional Riemann solver.
Their method has been employed as a baseline for subsequent schemes \citep{2005JCoPh.205..509G,2009JCoPh.228..952L,2013JCoPh.243..269L}.
Meanwhile, the solution of a multidimensional Riemann solver was derived for the staggered electric field \citep{2004JCoPh.195...17L,2010JCoPh.229.1970B,2012JCoPh.231.7476B,2015JCoPh.299..863A}.
}
Comparative studies between divergence-cleaning and divergence-free methods have been reported by \cite{2000JCoPh.161..605T}, \cite{2004ApJ...602.1079B}, and \cite{2011PFR.....6.1124M}.

Numerous studies have been devoted to develop novel numerical schemes so as to attain accurate and robust MHD simulations.
{A high-order scheme combined with a shock-capturing method improves the resolution of shocks and discontinuities, smooth structures such as vortices and waves, and their interaction driven by a high speed flow at a high Reynolds number \citep{2009SIAMR..51...82S}.
A divergence-cleaning method is combined with a shock-capturing method in a straightforward manner.
However, it might lead to a spurious solution for a stringent problem such as multiple shock interaction, in which a continuously-produced local divergence error affects globally through a cleaning method \citep{2004ApJ...602.1079B}.
{A recipe for combining a high-order (better than two) scheme, an MHD shock-capturing method, and a divergence-free method must be useful for space and astrophysical plasma simulations \citep{2007A&A...473...11D,2009JCoPh.228.2480B,2010JCoPh.229.7893L,2018JCoPh.375.1365F}.}

The purpose of this paper is to design a numerical scheme for ideal MHD simulations, specifically focusing on the high-order interpolation and the treatment for the divergence-free condition.
We build the scheme based on the finite difference scheme with several criteria: (1) the order of accuracy can be improved to an arbitrary level (in principle) through a dimension-by-dimension interpolation, (2) a variety of Riemann solvers are available, and (3) the multidimensional scheme is divergence-free and reduces to the base one-dimensional scheme for one-dimensional problems.
Section \ref{sec:finite-diff-schem} presents the scheme design.
Details of the finite difference scheme for hyperbolic conservation laws are introduced in Section \ref{sec:high-order-cdif}.
The high-order interpolation methods are described in Sections \ref{sec:nonl-interp}-\ref{sec:char-decomp}, the Riemann solver is briefly introduced in Section \ref{sec:riemann-solver}, and the treatment for the divergence-free condition is presented in Section \ref{sec:multi-dimension}.
Section \ref{sec:numerical-tests} details numerical simulation results of various benchmark tests so as to measure the performance of the present scheme.
We compare a variety of schemes to assess the effect of numerical techniques on the quality of solutions.
Finally, Section \ref{sec:summary} summarizes the paper.

\section{High-order finite difference schemes for ideal MHD}\label{sec:finite-diff-schem}
The normalized, fully compressible ideal MHD equations are
\begin{eqnarray}
&&\pdif{\rho}{t} + \nabla \cdot \left(\rho \vect{u}\right) = 0,\label{eq:1}\\
&&\pdif{\left(\rho \vect{u}\right)}{t} + \nabla \cdot \left[ \rho \vect{u}\vect{u} + \left(P+\frac{|\vect{B}|^2}{2}\right)\vect{\rm I} - \vect{B}\vect{B}\right]=0,\label{eq:2}\\
&&\pdif{\vect{B}}{t} + \nabla \times \vect{E} = 0,\label{eq:3}\\
&&\pdif{e}{t} + \nabla \cdot \left[\left(\frac{\rho u^2}{2} + \frac{\gamma P}{\gamma-1}\right)\vect{u} + \vect{E}\times\vect{B}\right] = 0,\label{eq:4}
\end{eqnarray}
where $\rho, \vect{u}=(u,v,w), \vect{B}=(B_x,B_y,B_z), \vect{E}=(E_x,E_y,E_z), e$ and $P$ are the fluid mass density, the velocity, the magnetic field, the electric field, the total energy density, and the gas pressure, respectively. 
The electric field and the gas pressure are determined from the Ohm's law and the equation of state,
\begin{eqnarray}
&&\vect{E} = -\vect{u} \times \vect{B},\label{eq:5}\\
&&P = \left(\gamma-1\right)\left(e-\frac{\rho |\vect{u}|^2}{2} - \frac{|\vect{B}|^2}{2}\right),\label{eq:6}
\end{eqnarray}
where $\gamma$ is the specific heat ratio.
The magnetic field should satisfy the divergence-free condition,
\begin{eqnarray}
 \nabla \cdot {\vect{B}} = 0.\label{eq:9}
\end{eqnarray}

To construct a finite difference scheme in a dimension-by-dimension fashion, we firstly consider the one-dimensional MHD equations in the conservative form as
\begin{eqnarray}
\pdif{\vect{U}}{t} + \pdif{ \vect{F}}{x} = 0,\label{eq:10} 
\end{eqnarray}
where
\begin{eqnarray}
\vect{U} = \left[
\begin{array}{c}
 \rho \\
 {\rho u} \\
 {\rho v} \\
 {\rho w} \\
 {B_y} \\
 {B_z} \\
 e \\
\end{array}
\right],
\vect{F} = \left[
\begin{array}{c}
 \rho u \\
 \rho u^2 + P + \left(B_y^2+B_z^2-B_x^2\right)/2 \\
 \rho v u - B_x B_y \\
 \rho w u - B_x B_z \\
 B_y u - B_x v \\
 B_z u - B_x w \\
\left(e+P+B^2/2\right)u - B_x\left(\vect{u}\cdot\vect{B}\right)\\
\end{array}
\right],\label{eq:12}
\end{eqnarray}
and $B_x = \rm{constant}$ is given as a constraint in one dimension.

We discretize the physical variables $\vect{U}$ at a time $t_n$ on a computational node $i$ as pointwise representation, $\vect{U}^n_i = \vect{U}(x_i,t_n)$.
Eq. (\ref{eq:10}) is discretized as
\begin{eqnarray}
\dif{\vect{U}^n_i}{t} = - \left.\pdif{\vect{F}}{x}\right|_i,\label{eq:7}
\end{eqnarray}
and is integrated in time with the third-order Strong Stability Preserving Runge-Kutta (SSPRK) method that is popular for the time integration of partial differential equations \citep{1988JCoPh..77..439S,1998MaCom..67...73G},
\begin{eqnarray}
 \vect{U}^{*}&=&\vect{U}^{n} - {\cal D}(\vect{F}^n) \Delta t,\nonumber \\
 \vect{U}^{**}&=&\frac{3}{4}\vect{U}^{n} + \frac{1}{4}\left(\vect{U}^{*}-{\cal D}(\vect{F}^{*}) \Delta t\right),\nonumber \\
 \vect{U}^{n+1}&=&\frac{1}{3}\vect{U}^{n} + \frac{2}{3}\left(\vect{U}^{**}-{\cal D}(\vect{F}^{**}) \Delta t\right),\label{eq:22}
\end{eqnarray}
where $\cal D$ stands for a difference operator and $\Delta t$ is a time step.
The quality of numerical solutions relies on the evaluation of the flux derivative as long as a temporal error is relatively small.

Conventional high-order finite difference schemes for hyperbolic conservation laws approximate the flux derivative as a two-point difference \citep{1994JCoPh.115..200L,1996JCoPh.126..202J,1999JCoPh.150..561J,2000JCoPh.160..405B,2010JCoPh.229.5896M,2014JCoPh.268..302C},
\begin{eqnarray}
\left.\pdif{\vect{F}}{x}\right|_i = \frac{\vect{\hat{F}}_{i+1/2}-\vect{\hat{F}}_{i-1/2}}{\Delta x} + O(\Delta x^q),\label{eq:8}
\end{eqnarray}
where $\vect{\hat{F}}_{i+1/2}$ is the pointwise numerical flux at a midpoint $i+1/2$ and $\Delta x$ is a grid size.
The numerical flux $\vect{\hat{F}}$ is evaluated so that its two-point difference is an approximation of the flux derivative with the desired order of accuracy $q$; it is not identical to the $q$th-order approximation of the physical flux $\vect{F}(\vect{U})$ itself for $q > 2$. 
As is shown by e.g. \cite{2010JCoPh.229.5896M}, $\vect{\hat{F}}_{i+1/2}$ is reconstructed from the pointwise physical flux $\vect{F}_{i}=\vect{F}(\vect{U}_{i})$ by the same reconstruction method used for finite volume schemes.

The example procedure of the conventional finite difference scheme is as follows:
(i) the physical flux $\vect{F}_{i}$ is calculated at the node, and is split into positive and negative fluxes for correct upwinding, $\vect{F}_{i} = \vect{F}^{+}_{i} + \vect{F}^{-}_{i}$, where ${\partial \vect{F}^{+}}/{\partial \vect{U}} > 0$ and ${\partial \vect{F}^{-}}/{\partial \vect{U}} < 0$,
(ii) the numerical fluxes are reconstructed to the left side at $x_{i+1/2}$ and the right side at $x_{i-1/2}$, $\vect{\hat{F}}^{+}_{i+1/2}$ and $\vect{\hat{F}}^{-}_{i-1/2}$, with the positive and negative fluxes $\vect{F}^{\pm}$ from Step (i) in an upwind-biased stencil $(i-r,\dots,i+r)$. 
For example, the optimal fifth-order $(r=2)$ upstream central scheme gives $\vect{\hat{F}}^{\pm}_{i\pm1/2} = (2\vect{F}^{\pm}_{i\mp2} -13\vect{F}^{\pm}_{i\mp1} +47\vect{F}^{\pm}_{i} +27\vect{F}^{\pm}_{i\pm1} -3\vect{F}^{\pm}_{i\pm2})/60$.
 To suppress numerical oscillation caused by linear reconstruction methods, one employs a nonlinear method such as the WENO scheme \citep{1996JCoPh.126..202J}, and
(iii) the numerical flux is obtained at the midpoint, $\vect{\hat{F}}_{i+1/2} = \vect{\hat{F}}^{+}_{i+1/2} + \vect{\hat{F}}^{-}_{i+1/2}$, and then Eq. (\ref{eq:7}) is integrated through Eq. (\ref{eq:8}).

\cite{1981JCoPh..40..263S} developed a flux-vector splitting method for hydrodynamic simulations, but their method cannot be utilized for MHD simulations because the MHD equations are not homogeneous of degree one, $\vect{F} \neq A \vect{U}$, where $A={\partial \vect{F}}/{\partial \vect{U}}$ is the Jacobian matrix. 
One may use the Lax-Friedrichs method, $\vect{F}^{\pm}_{i} = \left(\vect{F}_{i} \pm |\lambda|_{\rm max} \vect{U}_{i}\right)/2$, where $|\lambda|_{\rm max}$ is the maximum absolute speed among eigenmodes.
Although the Lax-Friedrichs method is applicable to any hyperbolic conservation laws, a solution suffers from numerical diffusion with a time scale of $\Delta x / |\lambda|_{\rm max}$ that may be much faster than a dynamical time scale \citep{2015ApJ...808..54M}.

\subsection{Differencing}\label{sec:high-order-cdif}
We adopt another type of high-order finite difference schemes, termed as the Weighted Compact Nonlinear Scheme \cite[WCNS;][]{2000JCoPh.165...22D}.
This scheme is built based on the compact finite difference scheme in \cite{1992JCoPh.103...16L}, which expresses a linear combination of flux derivatives around a node as a function of fluxes at midpoints (Eq. (4) in \cite{2000JCoPh.165...22D}).
For low computational cost and simplicity in parallel computation, we use an explicit form to approximate the flux derivative at the node $i$ as
\begin{eqnarray}
&& \left.\pdif{\vect{F}}{x}\right|_{i} = \sum_{k\geq0} d_{k} \frac{\vect{\tilde{F}}_{i+k+1/2}-\vect{\tilde{F}}_{i-k-1/2}}{\Delta x}+O(\Delta x^p) + O(\Delta x^q),\label{eq:11}
\end{eqnarray}
where we consider uniform grid spacing.
Here, the pointwise numerical flux $\vect{\tilde{F}}_{i+1/2}$ is a high-order approximation of the physical flux at the midpoint $\vect{F}(\vect{U}(x_{i+1/2}))$ itself.
The truncation error terms $O(\Delta x^p)$ and $O(\Delta x^q)$ come from the approximations of the numerical flux and the derivative, and the spatial accuracy depends on both.
Conservative property is guaranteed as long as the coefficients $d_k$ are constant.
We use the fourth-order linear central difference,
\begin{eqnarray}
{\cal D}_4(\vect{\tilde{F}})_i &=& \frac{27\left(\vect{\tilde{F}}_{i+1/2}-\vect{\tilde{F}}_{i-1/2}\right)-\left(\vect{\tilde{F}}_{i+3/2}-\vect{\tilde{F}}_{i-3/2}\right)}{24 \Delta x} \label{eq:13} \\ 
&=& \left.\pdif{\vect{F}}{x}\right|_{i} - \frac{3}{640}\left.\pdif[5]{\vect{F}}{x}\right|_{i} \Delta x^4 + O(\Delta x^6) +O(\Delta x^p).\nonumber
\end{eqnarray}
Sixth-, eighth- and tenth-order differences are found in \cite{2008JCoPh.227.7294Z}.
At the leftmost interior point $i=1$, one can use the third-order boundary scheme,
\begin{eqnarray}
{\cal D}_4(\vect{\tilde{F}})_1 &=& \frac{-\vect{\tilde{F}}_{7/2}+3\vect{\tilde{F}}_{5/2}+21\vect{\tilde{F}}_{3/2}-23\vect{\tilde{F}}_{1/2}}{24 \Delta x}\label{eq:24} \\
&=& \left.\pdif{\vect{F}}{x}\right|_{1} - \frac{1}{24}\left.\pdif[4]{\vect{F}}{x}\right|_{1} \Delta x^3 + O(\Delta x^4) +O(\Delta x^p).\nonumber
\end{eqnarray}
An expression for the rightmost interior point is symmetrically derived.
{The flux derivative is approximated by the numerical fluxes only similar to \cite{2000JCoPh.165...22D} and \cite{2007A&A...473...11D}.
Meanwhile, one may employ a combination of the physical fluxes at the nodes and the numerical fluxes at the midpoints that makes a stencil smaller \citep{2013JCoF.85.8,2016JCoPh.305..604C}.}

This scheme allows us to interpolate the physical variables \citep{2000JCoPh.165...22D} or the physical fluxes \citep{2008JCoPh.227.7294Z} for the calculation of the numerical fluxes at the midpoints, whereas the conventional finite difference scheme directly reconstructs the numerical fluxes at the midpoints from the physical fluxes at the nodes.
The present scheme adopts the following procedure:
(i) the physical variables are interpolated to the left side at $x_{i+1/2}$ and the right side at $x_{i-1/2}$, $\vect{\tilde{U}}^{L}_{i+1/2}$ and $\vect{\tilde{U}}^{R}_{i-1/2}$, in an upwind-biased stencil $(i-r,\dots,i+r)$. They should be a high-order approximation of $\vect{U}(x_{i\pm1/2})$,
(ii) using $\vect{\tilde{U}}^{L,R}_{i+1/2}$ from Step (i) as the initial left and right Riemann states, the numerical flux is evaluated at the midpoint by a Riemann solver, $\vect{\tilde{F}}_{i+1/2}$, which is a high-order approximation of $\vect{F}(\vect{U}(x_{i+1/2}))$ itself, and
(iii) Eq. (\ref{eq:7}) is integrated through $d\vect{U}^n_i/dt = -{\cal D}_4(\vect{\tilde{F}})_i$.

{Since the present scheme interpolates the physical variables to the midpoints, one can utilize a variety of Riemann solvers, some of which resolve multiple eigenmodes in the Riemann fan that are smeared in the Lax-Friedrichs method.}
This is a main advantage against the conventional finite difference scheme for MHD simulations.
A disadvantage of the scheme is the increase of the number of grid points necessary to update.
{The explicit WCNS scheme achieves the $(2r+1)$th order of accuracy by combining a $(2r+1)$th-order interpolation and a $(2r+2)$th-order difference.
The $(2r+2)$th-order flux difference at the node $i$ is approximated by the numerical fluxes at the midpoints $(i-1/2-r,\dots,i+1/2+r)$, and the numerical flux at the rightmost midpoint $(i+1/2+r)$ is reconstructed in the stencil $(i,\dots,i+2r+1)$.
Therefore, the $(2r+1)$th-order explicit WCNS scheme needs $(4r+3)$ grid points to update the physical variables at the node $i$.
The example grid spacing for $r=1$ is displayed in Figure \ref{fig:grid1d}.}
On the other hand, the $(2r+1)$th-order conventional finite difference scheme needs $(2r+3)$ grid points.

\subsection{Interpolation}\label{sec:nonl-interp}
In this subsection, we consider a scalar linear advection case in which $F$ is proportional to $U$, e.g., $F=aU$ for simplicity.
The scheme calculates the physical variable at the midpoint $i+1/2$ in an upwind-biased stencil $(i-r,\dots,i+r)$.
The optimal third-order interpolation of ${\tilde{U}}^{L}_{i+1/2}$ from the point values ${U}_i$ in the stencil $(i-1,\dots,i+1)$ is
\begin{eqnarray}
{\tilde{U}}^{L}_{i+1/2} &=& \frac{-{U}_{i-1}+6{U}_{i}+3{U}_{i+1}}{8} \label{eq:50} \\
&=& {U}_{i+1/2} +\frac{1}{16} \left.\pdif[3]{{U}}{x}\right|_{i} {\Delta x^3} + \frac{1}{128} \left.\pdif[4]{{U}}{x}\right|_{i} {\Delta x^4} + O(\Delta x^5).\nonumber 
\label{eq:15}
\end{eqnarray}
The equation for ${\tilde{U}}^{R}_{i-1/2}$ is symmetric with respect to $i$.
Assuming the advection velocity is $+1$, i.e., $F=U$, ${\tilde{F}}_{i+1/2}$ in Eq. (\ref{eq:13}) is replaced by ${\tilde{U}}^{L}_{i+1/2}$ and then Eq. (\ref{eq:13}) reduces to
\begin{eqnarray}
{\cal D}_4({\tilde{U}^{L}})_i = \left.\pdif{{U}}{x}\right|_{i} + \left.\frac{1}{16}\pdif[4]{{U}}{x}\right|_{i}\Delta x^3 - \left.\frac{9}{320}\pdif[5]{{U}}{x}\right|_{i} \Delta x^4 + O(\Delta x^5).\label{eq:16}
\end{eqnarray}
The combination of the third-order interpolation in Eq. (\ref{eq:50}) and the fourth-order difference in Eq. (\ref{eq:13}) has the third-order dissipation as a leading error.
Similarly, the fifth-order interpolation in the stencil $(i-2,\dots,i+2)$ is 
\begin{eqnarray}
{\tilde{U}}^{L}_{i+1/2} &=& \frac{3{U}_{i-2}-20{U}_{i-1}+90{U}_{i}+60{U}_{i+1}-5{U}_{i+2}}{128}\label{eq:17}\\
&=&{U}_{i+1/2}-\frac{3}{256} \left.\pdif[5]{{U}}{x}\right|_{i} {\Delta x^5} + O(\Delta x^6),\nonumber 
\end{eqnarray}
and Eq. (\ref{eq:13}) reduces to
\begin{eqnarray}
{\cal D}_4({\tilde{U}^{L}})_i = \left.\pdif{{U}}{x}\right|_{i} - \left.\frac{3}{640}\pdif[5]{{U}}{x}\right|_{i} \Delta x^4 - \left.\frac{3}{256}\pdif[6]{{U}}{x}\right|_{i} \Delta x^5 + O(\Delta x^6).\label{eq:18}
\end{eqnarray}
The combination of the fifth-order interpolation in Eq. (\ref{eq:17}) and the fourth-order difference in Eq. (\ref{eq:13}) has the fourth-order dispersion as a leading error.
One can improve the leading error to be fifth-order dissipation by using a sixth-order difference.

The fifth-order interpolation in Eq. (\ref{eq:17}) is not optimal for this scheme because the order of accuracy is four in spite of the five-point interpolation.
Instead, we propose the following fourth-order interpolation that is appropriate to Eq. (\ref{eq:13}),
\begin{eqnarray}
{\tilde{U}}^{L}_{i+1/2} &=& \frac{9{U}_{i-2}-56{U}_{i-1}+234{U}_{i}+144{U}_{i+1}-11{U}_{i+2}}{320}\label{eq:19}\\
&=&{U}_{i+1/2} +\frac{3}{640} \left.\pdif[4]{{U}}{x}\right|_{i} {\Delta x^4} - \frac{3}{256} \left.\pdif[5]{{U}}{x}\right|_{i} {\Delta x^5} + O(\Delta x^6).\nonumber 
\end{eqnarray}
Replacing ${\tilde{F}}_{i+1/2}$ in Eq. (\ref{eq:13}) by ${\tilde{U}}^{L}_{i+1/2}$, one find that the $O(\Delta x^4)$ error cancels and the leading error is fifth-order dissipation,
\begin{eqnarray}
{\cal D}_4({\tilde{U}})_i = \left.\pdif{{U}}{x}\right|_{i} - \left.\frac{9}{640}\pdif[6]{{U}}{x}\right|_{i} \Delta x^5 + O(\Delta x^6).\label{eq:20}
\end{eqnarray}
{In Appendix \ref{sec:sixth-order-finite}, we combine seventh- and sixth-order interpolation methods with a sixth-order difference, giving sixth- and seventh-order finite difference schemes.}

In order to confirm the dissipative and dispersive properties of the above interpolation methods in Eqs. (\ref{eq:50})-(\ref{eq:20}), we perform the Fourier analysis similar to \cite{1992JCoPh.103...16L}.
Using the periodic function for a scalar variable $U_i=\exp(j \omega x_i)$ (here $j$ is the imaginary unit), we interpolate it to the midpoints and calculate ${\cal D}_4(U)$.
Subsequently, we evaluate the modified wave number as
\begin{eqnarray}
\omega^* =  -j \exp(-j \omega x) {\cal D}_4(U).\label{eq:21}
\end{eqnarray}
Figure \ref{fig:wcns_fourier} shows the real and imaginary parts of the modified wave number as a function of the actual wave number. 
Evidently, the exact solution is ${\rm Re}(\omega^*)=\omega,{\rm Im}(\omega^*)=0$.
{The combinations of the third-, fourth-, and fifth-order interpolation methods (Eqs. (\ref{eq:50}), (\ref{eq:19}), and (\ref{eq:17})) with the fourth-order difference (Eq. (\ref{eq:13})) are referred to as 3I4D3, 4I4D5, and 5I4D4, where the first digit represents the order of interpolation, the second digit is the order of difference, and the last digit is the designed order of accuracy, respectively.}
The third- and fifth-order upstream central schemes (UP3 and UP5) are shown for comparison.
The imaginary part is not zero (dissipative) because the interpolation methods are upwind biased.
The numerical dispersion and dissipation are improved with increasing the order of accuracy of the interpolation.
The 4I4D5 scheme improves/degrades the numerical dispersion/dissipation against the 5I4D4 scheme, which can be explained by the fact that the absolute value of the imaginary part in the 4I4D5 scheme is 1.2 times larger than the 5I4D4 scheme, as is found in Eqs.~(\ref{eq:18}) and (\ref{eq:20}).
In terms of the modified wave number, the schemes are slightly better than the same-order upstream central schemes as is mentioned in \cite{2000JCoPh.165...22D}.

In practical compressible hydrodynamic simulations, high-order linear interpolation methods do not work due to severe numerical oscillation around discontinuities. 
Therefore, the WCNS scheme adopts the same methodology as the WENO scheme to suppress the numerical oscillation.
The interpolation is expressed as a convex combination of substencils,
\begin{eqnarray}
{\tilde{U}}^{L}_{i+1/2} = \sum_{k} w_k {\tilde{U}}^{L,k}_{i+1/2}, \;\;\; w_s = \frac{\alpha_s}{\sum_{k}\alpha_k}\label{eq:26}
\end{eqnarray}
where ${\tilde{U}}^{L,k}_{i+1/2}$ is the interpolated value in the $k$th substencil and the nonlinear weight $w_s$ is a function of the smoothness of the profile measured in the substencil.
The nonlinear weight is designed so as to be close to the optimal weight when the profile is sufficiently smooth in all substencils, but nearly zero in the substencil containing discontinuities.
For the third-order case (Eq. (\ref{eq:50})), the interpolations in two substencils $(i-1,i)$ and $(i,i+1)$ are 
\begin{eqnarray}
{\tilde{U}}^{L,0}_{i+1/2} &=& \frac{\left(-{U}_{i-1}+3{U}_{i}\right)}{2},\nonumber \\
 {\tilde{U}}^{L,1}_{i+1/2} &=& \frac{\left({U}_{i} + {U}_{i+1}\right)}{2}. \label{eq:28}
\end{eqnarray}
We employ the weighted function from the third-order energy stable WENO scheme \citep{2009JCoPh.228.3025Y},
\begin{eqnarray}
&& \alpha_s=c_s\left(1+\frac{\tau}{IS_s+\epsilon}\right),\nonumber \\
&& IS_0 = \left({U}_{i}-{U}_{i-1}\right)^2, IS_1 = \left({U}_{i+1}-{U}_{i}\right)^2, \tau = \left({U}_{i+1}-2{U}_i+{U}_{i-1}\right)^2,\nonumber \\
&& c_0 = \frac{1}{4}, c_1=\frac{3}{4},\label{eq:23}
\end{eqnarray}
where $\epsilon=10^{-40}$ is a positive real number to avoid the denominator becoming zero.
For the fifth- and fourth-order cases (Eqs. (\ref{eq:17}) and (\ref{eq:19})), the interpolations in three substencils $(i-2,i-1,i)$, $(i-1,i,i+1)$, and $(i,i+1,i+2)$ are found in \cite{2008JCoPh.227.7294Z} (Eqs. (2-16) - (2-18) therein),
\begin{eqnarray}
{\tilde{U}}^{L,0}_{i+1/2} &=& \frac{\left(3{U}_{i-2}-10{U}_{i-1}+15{U}_{i}\right)}{8},\nonumber \\
{\tilde{U}}^{L,1}_{i+1/2} &=& \frac{\left(-{U}_{i-1}+6{U}_{i} + 3{U}_{i+1}\right)}{8},\nonumber \\
{\tilde{U}}^{L,2}_{i+1/2} &=& \frac{\left(3{U}_{i}+6{U}_{i+1} - {U}_{i+2}\right)}{8}.\label{eq:25} 
\end{eqnarray}
The weighted function is the same as used in \cite{2008JCoPh.227.7294Z},
\begin{eqnarray}
&& \alpha_s=\frac{c_s}{\left(IS_s+\epsilon\right)^2},\nonumber \\
&& IS_0 = \frac{13}{12}\left({U}_{i-2}-2{U}_{i-1}+{U}_{i}\right)^2+\frac{1}{4}\left({U}_{i-2}-4{U}_{i-1}+3{U}_{i}\right)^2,\nonumber \\
&& IS_1 = \frac{13}{12}\left({U}_{i-1}-2{U}_{i}+{U}_{i+1}\right)^2+\frac{1}{4}\left({U}_{i-1}-{U}_{i+1}\right)^2,\nonumber \\
&& IS_2 = \frac{13}{12}\left({U}_{i}-2{U}_{i+1}+{U}_{i+2}\right)^2+\frac{1}{4}\left(3{U}_{i}-4{U}_{i+1}+{U}_{i+2}\right)^2,\nonumber \\
&& c_0 = \frac{1}{16},c_1 = \frac{5}{8}, c_2=\frac{5}{16}, \;\;\; ({\rm fifth\;order\;interpolation})\nonumber \\
&& c_0 = \frac{3}{40},c_1 = \frac{13}{20}, c_2=\frac{11}{40}, \;\;\; ({\rm fourth\;order\;interpolation})\label{eq:27}
\end{eqnarray}
where we use $\epsilon=10^{-40}$.
{The smoothness indicator $IS_s$ in Eq. (\ref{eq:27}) measures the average of the square of first- and second-order derivatives in a cell $[x_{i-1/2},x_{i+1/2}]$, whereas the indicator in \cite{2000JCoPh.165...22D} measures it at a point $x_i$.
Their difference is very small (factor $13/12$ is replaced by 1 in \cite{2000JCoPh.165...22D}) and does not affect numerical simulation results in this paper.}
The accuracy may be further improved by adopting sophisticated weighted functions such as in \cite{2005JCoPh.207..542H}, \cite{2008JCoPh.227.3191B}, \cite{2013JCoPh.232...68H}, and \cite{2014JCoPh.269..329F}.

Figure \ref{fig:adv1d_scalar} shows the numerical solutions of the one-dimensional advection $\partial U / \partial t + \partial U / \partial x = 0$ over ten periods.
We set the initial condition of a combination of rectangular and triangle waves, a Gaussian profile, and a half ellipse, and use 200 grid points and a CFL number of 0.4; {this is the same condition used in \cite{1996JCoPh.126..202J} and \cite{1997JCoPh.136...83S}}.
From top to bottom, we employ a second-order scheme based on MUSCL reconstruction with Monotonized Central flux limiter (MUSCL-MC), the Weighted 3I4D3 scheme (W3I4D3, Eqs. (\ref{eq:28}) and (\ref{eq:23})), and the Weighted 4I4D5 scheme (W4I4D5, Eqs. (\ref{eq:25}) and (\ref{eq:27})).
All schemes effectively suppress numerical oscillation that will arise around discontinuities in linear schemes.
The MUSCL-MC scheme shows an asymmetric profile toward the direction of advection.
The weighted high-order schemes presented here improve numerical dispersion, and then remedy the asymmetry.
The W3I4D3 scheme considerably dissipates a profile extremum like TVD schemes (e.g., MUSCL-MC).
Using the five-point interpolation, the W4I4D5 scheme succeeds in distinguishing the extremum from the discontinuity and preserving it similar to fifth-order schemes \cite[e.g.,][]{1997JCoPh.136...83S}.
Note that the solution of the W4I4D5 scheme is very close to that of the fifth-order WENO scheme (Fig. 4.2. in \cite{1997JCoPh.136...83S}).

\subsection{Characteristic decomposition}\label{sec:char-decomp}
Upwind schemes are developed based on the property of the linear advection equation. 
When one applies the scheme to nonlinear hyperbolic conservation laws, it is more appropriate to interpolate characteristic variables rather than conservative or primitive variables \citep{2002JCoPh.183..187Q,2009SIAMR..51...82S}.
For the primitive variables in MHD, $\vect{V}=(\rho,u,v,w,B_y,B_z,P)^{T}$, the non-conservative form of Eq. (\ref{eq:10})
\begin{eqnarray}
 \pdif{\vect{V}}{t} + \vect{A}_p \pdif{\vect{V}}{x} = 0,\;\;\; \vect{A}_p=\left(\pdif{\vect{U}}{\vect{V}}\right)^{-1} \vect{A} \left(\pdif{\vect{U}}{\vect{V}}\right),\;\;\; \vect{A}=\pdif{\vect{F}}{\vect{U}},\label{eq:29}
\end{eqnarray}
is diagonalized by left and right matrices $\vect{L}$ and $\vect{R}$, and then the equations reduce to nonlinear advection equations for characteristic variables $d \vect{Q}=\vect{L} \cdot d \vect{V}$. The matrices consist of left and right eigenvectors of $k$th characteristic variables $\vect{l}^k$ and $\vect{r}^k$,
\begin{eqnarray}
 \vect{L} = \left(
\begin{array}{c}
 \vect{l}^1\\
 \vect{l}^2\\
 \vect{l}^3\\
 \vect{l}^4\\
 \vect{l}^5\\
 \vect{l}^6\\
 \vect{l}^7\\
\end{array}
\right),
\vect{R} = \left(\vect{r}^1,\vect{r}^2,\vect{r}^3,\vect{r}^4,\vect{r}^5,\vect{r}^6,\vect{r}^7\right),\vect{L}\vect{R}=1.\label{eq:30}
\end{eqnarray}
Using the eigenvectors given by \cite{2008ApJS..178..137S}, the interpolation is carried out by the following steps: (i) $\vect{V}$ is calculated from $\vect{U}$, and then $\vect{L},\vect{R}$ are calculated from $\vect{V}$, (ii) $\vect{V}$ is converted to $\vect{Q}$ in each stencil through $\vect{Q}_{i+s} = \vect{L}_{i} \cdot \vect{V}_{i+s}$ where $s=-r,\dots,r$ covers the stencil, (iii) the characteristic variables at the midpoints, $\vect{\tilde{Q}}^{L}_{i+1/2},\vect{\tilde{Q}}^{R}_{i-1/2}$, are interpolated from $\vect{Q}_{i+s}$, and (iv) $\vect{Q}$ is converted to $\vect{V} $ through $\vect{\tilde{V}}^{L}_{i+1/2} = \vect{R}_i \cdot \vect{\tilde{Q}}^{L}_{i+1/2}$, $\vect{\tilde{V}}^{R}_{i-1/2} = \vect{R}_i \cdot \vect{\tilde{Q}}^{R}_{i-1/2}$, and then $\vect{V}$ to $\vect{U}$.

\subsection{Riemann solver}\label{sec:riemann-solver}
The physical variables are prepared at the midpoints to evaluate the numerical fluxes. 
A common strategy for the calculation of the numerical flux in hyperbolic conservation laws is to solve the Riemann problem for $\vect{\tilde{U}}^{L,R}_{i+1/2}$ as the initial left and right states.
A number of linearized and nonlinear approximate Riemann solvers have been proposed and succeeded in practical MHD simulation studies \citep{1988JCoPh..75..400B,1998ApJS..116..119B,2005JCoPh.203..344L,2005JCoPh.208..315M} .
Among them, we adopt the HLL-type approximate Riemann solvers because of their simplicity, good performance, and widespread acceptance \citep{2011ApJ...737...13K}.
The HLL Riemann solver is one of the simplest solver, in which the solution of the Riemann problem is approximated as a single state bounded by the fastest and slowest waves \citep{1983SIAMrev.25..35M}.
For MHD simulations, we employ the HLLD Riemann solver developed by \cite{2005JCoPh.208..315M} to improve the accuracy of {\Alfven} and entropy waves against the HLL Riemann solver.

\subsection{Multi-dimension}\label{sec:multi-dimension}
The finite difference scheme designed to solve hyperbolic conservation laws in a divergence form is extended to multi-dimension by a dimension-by-dimension fashion.
Applying the scheme to the MHD equations, the induction equation (\ref{eq:3}) may be written into the divergence form, $\partial \vect{B}/ \partial t + \nabla \cdot (\vect{u}\vect{B}-\vect{B}\vect{u}) = 0$.
However, this form does not numerically guarantee $\nabla \cdot \vect{B}=0$ in multi-dimension.
 The error is accumulated, and could finally crash a simulation run.
\cite{1988ApJ...332..659E} proposed the Constrained Transport (CT) method to avoid this problem, in which the induction equation is discretized on staggered grid points and is solved in the curl form to preserve \divb within a machine precision.
The present two-dimensional scheme employs the same grid spacing as the CT method in Figure \ref{fig:grid2d}(a) (extension to three dimension is straightforward).
The in-plane magnetic field and the out-of-plane electric field components are defined at the staggered grid points as pointwise representation (not face-averaged or line-averaged ones),
\begin{eqnarray}
\left\{
\begin{array}{l}
B_{x,i+1/2,j}=B_{x}(x_{i+1/2},y_j),\\
B_{y,i,j+1/2}=B_{y}(x_i,y_{j+1/2}),\\
E_{z,i+1/2,j+1/2}=E_{z}(x_{i+1/2},y_{j+1/2}).
\end{array}
\right.\label{eq:68}
\end{eqnarray}
A conventional two-point difference formula for the induction equation preserves \divb in second-order accuracy.
To ensure \divbzero in the spatial accuracy higher than two, we adopt a similar methodology in \cite{2004JCoPh.195...17L} and \cite{2007A&A...473...11D} to the present scheme.
As they mentioned, we should measure the derivative in \divb by the same algorithm used for the flux derivative.
Therefore, the present scheme measures \divb from the staggered magnetic field as
\begin{eqnarray}
\left(\nabla \cdot \vect{B}\right)_{i,j} = {\cal D}_{4,x} (B_x)_{i,j} + {\cal D}_{4,y} (B_y)_{i,j},\label{eq:33}
\end{eqnarray}
where ${\cal D}_{4,x}$ and ${\cal D}_{4,y}$ stand for the fourth-order difference (Eq. (\ref{eq:13})) in the $x$ and $y$ directions.
It is obvious that the above \divb is preserved in a discretized manner when the magnetic field is updated through
\begin{eqnarray}
 \dif{B_{x,i+1/2,j}}{t} &=& - {\cal D}_{4,y} (E_z)_{i+1/2,j},\label{eq:36}\\
 \dif{B_{y,i,j+1/2}}{t} &=& {\cal D}_{4,x} (E_z)_{i,j+1/2}.\label{eq:37}
\end{eqnarray}
{This is an explicit formula and is essentially same as the method used in \cite{2007A&A...473...11D}.}
On the other hand, the finite difference scheme in \cite{2004JCoPh.195...17L} solves the induction equation for ``primary'' magnetic field $\vect{\hat{B}}$, not $\vect{B}$ itself, which approximates the first derivative as a two-point difference irrespective of the order of accuracy.
Since they are related implicitly, their scheme needs a matrix inversion to obtain $\vect{B}$ from $\vect{\hat{B}}$.
See their Appendix A for detail.

A critical issue is how to calculate the staggered electric field $E_{z,i+1/2,j+1/2}$, being consistent with the solution of the Riemann problem.
A straightforward strategy is to approximate it from the solutions of the one-dimensional Riemann problem at the neighboring midpoints.
The one-dimensional procedure (Sections \ref{sec:nonl-interp}-\ref{sec:riemann-solver}) returns the numerical fluxes in the $x$ and $y$ directions, $\tilde{\vect{F}}_{i+1/2,j}$ and $\bar{\vect{G}}_{i,j+1/2}$, {where the tilde (bar) notation means that a variable is interpolated in the $x$ $(y)$ direction}.
A simple calculation method for the staggered electric field is an arithmetic average proposed by \citet{1999JCoPh.149..270B}, termed as the FluxCT method.
A second-order accurate solution is
\begin{eqnarray}
\bar{\tilde{E}}_{z,i+1/2,j+1/2} &=& \frac{\bar{G}^{B_x}_{i,j+1/2}+\bar{G}^{Bx}_{i+1,j+1/2}-\tilde{F}^{B_y}_{i+1/2,j}-\tilde{F}^{B_y}_{i+1/2,j+1}}{4},\label{eq:38}
\end{eqnarray}
where $\tilde{F}^{B_y}$ and $\bar{G}^{B_x}$ stand for the numerical fluxes of $B_y$ and $B_x$ in the $x$ and $y$ directions, respectively.
A high-order extension of the FluxCT method is provided by \citet{2009JCoPh.228..952L}.

The numerical fluxes obtained from the Riemann solver $F,G$ would be expressed as a sum of the non-dissipative terms ${\cal F},{\cal G}$, and the upwind-correction terms ${\Phi},{\Gamma}$ in the $x$ and $y$ directions that contain numerical diffusion to stabilize a simulation, 
\begin{eqnarray}
\left\{
\begin{array}{l}
 \tilde{F}^{B_y}_{i+1/2,j} = \tilde{\cal F}^{B_y}_{i+1/2,j}+\tilde{\Phi}^{B_y}_{i+1/2,j},\\
 \bar{G}^{B_x}_{i,j+1/2} = \bar{\cal G}^{B_x}_{i,j+1/2}+\bar{\Gamma}^{B_x}_{i,j+1/2}.\label{eq:60}
\end{array}
\right.
\end{eqnarray}
Substituting Eq. (\ref{eq:60}) into Eq. (\ref{eq:38}), one obtain
\begin{eqnarray}
\bar{\tilde{E}}_{z,i+1/2,j+1/2} &=& \frac{\bar{\cal G}^{B_x}_{i,j+1/2}+\bar{\cal G}^{B_x}_{i+1,j+1/2}-\tilde{\cal F}^{B_y}_{i+1/2,j}-\tilde{\cal F}^{B_y}_{i+1/2,j+1}}{4}\nonumber \\
&&-\frac{\tilde{\Phi}^{B_y}_{i+1/2,j}+\tilde{\Phi}^{B_y}_{i+1/2,j+1}}{4}+\frac{\bar{\Gamma}^{B_x}_{i,j+1/2}+\bar{\Gamma}^{B_x}_{i+1,j+1/2}}{4}.\label{eq:54}
\end{eqnarray}
A solution of one-dimensional problems in two dimension with the above two-dimensional method is not equivalent to one with the base one-dimensional scheme.
Following \cite{2005JCoPh.205..509G}, let us consider a one-dimensional problem aligned in the $x$-direction that ignores $j$-dependence, $\bar{\cal G}^{Bx}_{i,j+1/2}= {\cal G}^{Bx}_{i,j}= -{F}^{B_y}_{i,j}$, and the numerical diffusion in the $y$ direction, $\bar{\Gamma}^{B_x}=0$.
Using a piecewise-constant approximation, e.g., $\tilde{\cal F}^{B_y}_{i+1/2}=\left( {F}^{B_y}_{i}+ {F}^{B_y}_{i+1}\right)/2$, Eq. (\ref{eq:54}) becomes
\begin{eqnarray}
{\tilde{E}}_{z,i+1/2} = -\frac{1}{2}\left[\left({F}^{B_y}_{i}+\frac{\tilde{\Phi}^{B_y}_{i+1/2}}{2}\right) + \left({F}^{B_y}_{i+1}+\frac{\tilde{\Phi}^{B_y}_{i+1/2}}{2}\right)\right].\label{eq:65}
\end{eqnarray}
Owing to the arithmetic average in Eq. (\ref{eq:38}), the upwind-correction terms in Eq. (\ref{eq:65}) are halved from the base numerical flux in Eq. (\ref{eq:60}).
The reduction of upwind property might lead to a solution more oscillatory.

{Variants of the FluxCT method have been proposed to take into account for the full upwind property.
\cite{1999JCoPh.149..270B} themselves considered a directional biasing with respect to the gas pressure gradient rather than the arithmetic average in Eq. (\ref{eq:38}).
\citet{2005JCoPh.205..509G} modified Eq. (\ref{eq:38}) by introducing the first-order derivative of the electric field.
With proper approximation of the derivative, their scheme reduces to the base one-dimensional scheme for one-dimensional problems.
\cite{2013JCoPh.243..269L} incorporated an upwind biasing with respect to the entropy mode into the high-order FluxCT method.
}

{
In this paper, we present a simple modification of the high-order FluxCT method so as to be consistent with the base one-dimensional scheme.
We explicitly calculate the non-dissipative terms in the numerical flux of the magnetic field (Eq. (\ref{eq:60})) through a linear interpolation,
\begin{eqnarray}
 \left(B_y u\right)^{L}_{i+1/2,j},\left(B_y u\right)^{R}_{i-1/2,j} &\leftarrow& \left(B_y u\right)_{i-r,j},\dots,\left(B_y u\right)_{i+r,j},\nonumber \\
{v}^{L}_{i+1/2,j},{v}^{R}_{i-1/2,j} &\leftarrow& v_{i-r,j},\dots,v_{i+r,j},\nonumber \\
 \left(B_x v\right)^{B}_{i,j+1/2},\left(B_x v\right)^{T}_{i,j-1/2} &\leftarrow& \left(B_x v\right)_{i,j-r},\dots,\left(B_x v\right)_{i,j+r},\nonumber \\
{u}^{B}_{i,j+1/2},{u}^{T}_{i,j-1/2} &\leftarrow& u_{i,j-r},\dots,u_{i,j+r},\nonumber \\
\tilde{\cal F}^{B_y}_{i+1/2,j} &=& \frac{\left(B_y u\right)^{L}_{i+1/2,j}+\left(B_y u\right)^{R}_{i+1/2,j}}{2}-B_{x,i+1/2,j}\frac{\left( {v}^{L}_{i+1/2,j}+{v}^{R}_{i+1/2,j}\right)}{2},\nonumber \\
\bar{\cal G}^{B_x}_{i,j+1/2} &=& \frac{\left(B_x v\right)^{B}_{i,j+1/2}+\left(B_x v\right)^{T}_{i,j+1/2}}{2}-B_{y,i,j+1/2}\frac{\left({u}^{B}_{i,j+1/2}+{u}^{T}_{i,j+1/2}\right)}{2},\label{eq:53}\nonumber \\
\end{eqnarray}
where the superscripts $B,T$ denote the bottom and top sides at the midpoint, and the left arrow stands for the linear interpolation method (e.g., Eq. (\ref{eq:19})) with the same order of accuracy as the nonlinear interpolation method used before.
The upwind-correction terms are readily obtained from Eq. (\ref{eq:60}).
Subsequently, we interpolate the non-dissipative and upwind-correction terms toward the edge along the orthogonal direction (shown in Figure \ref{fig:grid2d}(b)),
\begin{eqnarray}
 \tilde{\bar{\cal G}}^{B_x,L}_{i+1/2,j+1/2}, \tilde{\bar{\cal G}}^{B_x,R}_{i-1/2,j+1/2}, &\leftarrow& {\bar{\cal G}}^{B_x}_{i-r,j+1/2},\dots,{\bar{\cal G}}^{B_x}_{i+r,j+1/2},\nonumber \\
 \tilde{\bar{\Gamma}}^{B_x,L}_{i+1/2,j+1/2}, \tilde{\bar{\Gamma}}^{B_x,R}_{i-1/2,j+1/2}, &\leftarrow& {\bar{\Gamma}}^{B_x}_{i-r,j+1/2},\dots,{\bar{\Gamma}}^{B_x}_{i+r,j+1/2},\nonumber \\
 \bar{\tilde{\cal F}}^{B_y,B}_{i+1/2,j+1/2}, \bar{\tilde{\cal F}}^{B_y,T}_{i+1/2,j-1/2} &\leftarrow& {\tilde{\cal F}}^{B_y}_{i+1/2,j-r},\dots,{\tilde{\cal F}}^{B_y}_{i+1/2,j+r},\nonumber \\
 \bar{\tilde{\Phi}}^{B_y,B}_{i+1/2,j+1/2}, \bar{\tilde{\Phi}}^{B_y,T}_{i+1/2,j-1/2} &\leftarrow& {\tilde{\Phi}}^{B_y}_{i+1/2,j-r},\dots,{\tilde{\Phi}}^{B_y}_{i+1/2,j+r},\label{eq:61}
\end{eqnarray}
where the interpolation method is same as used in Eq. (\ref{eq:53}).
The staggered electric field is calculated through their arithmetic average,
\begin{eqnarray}
\bar{\tilde{E}}_{z,i+1/2,j+1/2} &=& \frac{\tilde{\bar{\cal G}}^{B_x,L}_{i+1/2,j+1/2}+\tilde{\bar{\cal G}}^{B_x,R}_{i+1/2,j+1/2}-\bar{\tilde{\cal F}}^{B_y,B}_{i+1/2,j+1/2}-\bar{\tilde{\cal F}}^{B_y,T}_{i+1/2,j+1/2} }{4}\nonumber \\
&-&\frac{\bar{\tilde{\Phi}}^{B_y,B}_{i+1/2,j+1/2}+\bar{\tilde{\Phi}}^{B_y,T}_{i+1/2,j+1/2}}{2} + \frac{\tilde{\bar{\Gamma}}^{B_x,L}_{i+1/2,j+1/2}+\tilde{\bar{\Gamma}}^{B_x,R}_{i+1/2,j+1/2}}{2}\label{eq:62}.
\end{eqnarray}
The first term on the right-hand side of Eq. (\ref{eq:62}) corresponds to the non-dissipative approximation of the electric field.
The second and third terms are the upwind-correction terms in the $x$ and $y$ directions, which are twice as large as the FluxCT method in Eq. (\ref{eq:54}) and thus give the same amount of numerical diffusion as the base numerical flux in Eq. (\ref{eq:60}).

\cite{2013JCoPh.243..269L} pointed out that a simple arithmetic average for the staggered electric field might lead to spurious oscillations around discontinuous regions, even though the proper amount of numerical diffusion is recovered like Eq. (\ref{eq:62}).
This is observed, for example, when a two-dimensional object flows along the direction almost parallel to one of the coordinate axes.
In order to deal with this problem, we modify Eq. (\ref{eq:62}) by taking a direction-biased average with respect to the {\Alfven} mode,
\begin{eqnarray}
 \bar{\tilde{E}}_{z,i+1/2,j+1/2} &=& \Theta_{i+1/2,j+1/2} \frac{\tilde{\bar{\cal G}}^{B_x,L}_{i+1/2,j+1/2}+\tilde{\bar{\cal G}}^{B_x,R}_{i+1/2,j+1/2}}{2}\nonumber \\
&-& \left(1-\Theta_{i+1/2,j+1/2}\right)\frac{\bar{\tilde{\cal F}}^{B_y,B}_{i+1/2,j+1/2}+\bar{\tilde{\cal F}}^{B_y,T}_{i+1/2,j+1/2} }{2}\nonumber \\
&-&\frac{\bar{\tilde{\Phi}}^{B_y,B}_{i+1/2,j+1/2}+\bar{\tilde{\Phi}}^{B_y,T}_{i+1/2,j+1/2}}{2} + \frac{\tilde{\bar{\Gamma}}^{B_x,L}_{i+1/2,j+1/2}+\tilde{\bar{\Gamma}}^{B_x,R}_{i+1/2,j+1/2}}{2},\label{eq:63}
\end{eqnarray}
where
\begin{eqnarray}
 \Theta = \frac{|u|+|B_x|/\mart{\rho}+\delta}{|u|+|v|+(|B_x|+|B_y|)/\mart{\rho}+2\delta},\label{eq:64}
\end{eqnarray}
 and $\delta=10^{-6}$ is a small positive number to avoid the denominator becoming zero.
We take a second-order arithmetic average for the primitive variables to calculate Eq. (\ref{eq:64}) at the edge for simplicity, e.g., $\rho_{i+1/2,j+1/2}=(\rho_{i,j}+\rho_{i+1,j}+\rho_{i,j+1}+\rho_{i+1,j+1})/4$.

The present method has two main differences from the previous CT variants proposed by \cite{2005JCoPh.205..509G} and \cite{2013JCoPh.243..269L}:
the method is able to extend to an arbitrary high order of accuracy whereas the order of accuracy is limited to be two in \cite{2005JCoPh.205..509G} (see also \cite{2018JCoPh.375.1365F}), and the method reduces to the base one-dimensional scheme for one-dimensional problems whereas it is not always satisfied in \cite{2013JCoPh.243..269L}.
Hereafter, we refer to Eq. (\ref{eq:63}) as the Central Upwind CT (CUCT) method.
}

\subsection{Summary of numerical procedure}\label{sec:summ-numer-proc}
Here, we summarize the numerical procedure for the present scheme. Initially, we define the fluid conservative variables $(\rho,\rho u,\rho v,\rho w,e)$ at each node $(i,j,k)$, and the staggered magnetic field at the midpoint $(B_{x,i+1/2,j,k},B_{y,i,j+1/2,k},B_{z,i,j,k+1/2})$ as pointwise representation.
The magnetic field satisfies the divergence-free condition in a discretized manner, $(\nabla \cdot \vect{B}) = {\cal D}_{4,x}(B_x)+{\cal D}_{4,y}(B_y) + {\cal D}_{4,z}(B_z) = 0$.
The magnetic field is calculated at the node from the midpoints by a central interpolation with the same order of accuracy of the scheme. 
For example, the fourth-order interpolation gives \cite[e.g.,][]{1992JCoPh.103...16L}
\begin{eqnarray}
B_{x,i} = [9(B_{x,i-1/2}+B_{x,i+1/2})-(B_{x,i-3/2}+B_{x,i+3/2})]/16,\label{eq:51}
\end{eqnarray}
which can improve the accuracy of $B_{x,i}$ against the two-point average.
Then, we perform the following steps in each substep of the third-order SSPRK time integration (Eq. (\ref{eq:22})).
\begin{enumerate}
\item  Primitive variables $\vect{V}=(\rho,u,v,w,B_x,B_y,B_z,P)^{T}$ are calculated from the conservative variables $\vect{U}$ at each node.
\item Characteristic variables $\vect{Q}_{i,j,k}$ are converted from the primitive variables in the stencil $(i-r,\dots ,i+r)$ along the $x$ direction (Section \ref{sec:char-decomp}). 
They are interpolated to the left and right sides at the right and left midpoints by the selected method (Section \ref{sec:nonl-interp}), $\vect{\tilde{Q}}^{L}_{i+1/2,j,k}$ and $\vect{\tilde{Q}}^{R}_{i-1/2,j,k}$, and then are converted to the primitive variables.
\item The approximate HLLD Riemann solver calculates the numerical flux in the $x$ direction, $\tilde{\vect{F}}_{i+1/2,j,k}$ (Section \ref{sec:riemann-solver}).
\item The above one-dimensional procedure (Step 2-3) is performed along the $y$ and $z$ directions.
\item The staggered electric field $E_{z,i+1/2,j+1/2,k}$ is calculated by the selected CT method (Section \ref{sec:multi-dimension}).
The $x$ and $y$ components, $E_{x,i,j+1/2,k+1/2}$ and $E_{y,i+1/2,j,k+1/2},$ are obtained in the same manner.
\item The flux derivative is approximated by the fourth-order central difference (Eq. (\ref{eq:13})), and then the fluid conservative variables are updated in the divergence form (Eqs. (\ref{eq:1}), (\ref{eq:2}), (\ref{eq:4})).
The staggered magnetic field is updated in the curl form (Eq. (\ref{eq:3})), and then it is interpolated to the node by Eq. (\ref{eq:51}).
\end{enumerate}

A higher-order interpolation method tends to make a solution more oscillatory that might lead to negative density or pressure and crash a simulation run.
{In order to suppress the numerical oscillation around shocks, we employ a lower-order method (e.g., the third-order method in Eqs. (\ref{eq:28}) and (\ref{eq:23}) as opposed to the fifth-order method in Eqs. (\ref{eq:25}) and (\ref{eq:27})) to the left and right state interpolations at the right and left midpoints (e.g., $\vect{\tilde{Q}}^L_{i+1/2,j,k}$ and $\vect{\bar{Q}}^R_{i-1/2,j,k}$), when a node $(i,j,k)$ is judged as a shocked-node based on $(\nabla \cdot \vect{u})_{i,j,k} < 0$.
We apply this hybrid method to the Orszag-Tang vortex problem in Section \ref{sec:orszag-tang-vortex} and a blast wave problem in Section \ref{sec:blast-wave-strongly}.}

\section{Numerical tests}\label{sec:numerical-tests}
This section presents numerical simulation results of MHD test problems including propagation, shock, and fundamental plasma phenomena to discuss the capability of the present scheme.
In the following benchmark tests, we employ the W4I4D5 scheme in Eqs. (\ref{eq:25}) and (\ref{eq:27}), the HLLD Riemann solver, and the CUCT method in Eq. (\ref{eq:63}) for the divergence-free treatment as a fiducial scheme, referred to as the W4I4D5-HLLD-CUCT scheme.
In Sec. \ref{sec:field-loop-advection} and \ref{sec:magn-reconn}, we also use the CUCT-AA method that adopts the arithmetic average in Eq. (\ref{eq:62}) as opposed to the direction-biased average in Eq. (\ref{eq:63}).
We solve two- or three-dimensional ideal MHD equations with a specific heat ratio $\gamma = 5/3$ and a CFL number of 0.4 unless otherwise stated.

\subsection{Circularly polarized {\Alfven} waves}\label{sec:circ-polar-alfv}
The circularly polarized {\Alfven} wave is utilized to check the accuracy of the designed code since the wave is an analytic solution for the MHD equations.
Improving the accuracy of the {\Alfven} wave solution is of great importance to tackle with MHD turbulent flows.
We test a problem similar to \cite{2000JCoPh.161..605T}. The two-dimensional computational domain is periodic with $0 \leq x<1/\cos\alpha$ and $0 \leq y<1/\sin\alpha$, where $\alpha=30^{\circ}$ is the wave propagation angle relative to the $x$ axis.
The domain is resolved by $N \times N$ grid points and the grid number increases as $N=16,32,64,128,256,512$.
The initial condition is
\begin{eqnarray}
\left[
\begin{array}{c}
\rho \\
u_{\parallel} \\
u_{\perp} \\
w \\
B_{\parallel} \\
B_{\perp} \\
B_{z} \\
P \\
\end{array}
\right] = \left[
\begin{array}{c}
1 \\
0 \\
\delta V \sin(2\pi x_{\parallel}) \\
\delta V \cos(2\pi x_{\parallel}) \\
1 \\
\delta B \sin(2\pi x_{\parallel}) \\
\delta B \cos(2\pi x_{\parallel}) \\
0.05 \\
\end{array}
\right],
\end{eqnarray}
where the subscripts $\parallel,\perp$ stand for the parallel and perpendicular directions relative to the wave propagation.
They are related to the variables in an original coordinate system through $x_{\parallel}=x \cos\alpha+y \sin \alpha,(u,B_x)=(u_{\parallel},B_{\parallel})\cos\alpha-(u_{\perp},B_{\perp})\sin\alpha,$ and $(v,B_y)=(u_{\parallel},B_{\parallel})\sin\alpha+(u_{\perp},B_{\perp})\cos\alpha$.
The wave amplitude is $\delta V=\delta B=0.01$ so that the wave propagates in negative direction with the {\Alfven} speed $V_A=B_{\parallel}/\mart{\rho}=1$.
The simulations are conducted by five schemes with different interpolation methods, the first order, MUSCL-MC, W3I4D3, W4I4D5, and W5I4D4.
The first-order Euler time integration is employed for the first-order scheme, the second-order SSPRK time integration for the MUSCL-MC scheme, and the third-order SSPRK time integration for the remaining schemes.
In order to assess the numerical error attributed to the spatial discretization, the CFL number is changed as ${\rm CFL}= 16\mart 3/5N$ $(\Delta t \propto \Delta x^2)$ so that the temporal discretization errors in the five schemes are negligible compared to the spatial discretization errors.

We measure numerical errors of the in-plane and out-of-plane magnetic fields, $B_{\perp}$ and $B_{z}$, at $t=2$.
Table \ref{tab:alf_errors} summarizes the $L_1$ errors and the orders in parenthesis. 
Except for the W3I4D3 scheme, the four schemes converge at the designed order of accuracy.
The W3I4D3 scheme is slightly better than the MUSCL-MC scheme, but it does not achieve the optimal third order of accuracy even with the finest grid points.
The nonlinear weight for this scheme (Eq. (\ref{eq:23})) may not converge to the optimal weight, even though the unit wavelength is resolved by 512 grid points.
The W3I4D3 scheme requires 1.5 times longer computational time (due to the order of the Runge-Kutta method) and a larger stencil than the MUSCL-MC scheme.

The errors in the W4I4D5 and W5I4D4 schemes are remarkably improved from the three lower-order schemes.
The W4I4D5 scheme achieves fifth order of accuracy in the out-of-plane $(B_{z})$ case whereas W5I4D4 scheme converges at fourth order in both the in-plane $(B_{\perp})$ and out-of-plane cases, because the W4I4D5 scheme is designed so as to satisfy the fifth order for the linear advection problem (Eq. (\ref{eq:20})).
In the in-plane case, the scheme calculates the staggered electric field by averaging the fourth-order accurate solutions of the one-dimensional Riemann solver in the two-dimensional plane (Eq. (\ref{eq:63})).
This limits the order of accuracy of the W4I4D5 scheme to be four in the in-plane case (and in a three-dimensional case as well).
Nevertheless, the error in the W4I4D5 scheme is slightly smaller than the W5I4D4 scheme at least in this problem.
The computational times of the two schemes are almost equal, and about two times longer than the MUSCL-MC scheme.

\subsection{Kelvin-Helmholtz instability}\label{sec:kelv-helmh-inst}
The boundary layer is a common feature in a variety of plasma environments.
For example, the magnetopause in the planetary magnetosphere is a boundary between the magnetospheric and the solar wind plasmas.
The Kelvin-Helmholtz instability is thought to contribute to the transport of the tailward-flowing solar wind plasma into the Earth magnetosphere \cite[e.g.,][]{2004Natur.430..755H}.
{We simulate the linear growth of the Kelvin-Helmholtz instability so as to assess the capability to resolve a shear flow.
The two-dimensional computational domain ranging $0 \leq x<L$ and $-L/2 \leq y<L/2 \; (L=14.0)$ is resolved by $N \times N$ grid points.
The boundary condition is  periodic and symmetric in the $x$ and $y$ directions.
The initial condition has a velocity shear, $u=-(V_0/2)\tanh(y/\lambda)$, uniform density and pressure $\rho=\rho_0=1.0,P=P_0=500$, and a uniform out-of-plane magnetic field $B_z=B_0=1.0$, where $\lambda=1.0$ and $V_0=1.0$ is normalized by the {\Alfven} speed.
{We use $\gamma=2.0$.}
The system is essentially hydrodynamic because the in-plane magnetic field is kept zero.
To initiate the instability, we impose a small $(1\%)$ perturbation to the $y$-component of the velocity around the boundary with a wavelength equal to $L$, which corresponds to the fastest growing mode under the adopted initial condition \citep{1982JGR....87.7431M}.
The simulations are conducted by the MUSCL-MC and the W4I4D5 schemes.
}

{
Figure \ref{fig:kh_lgrowth} shows the linear growth of the fastest growing mode from the simulations and the theory. 
The theoretical growth rate, $\omega_{i}=0.095V_0/\lambda$, is obtained by numerically solving {the linearized equations for the velocity and the total pressure $P_t = P+B_z^2/2$ as an eigenvalue problem,
\begin{eqnarray}
\omega_{i} u &=& -i k_x \frac{V_0}{2}\tanh\left(\frac{y}{\lambda}\right) u - \frac{V_0}{2 \cosh^2(y/\lambda)}v - \frac{i k_x}{\rho_0}P_t,\nonumber \\
\omega_{i} v &=& -i k_x \frac{V_0}{2}\tanh\left(\frac{y}{\lambda}\right) v - \frac{1}{\rho_0} \pdif{P_t}{y},\nonumber \\
\omega_{i} P_t &=& -i k_x \frac{V_0}{2}\tanh\left(\frac{y}{\lambda}\right) P_t - 2P_{t0} \left(i k_x u + \pdif{v}{y}\right),\label{eq:69}
\end{eqnarray}
where $P_{t0} = P_0+B_0^2/2=500.5$, $k_x=2 \pi/L$, and $i$ is the imaginary unit.
} 
The W4I4D5 scheme converges to the theory at resolutions $N \geq 64$, while the MUSCL-MC scheme requires $N=128$ for convergence.
The computational time of the W4I4D5 scheme with $N=64$ is about 4 times faster than the MUSCL-MC scheme with $N=128$.
This problem demonstrates that the present high-order scheme is highly cost effective for the shear flow against the low-order scheme.

}

\subsection{Orszag-Tang vortex}\label{sec:orszag-tang-vortex}
{
We conduct the so-called Orszag-Tang vortex problem, which is a well-known two-dimensional MHD test so as to verify the capability to capture multiple interactions of shock waves and vortices \citep{1979JFM....90..129O}.
The periodic computational domain with $0<x<2\pi$ and $0<y<2\pi$ is resolved by $N\times N$ grid points.
The initial condition is $(\rho,u,v,w,B_x,B_y,B_z,P)=(\gamma^2,-\sin(y),\sin(x),0,-\sin(y),\sin(2x),0,\gamma)$.
{We solve the problem with the MUSCL-MC scheme, the hybrid W3I4D3-W4I4D5 scheme (see Section \ref{sec:summ-numer-proc}), and the W4I4D5 scheme at the resolution of $N=100,200,400,800$.}
The solution with the W4I4D5 scheme at $N=1600$ is employed as a reference.

Figure \ref{fig:ot_vortex} (a)-(d) compares the two-dimensional profile of the temperature $T=2P/\rho$ at $t=\pi$ obtained from the three schemes with $N=200$ and the reference solution.
The overall structure, such as cold and hot regions and the position of discontinuities, agrees well with previous works \citep[e.g., Fig. 12 in ][]{2005JCoPh.208..315M}.
The three schemes give almost identical solutions at this period, implying that the spatial order of accuracy is degraded and the quality of solution less depends on the designed optimal order of accuracy when the discontinuities become dominant.
{Panels (e) and (f) compare the one-dimensional profiles at $y=0.64\pi$ and $y=\pi$ among different resolutions with the W4I4D5 scheme.
Compared to the reference, the profile at $N=100$ (green cross) is degraded especially around spike regions (e.g., $(x,y)=(0.4,0.64\pi)$ in panel (e)), and shows a spurious oscillation around $(x,y)=(0,\pi)$ in panel (f).
These errors are found also in the MUSCL-MC and hybrid W3I4D3-W4I4D5 schemes, and are largely improved at resolutions $N\geq 200$.}

We measure relative differences of the temperature profile at different resolutions from the reference one $T_{\rm ref}$, defined as $\delta T = \sum |T-T_{\rm ref}|/\sum T_{\rm ref}$. 
Table \ref{tab:ot_table} summarizes the relative difference and the rate of convergence in parenthesis at two different periods $t=0.2\pi$ and $t=\pi$.
The MUSCL-MC and W4I4D5 schemes converge at the designed order of accuracy when the profile is smooth at $t=0.2\pi$.
{In contrast, the rate of convergence of the hybrid W3I4D3-W4I4D5 scheme is worse than the optimal fourth order at $t=0.2\pi$.
As is presented in Section \ref{sec:summ-numer-proc}, this hybrid method adopts the low-order interpolation when the flow is divergent irrespective of the degree of compressibility: this is a crude criterion.
Actually, the flow is compressible in $38 \%$ of the domain at this period that would limit the overall convergence.
To sophisticate the method, for example, we may modify the criterion based on the strength of shocks.}
When the discontinuities are dominant at $t=\pi$ in Figure \ref{fig:ot_vortex}, their relative differences are close and convergence rate drops to less than two.}


\subsection{Oblique Shock}\label{sec:oblique-shock}
The shock tube problem is a standard test for hydrodynamic simulations so as to examine the accuracy and robustness for discontinuities and rarefaction waves.
In addition, divergence-free preservation is a critical issue for multidimensional MHD shocks since it tends to be violated especially at shocks in non-divergence-free schemes.
We test the familiar Brio-Wu shock tube problem \citep{1988JCoPh..75..400B} in two dimensions.
The initial left and right states are $[\rho,u_{\parallel},u_{\perp},w,B_{\perp},B_{z},P]=[1,0,0,0,1,0,1]$ and $[0.125,0,0,0,-1,0,0.1]$, where the subscripts have the same meaning in Section \ref{sec:circ-polar-alfv}.
We use  $B_{\parallel}=0.75$ and $\gamma=2.0$.
The shock tube is tilted by angles $\alpha=\tan^{-1}(1/2)=26.6^{\circ}$ and $45^{\circ}$ relative to the $x$ axis so as to assess the divergence-free condition.
The domain of the shock tube $0 \leq x<\cos\alpha$ and $0 \leq y<\sin\alpha$ is resolved by $N \times N/4$ (for $\alpha=26.6^{\circ}$) and $N \times N$ (for $\alpha=45^{\circ}$) grid points where $N=200$.
In order to apply the periodic boundary condition, the computational domain is extended to $0 \leq x<2/\cos\alpha$ and $0 \leq y<2/\sin\alpha$, and the whole domain is resolved by $mN \times mN$ grid points where $m=2.5$ for $\alpha=26.6^{\circ}$ and $4$ for $\alpha=45^{\circ}$ \citep{2013JCoPh.251..292K}.
The initial condition is set to be the left state in $0 \leq x_{\parallel}<0.5,1.5 \leq x_{\parallel}<2.5,3.5 \leq x_{\parallel}<4$ ($x_{\parallel}=x \cos\alpha+y \sin \alpha$), and the right state otherwise.
The initial jump condition is given as a hyperbolic tangent function with a finite thickness of $0.5 \Delta x / \cos\alpha$.

Figure \ref{fig:bwrshock_uct} shows the primitive variables at angles of $\alpha=26.6^{\circ}$  (blue) and $45^{\circ}$ (red) solved by the HLLD-CUCT scheme at $t=0.1$. 
The both cases capture the expected discontinuities and rarefaction waves without numerical instabilities (from left to right; the fast rarefaction wave, the slow compound wave, the contact discontinuity, the slow shock, and the fast rarefaction wave), and agree well with the one-dimensional reference solution with 2000 grid points (black).
Note that the two-dimensional scheme for $\alpha = 0^{\circ}$ reduces to the base one-dimensional scheme.
The bottom left panel shows the deviation of the parallel magnetic field component from the initial value, $\Delta B_{\parallel}$.
The parallel magnetic field is preserved extremely well within an error of $\sim 10^{-8}$ at this period.
We find that the \divb error is preserved within an error of $\sim 10^{-11}$ (not shown here).

In order to compare the divergence-free and the non-divergence-free schemes, we combine the HLLD scheme with the hyperbolic divergence cleaning method \citep[GLM;][]{2002JCoPh.175..645D}.
This scheme introduces a scalar variable $\psi$ to diminish the \divb error from a simulation domain through propagation and diffusion.
The in-plane magnetic field and $\psi$ obey the following equations,
\begin{eqnarray}
\pdif{\vect{B}}{t} + \nabla \times \vect{E} + \nabla \psi = 0,\label{eq:66}\\
\pdif{\psi}{t} + c_{h}^2 \nabla \cdot \vect{B} = -\frac{c_h^2}{c_p^2} \psi,\label{eq:67}
\end{eqnarray}
where $c_h$ is the propagation speed of the additional eigenmode and $c_p^2$ is the diffusion coefficient.
In Figure \ref{fig:bwrshock_glm}, this scheme also captures discontinuities and rarefaction waves as the HLLD-CUCT scheme.
The main difference is the parallel magnetic field (bottom left panel) with an error of $\sim 1 \times 10^{-2}$ mainly at the front of the compound shock and the slow shock. 
The error comes from the equation of the parallel magnetic field,
\begin{eqnarray}
\pdif{B_{\parallel}}{t} &=&  \pdif{B_x}{t} \cos\alpha + \pdif{B_y}{t} \sin\alpha\nonumber \\
&=& \left(-\pdif{E_z}{y}-\pdif{\psi}{x}\right) \cos\alpha + \left(\pdif{E_z}{x}-\pdif{\psi}{y}\right) \sin \alpha.\label{eq:48}
\end{eqnarray}
The scalar variable $\psi$ may have a finite value at $\nabla \cdot \vect{B} \neq 0$.
Since the grid spacing $(\Delta x,\Delta y) = (2/mN\cos\alpha,2/mN\sin\alpha)$ is set so that the shock surface is uniform and parallel to the diagonal axis in the cell $[x_{i-1/2},x_{i+1/2}] \times [y_{j-1/2},y_{j+1/2}]$, the variables are the left state at $(i-1/2,j),(i,j-1/2)$ and the right state at $(i+1/2,j),(i,j+1/2)$.
Therefore, Eq. (\ref{eq:48}) is discretized as
\begin{eqnarray}
\pdif{B_{\parallel}}{t} &=&  \left(\frac{E^{L}_{z}-E^{R}_{z}}{\Delta y} + \frac{\psi^L-\psi^R}{\Delta x}\right)\cos\alpha + \left(\frac{E^{R}_{z}-E^{L}_{z}}{\Delta x} + \frac{\psi^L-\psi^R}{\Delta y}\right)\sin\alpha\nonumber \\
&=&mN\left(\psi^L-\psi^R\right).\label{eq:46}
\end{eqnarray}
It may be finite around discontinuities.
In Appendix \ref{sec:c_h-dependence-brio}, we examine the dependence of the solution on the parameter in the hyperbolic divergence cleaning method ($c_h$ and $c_p$).

\subsection{Blast wave in strongly magnetized medium}\label{sec:blast-wave-strongly}
This test problem treats the propagation of strong MHD shocks in a two-dimensional strongly magnetized medium so as to assess the robustness of a scheme.
Many similar problems have been tested as a simple example of astrophysical phenomena \citep{2000ApJ...530..508L,2004ApJ...602.1079B,2007ApJS..170..228M,2008ApJS..178..137S}.
The periodic computational domain with $-2<x<2$ and $-2<y<2$ is resolved by $1024\times1024$ grid points.
We adopt the initial condition used in \cite{2000ApJ...530..508L}: the ambient medium $(\rho,u,v,w,B_x,B_y,B_z,P)=(1,0,0,0,B_0 \sin(\theta),B_0 \cos(\theta),0,1)$ where $B_0=10$, $\theta=45^{\circ}$, and then a high pressure cylinder imposed at the center of the domain, $P=100$ for $\mart{x^2+y^2} < 0.125$.
The plasma beta in the ambient medium is very low, $\beta=2P/|\vect{B}|^2=0.02$, thus this is a stringent condition for conservative schemes.
We compare the CUCT method with the GLM method (Eqs. (\ref{eq:66}) and (\ref{eq:67})) and the high-order FluxCT method (Eq. (\ref{eq:62}) with replacing the denominator in the second and third terms on the right-hand side by 4), so as to examine the effects of the finite \divb error and the amount of numerical diffusion for the CT method on the quality of numerical solution in the presence of strong low-$\beta$ shocks.
{We use the hybrid W3I4D3-W4I4D5 scheme (Sec. \ref{sec:summ-numer-proc}), because the W4I4D5 scheme is found not to be enough to suppress the numerical oscillation in this stringent problem irrespective of the divergence-free treatment. }

Figure \ref{fig:blast_summary} (a)-(c) shows the profile of the magnetic energy $|\vect{B}|^2/2$ at $t=0.02$ for the HLLD-GLM, the HLLD-FluxCT, and the HLLD-CUCT schemes.
The three schemes capture complex shocks and rarefaction waves without numerical instabilities, and they are in good agreement with the previous work (Fig. 13 in \cite{2000ApJ...530..508L}), indicating that the present schemes well preserve the axial symmetry even though the ambient magnetic field is oblique relative to the coordinate axis.
At this period, the three schemes give the similar result.

The difference among the schemes becomes prominent as the system evolves.
Figure \ref{fig:blast_summary} (d)-(f) shows the gas pressure profile at $t=0.1$ for the three schemes.
The gas pressure is susceptible to a numerical error of the magnetic field especially at low-$\beta$ shock regions, since the present schemes calculate it by subtracting the magnetic energy from the total energy density (Eq. (\ref{eq:6})).
The HLLD-GLM scheme (d) shows spurious profiles at fast shock fronts.
In panel (g), we check the one-dimensional profile along the outermost fast shock front (denoted by dashed lines in panels (d)-(f)).
The HLLD-GLM scheme suffers from the oscillation in both the downstream and upstream regions of the shock, at which a \divb error of $O(1)$ is observed.
The oscillatory profile in the gas pressure is attributed to the finite \divb production at the shock and its propagation through the GLM method.
The oscillation is seen at the downstream region also in the two CT schemes, but its amplitude is reduced especially in the HLLD-CUCT scheme compared to the HLLD-GLM scheme.
We then suggest that the CT methods are better suited for low-$\beta$ shocks than the GLM method.

The HLLD-FluxCT scheme in Figure \ref{fig:blast_summary} (e) shows spurious profiles at the slow shock front (along the dot-dashed line).
In Figure \ref{fig:blast_summary} (h), we check the one-dimensional profile along this shock front.
The HLLD-FluxCT scheme suffers from an intolerable level of numerical oscillation that leads to a negative pressure and finally crashes the simulation.
The HLLD-CUCT scheme largely suppresses this oscillation due to the recovery of the proper amount of numerical diffusion for the induction equation against the HLLD-FluxCT scheme.
The \divb error in the HLLD-GLM scheme is relatively small, $O(10^{-1})$, compared to that at the fast shock, thus it would give a less oscillatory solution as the HLLD-CUCT scheme.
This problem demonstrates that the CUCT method can improve the robustness of the divergence-free MHD scheme for low-$\beta$ flows.

\subsection{Field loop advection}\label{sec:field-loop-advection}
We conduct a stringent problem of the advection of a weakly magnetized loop proposed by \citet{2005JCoPh.205..509G}, so as to assess the capability to capture a tangential discontinuity in a multidimensional flow.
The two-dimensional computational domain is periodic with $-1 \leq x<1$ and $-0.5 \leq y<0.5$, and is resolved by $256\times128$ grid points.
The initial condition is 
\begin{eqnarray}
 \left[
\begin{array}{c}
\rho \\
u \\
v \\
w \\
B_x \\
B_y \\
B_z \\
P \\
\end{array}
\right] = \left[
\begin{array}{c}
1 \\
v_0 \cos\theta \\
v_0 \sin\theta \\
v_0 \\
\partial A(r)/\partial y \\
-\partial A(r)/\partial x \\
0 \\
P_0 \\ 
\end{array}
\right], \;\;\; A(r) = \left\{
\begin{array}{l}
A_0\left(R-r\right) \;\;\; {\rm for} \;\;\; r \leq R,\\
0 \;\;\; {\rm otherwise},
\end{array}
\right.
\end{eqnarray}
where $v_0=\mart 5,A_0=10^{-3},R=0.3,r=\mart{x^2+y^2}$.
The in-plane magnetic field is initialized by numerically calculating the curl of the vector potential through Eq. (\ref{eq:13}) so as to satisfy \divbzero in a discretized form (Eq. (\ref{eq:33})).
The ambient gas pressure $P_0$ is set to be much higher than the magnetic pressure in the loop so that the loop behaves almost like a passive scalar.
The loop flows toward the positive direction with an angle $\theta=\tan^{-1}(1/2)$ relative to the $x$ axis.
The simulations are conducted by three different choices of the Riemann solver and the divergence-free treatment, the HLLD-GLM, the HLL-CUCT, and the HLLD-CUCT schemes.

Figure \ref{fig:floop_p=1e0} shows the magnetic pressure $|\vect{B}|^2/2$ for the case of $P_0=1$ at $t=8$.
The in-plane flow is supersonic, $v_0 > c_s=\mart{\gamma P/\rho}=1.3$.
All schemes successfully preserve the initial symmetric loop structure.
Numerical errors are mainly observed around the leading edge of the loop at $(x,y) = (0.2,0.2)$.
One-dimensional profile at $x=0.2$ in panel (d) shows that the HLLD-CUCT scheme slightly distorts the profile at $y=0.15$.
The thickness of the leading edge is nearly equal among the three schemes, but it is broadened at the side edge of the loop $(x,y) = (0.2,-0.2)$ in the HLLD-GLM scheme.

\cite{2008JCoPh.227.4123G} and \citet{2009JCoPh.228..952L} argued that the in-plane geometry could be distorted by the \divb error through $\partial B_z / \partial t \propto w \nabla \cdot \vect{B}$.
The HLL-CUCT and HLLD-CUCT schemes keep \divb within $\sim 10^{-14}$, and the amplitude of the out-of-plane magnetic field within $\sim 10^{-10}$.
On the other hand, the \divb error and the amplitude of the out-of-plane magnetic field are around $1\times 10^{-4}$ and $2\times10^{-5}$ (2 \% of the in-plane magnetic field) at $t=32$, and continue to increase in the HLLD-GLM scheme.

The problem is more stringent for a higher plasma beta (weaker compressibility) case.
Figure \ref{fig:floop_p=1e4} shows the case of $P_0=10^4$ so that the in-plane flow is subsonic $(v_0 \ll c_s=130)$.
The loop structure is seriously distorted in the HLLD-GLM and the HLL-CUCT schemes owing to the overestimation of the amount of numerical diffusion.
The time scale of the numerical diffusion is proportional to the transit time of the fast mode wave in the HLL-CUCT scheme \citep{2015ApJ...808..54M}.
Therefore, the magnetic loop quickly diffuses outward and inward in the course of the flow, and the inward diffusion leads to the cancellation of the anti-parallel field at the center of the loop.
As a result, the dilute ring-shaped profile is observed in the HLL-CUCT scheme (Figure \ref{fig:floop_p=1e4} (b)).

{The unphysical solution in the HLLD-GLM scheme (Figure \ref{fig:floop_p=1e4} (a)) is attributed to the numerical flux for the additional equations (Eqs. (\ref{eq:66}) and (\ref{eq:67})), which is evaluated by the Lax-Friedrichs flux splitting method (Eq. (42) in \cite{2002JCoPh.175..645D}).
The method introduces the numerical diffusion of $B_x$ $(B_y)$ in the $x$ $(y)$ direction, and its coefficient is proportional to $c_h$.}
The propagation speed $c_h$ is determined from the CFL condition so as to quickly diminish the \divb error, and thus it is much faster than the flow speed in this problem that leads to rapid numerical diffusion of $B_x$ in the $x$ direction observed as the horizontal structure.
Similarly, the vertical structure corresponds to the numerical diffusion of $B_y$ in the $y$ direction.
To avoid this unphysical solution, one has to carefully switch the Lax-Friedrichs flux splitting to a simple average so that the spatial derivatives in Eqs. (\ref{eq:66}) and (\ref{eq:67}) reduce to the central difference.

In contrast to the above two schemes, the HLLD-CUCT scheme (Figure \ref{fig:floop_p=1e4} (c)) can preserve the symmetric loop structure.
The thickness of the discontinuity is nearly unchanged from the former case with $P_0=1$ (see panel (d) in Figures \ref{fig:floop_p=1e0} and \ref{fig:floop_p=1e4}).
Although the solutions of the HLLD-GLM and the HLL-CUCT schemes are slightly better in the supersonic case (Figure \ref{fig:floop_p=1e0}), the HLLD-CUCT scheme has a capability to capture the tangential discontinuity in the multidimensional flow irrespective of the plasma beta (compressibility) by virtue of the combination of the multi-state Riemann solver and the divergence-free method.

{Additionally, we conduct a small angle advection problem for a three-dimensional loop proposed by \cite{2013JCoPh.243..269L}.
The three-dimensional periodic computational domain with $-0.5\le x<0.5,-0.5 \le y<0.5$, and $-1 \le z<1$ is resolved by $100 \times 100 \times 200$ grid points.
We adopt the initial condition of the magnetic field used in \cite{2008JCoPh.227.4123G}:
the loop is tilted around the $y$ axis by an angle of $\omega=\tan^{-1} (1/2)$, and its magnetic field components are described by the curl of the vector potential in the coordinate system $(x_1,x_2,x_3)=(x \cos \omega + z \sin \omega, y, -x \sin \omega + z \cos \omega)$,
\begin{eqnarray}
&& B_1 = \pdif{A(r)}{x_2}, \; B_2 = -\pdif{A(r)}{x_1}, \; B_3=0, \nonumber \\
&& B_x = B_1 \cos \omega-B_3 \sin \omega, \; B_y = B_2, \; B_z = B_1 \sin \omega + B_3 \cos \omega,
\end{eqnarray}
where $r=\mart{x_1^2+x_2^2}$.
For fluid variables, we set $\rho=P=1$, $(u,v,w)=(\cos\theta, \sin\theta, 2)$ and $\theta=\tan^{-1}(0.01)$ so that the loop flows almost in the $x$-$z$ plane.
Figure \ref{fig:3dloop_p=1e0_smallangle} compares the result at $t=1.0$ with two different CUCT methods; (a) the arithmetic average (Eq. (\ref{eq:62}), the HLLD-CUCT-AA scheme), and (b) the direction-biased average (Eq.~(\ref{eq:63}), the HLLD-CUCT scheme).
Obviously, the HLLD-CUCT-AA scheme suffers from the numerical oscillations in the $x$-$y$ and $y$-$z$ planes that arise at the side edge of the loop and then propagate inward.
As is pointed out by \cite{2013JCoPh.243..269L}, this is because the scheme takes the electric field average across the discontinuity in the $y$ direction (first term in Eq. (\ref{eq:62})), even though the numerical diffusion is biased in the flow direction ($x$ or $z$).
The oscillation is largely suppressed by the HLLD-CUCT scheme, because the scheme biases the electric field with respect to the direction of the {\Alfven} mode in each two-dimensional plane.
}

\subsection{Richtmyer-Meshkov instability}\label{sec:richtmy-meshk-inst}
The impact of a shock against a corrugated contact discontinuity can induce a shear flow at the interface and drive the so-called Richtmyer-Meshkov instability \citep[RMI,][]{1960CPAM....13..297R,1972FlDy....4..101M}.
Such a situation is expected in astrophysical environments, for example, the propagation of a supernova shock in the inhomogeneous interstellar medium (ISM).
The RMI in magnetized plasmas has a potential to amplify the magnetic field beyond the compression by a single shock, and thus it is of great interest to astrophysics.
\cite{2012ApJ...758..126S,2013PhRvL.111t5001S} conducted two-dimensional MHD simulations and argued that the RMI can be a powerful candidate for the mechanism of the magnetic field amplification at supernova remnants, at which strong field regions are localized and their maximum strength reaches $100$ times larger than the typical value in the ISM.
For practical application of the present scheme, we simulate the nonlinear evolution of the RMI.
The two-dimensional computational domain is taken in the rest frame of a shock at $y=0$.
The initial state in the upstream region $y>0$ is $(\rho,u,v,w,B_x,B_y,B_z,P)=(1.0,0,-1.0,0,0.00034641,0,0,0.006)$ so that the Mach number $M=|\vect{u}|/c_s=10$ and the plasma beta $\beta=2P/|\vect{B}|^2=10^5$.
The initial state in the downstream region $y<0$ is calculated from the Rankine-Hugoniot condition for perpendicular MHD shocks, $(\rho,u,v,w,B_x,B_y,B_z,P)=(3.8835,0,-0.25750,0,0.0013453,0,0,0.74850)$. 
A corrugated contact discontinuity is imposed in the upstream region, $y_{\rm cd}=Y_{0} + \psi_0 \cos (2 \pi x/\lambda)$, where $Y_{0} = 1.0$, $\psi_0=0.1$ is a corrugation amplitude, and $\lambda=1.0$ is the wavelength.
We consider that the density is decreased to $\rho=0.01$ behind the contact discontinuity $y>y_{\rm cd}$ (see Figure \ref{fig:rmi_summary}(a)).
The domain is extended in the $y$ direction so as to ignore the boundary effect, $0 \leq x < \lambda$ and $-80 \lambda \leq y < 80 \lambda$, and is resolved by $N \times 160N$ grid points.
We use $N=64$ for a fiducial run.
The boundary condition is periodic in the $x$ direction, and is fixed to be the initial state in the $y$ direction.

The contact discontinuity flows in the negative $y$ direction and strikes the shock at $t=1$.
Then, a transmitted shock and a reflected rarefaction wave propagate in the positive and negative $y$ directions.
Due to the corrugation of the contact discontinuity, the surfaces of these waves are also rippled that induce the shear flow at the interface.
Subsequently, the RMI grows and the interface develops nonlinearly.
Figure \ref{fig:rmi_summary}(b) shows the density profile at $t=20$.
Arrows represent the direction of the rotational velocity, which is obtained by decomposing the velocity field into rotational, compressible, and mean components, $\vect{u}_{\rm R}, \vect{u}_{\rm C}, \vect{u}_{M}$.
They satisfy
\begin{eqnarray}
\left\{
\begin{array}{l}
\vect{u}_{\rm R} + \vect{u}_{\rm C} + \vect{u}_{M} = \vect{u},\\
\nabla \cdot \vect{u}_{\rm R} = 0,\\
\nabla \times \vect{u}_{\rm C} = 0,\\
\overline{\vect{u}_{\rm R}}=\overline{\vect{u}_{\rm C}}=0,
\end{array}
\right.\label{eq:70}
\end{eqnarray}
where an overline means spatial averaging.
The mushroom-shaped spike develops with a height of $20$ times the initial corrugation amplitude.
The primary vortex in $9.5<y<11$ forms the spike, and its velocity is approximately $10-20 \%$ of the upstream velocity.
In addition, the secondary vortex and the associated ring-shaped profile are visible in $10<y<10.5$.
The rotational flow is observed only around the spike.
Compared to the rotation, the compressible flow does not show a significant feature around the spike, and its velocity is approximately 10 times slower than the rotational one (not shown here).

The magnetic field can be amplified by stretching and compression \citep{2012ApJ...758..126S}.
Figure \ref{fig:rmi_summary}(c) shows the profile of the magnetic field strength $|\vect{B}|$ normalized by the upstream field strength.
The strong field regions are observed as filamentary structures along the interface, e.g., $y=0.2,0.8$, and the field strength reaches 90 times the upstream value.
They correlate well with the rotational flow, indicating that the field is amplified mainly by the stretching.
The compression would less contribute to the field amplification because the compressible velocity is much slower than the rotational velocity around the spike.
Figure \ref{fig:rmi_summary}(d) shows the time profile of the rotational energy $\rho |\vect{u}_{R}|^2/2$ and the magnetic energy increased by the RMI $|\vect{B}|^2/2-|\vect{B}_{\rm ref}|^2/2$ integrated over space, where $\vect{B}_{\rm ref}$ is obtained from the run without the corrugation.
Just after the impact of the shock against the contact discontinuity, the rotational energy increases impulsively, followed by the gradual decrease of the rotational energy and increase of the magnetic energy.
The magnetic field behaves passively and is continuously amplified as long as its energy density is much smaller than the kinetic energy density.
However, the magnetic energy saturates and then begins to decrease after it exceeds the rotational energy at $t=36$, which can be interpreted as the suppression by the amplified magnetic field itself when the Lorentz force becomes strong enough to prevent the flow \citep{2008PhPl...15d2102C}.

In order to examine the resolution dependence of the RMI, we conduct simulation runs with different interpolation methods (MUSCL-MC and W4I4D5), different Riemann solvers (HLL and HLLD), and different divergence-free treatments (GLM and CUCT) at different resolutions $(N=32,64,128)$.
Figure \ref{fig:rmi_bmaxprof} compares the time profile of the maximum of the magnetic field strength $|\vect{B}|_{\rm max}$ among different runs.
Both the growth speed and the saturation level increase with increasing the resolution, as is reported by \cite{2012ApJ...758..126S} for weak initial field cases.
The maximum field strength exceeds 100 times the upstream value in the high resolution run.
The W4I4D5-HLLD-CUCT scheme exhibits higher saturation levels than the other schemes, indicating better accuracy of the solution.
In particular, the HLLD scheme obtains a solution comparable to the HLL scheme with a half grid size in terms of the field amplification.
The low saturation level in the HLL scheme is due to the overestimation of the amount of numerical diffusion in the high beta plasma (mentioned in Sec \ref{sec:field-loop-advection}).
The plasma beta value at the maximum field region is $\sim 10^2$ in this run.
However, the value is decreased to $\sim 10$ and the discrepancy of the saturation level between the HLL and HLLD schemes is reduced in the run with an initial plasma beta of $10^3$ (not shown here).

\subsection{Magnetic reconnection}\label{sec:magn-reconn}
Magnetic reconnection is a fundamental energy release process in space and terrestrial environments such as the solar corona and the planetary magnetosphere.
Since it intrinsically contains a hierarchical structure, a variety of models have been utilized so as to understand its nature from the fully kinetic scale to the MHD scale \cite[and collabolation papers]{2001JGR...106.3715B}.
Among them, the resistive MHD is frequently adopted to model the macroscopic dynamics of the reconnection and its associated phenomena far beyond the kinetic scale.
Using the ideal (not resistive) MHD equations, we simulate the evolution of an anti-parallel magnetic field configuration that is subject to reconnect numerically, so as to examine the effect of numerical diffusion and the robustness of a scheme similar to \cite{2005JCoPh.205..509G} and \cite{2009JCoPh.228..952L}.
The two-dimensional computational domain with $-L_x/2 \leq x<L_x/2 \; (L_x=512)$ and $-L_y/2 \leq y \leq L_y/2 \; (L_y=16)$ is resolved by $N \times N/32$ grid points.
The boundary condition is periodic and symmetric in the $x$ and $y$ directions.
The initial condition is a Harris current sheet configuration, $\rho=\rho_0,\vect{u}=0,\vect{B}=B_0 \tanh(y/\lambda) \vect{e}_x,P=[(1+\beta)B_0^2-|\vect{B}|^2]/2$, where $\rho_0=B_0=1.0$, $\lambda=1.0$ is the current sheet thickness, and $\beta=0.2$ is the plasma beta in the upstream region, respectively.
We impose a localized perturbation to the flux function $\delta A_z = 0.05 \lambda \exp [-(x^2+y^2)/(2\lambda)^2]$ and a small $(1\%)$ uniform random perturbation to $v$ inside the current sheet.
The localized perturbation transfers magnetic flux toward the positive and negative $x$ directions, and then forms a thin current sheet.
The resistivity plays a critical role for the reconnection that is imposed numerically in this problem.
The simulations are conducted by the HLL-CUCT, the HLLD-CUCT-AA, and the HLLD-CUCT schemes.

{Figures \ref{fig:mrxjz_hlluct} through \ref{fig:mrxjz_hlldcuct} show the time evolution of the out-of-plane current density solved by the three schemes at $N=8192$. 
After the passage of the initial perturbation, the current sheet gets thin down to the grid scale until $t \sim 100$.
Subsequently, the numerical resistivity switches on and the system evolves differently depending on the scheme.
The current sheet evolves in the Y-shape structure, and large-scale plasmoids are attached there in the HLL-CUCT scheme (Figure \ref{fig:mrxjz_hlluct}).
In contrast, the HLLD-CUCT-AA scheme shows the uniform current sheet elongated in the $x$ direction.
In the current sheet, multiple plasmoids grow and then merge each other (Figure \ref{fig:mrxjz_hlltcuctctw0}).
The HLLD-CUCT scheme also shows the elongated current sheet, but the subsequent plasmoid formation is modest compared to the HLLD-CUCT-AA scheme (Figure \ref{fig:mrxjz_hlldcuct}).
The growth of the plasmoid is considerably suppressed with decreasing the resolution to $N=4096$ in the HLLD-CUCT(-AA) schemes (not shown here).

Figure \ref{fig:mrx_flux} shows the time profile of the spatially-integrated magnetic energy for the three schemes at $N=4096,8192$.
The rate of dissipation of the magnetic energy is the fastest in the HLL-CUCT scheme, and the slowest in the HLLD-CUCT scheme.
The magnetic energy dissipates faster in the higher resolution run with the HLLD-CUCT-AA scheme, implying that the plasmoid dynamics dominates the dissipation of the current sheet.
{Since the HLLD Riemann solver exactly captures an isolated and stationary tangential discontinuity without numerical diffusion \citep{2005JCoPh.208..315M}, the rate of dissipation remains slow in the low resolution run when the plasmoid is absent and hence the current sheet is  approximately static.
The aspect ratio of the current sheet (length/thickness) increases with increasing the resolution that would increase the number of plasmoids as is demonstrated in the resistive MHD \citep{2007PhPl...14j0703L,2009PhPl...16k2102B}.}
The HLLD-CUCT scheme shows the similar tendency as the HLLD-CUCT-AA scheme, although the rate of dissipation is slower owing to the weaker plasmoid activity.
In contrast to the HLLD-CUCT(-AA) schemes, the rate of dissipation is faster in the lower resolution run with the HLL-CUCT scheme.
}

{The overall dynamics is quite different between the HLL-CUCT and the HLLD-CUCT(-AA) schemes.
The difference is already found in an early stage of the simulation run.
Figure \ref{fig:mrx_plt2d_ux_beta1d}(a)-(b) compares the $x$-component of the velocity at $t=100$ with the HLL-CUCT and the HLLD-CUCT schemes.
The HLLD-CUCT-AA scheme obtains an almost identical result to the HLLD-CUCT scheme until this period, thus it is not shown here.
The outflow jet driven by the initial perturbation propagates rightward.
The width of the outflow jet is unchanged from the center to the edge of the perturbation $(x=8)$ in the HLLD-CUCT scheme.
On the other hand, the outflow jet broadens toward the downstream and forms a crab-claw structure in the HLL-CUCT scheme, and the maximum jet velocity $u = 0.7$ is much faster than that in the HLLD-CUCT scheme $u=0.13$.
This structure, along with the Y-shaped structure in Figure \ref{fig:mrxjz_hlluct}, is a resemblance to the Petscheck-type reconnection triggered by an ad-hoc localized resistivity \citep{2011PhPl...18b2105Z,2015PhPl...22c2114Z}.
This implies that the HLL-CUCT scheme tends to artificially localize the diffusion region under the adopted initial condition.
Figure \ref{fig:mrx_plt2d_ux_beta1d}(c) compares the plasma beta value averaged over the current sheet $(-1<y<1)$ between the two schemes.
The HLL-CUCT scheme shows that the plasma beta gradually increases to $\sim 100$ in the course of the flow from the center to $x=9$.
As is discussed in Section \ref{sec:field-loop-advection}, rapid numerical diffusion of the magnetic field occurs in the high beta region with the HLL-CUCT scheme, and consequently, it increases the gas pressure since we use the conservative scheme.
The increased gas pressure helps to open the jet, and in turn, localize the diffusion region.
This numerical artifact is not observed in the HLLD-CUCT(-AA) schemes because the HLLD Riemann solver is designed to capture a tangential discontinuity.
The plasma beta is nearly uniform in the current sheet.
}

{
The different plasmoid evolution between the HLLD-CUCT-AA and the HLLD-CUCT schemes (Figures \ref{fig:mrxjz_hlltcuctctw0} and \ref{fig:mrxjz_hlldcuct}) may be attributed to the difference in the numerical Lundquist number that is proportional to the length of the current sheet.
The current sheet is elongated to $|x|=30$ in the HLLD-CUCT-AA scheme, but it is shortened by $|x|=10$ in the HLLD-CUCT scheme at $t=300$ (middle panel in Figures \ref{fig:mrxjz_hlltcuctctw0} and \ref{fig:mrxjz_hlldcuct}).
This discrepancy must be owing to the difference in the calculation of the electric field.
While the CUCT-AA method takes the arithmetic average in Eq. (\ref{eq:62}), the CUCT method takes the directional biasing with respect to the {\Alfven} mode in Eq. (\ref{eq:63}), and its nonlinear coefficient $\Theta$ depends on the position (Eq. (\ref{eq:64})).
This nonlinear effect may impose a numerical resistivity varying on the $x$ position, and eventually, it may discourage the elongation of the current sheet.
{The nonlinearity of the numerical resistivity in the CUCT-AA method would be relatively weak, and it results in the elongated current sheet and the subsequent plasmoid formation that is a resemblance to the tearing instability for a uniform resistivity \citep{1986PhFl...29.1520B}.}
In order to test this hypothesis, we apply the schemes to the visco-resistive MHD equations \citep{2016PhPl...23g2122M}, and solve the same problem.
Kinematic viscosity and resistivity coefficients, $\nu$ and $\eta$, are assumed constant and uniform, and the corresponding kinetic and magnetic Reynolds numbers are $V_{A} \lambda/\nu=V_{A} \lambda/\eta=10^{3}$.
The coefficients employed here are larger than numerical ones in this problem.
Figure \ref{fig:vrmrx} confirms the formation of multiple plasmoids and their nonlinear evolution in the visco-resistive MHD simulation with the HLLD-CUCT scheme (and a similar result is obtained with the HLLD-CUCT-AA scheme).
}

\section{Summary}\label{sec:summary}
We have designed a conservative finite difference scheme for ideal MHD simulations based on the Weighted Compact Nonlinear Scheme \cite[WCNS;][]{2000JCoPh.165...22D,2008JCoPh.227.7294Z}, which is easier to increase the order of accuracy in multi-dimension than finite volume schemes.
The scheme reconstructs a high-order accurate numerical flux and approximates its spatial derivative by a high-order linear central difference.
We have proposed the fourth-order upwind-biased nonlinear interpolation method (Eqs. (\ref{eq:25}) and (\ref{eq:27})) to combine with the fourth-order central difference (Eq. (\ref{eq:13})), which achieves optimal fifth order of accuracy for linear problems.
The scheme can use the physical variables for the high-order interpolation and the calculation of the numerical flux at the midpoint, allowing us to employ a multi-state Riemann solver for MHD.
Especially, the scheme employs the HLLD Riemann solver in order to robustly capture rotational, tangential, and contact discontinuities as well as fast mode shocks \citep{2005JCoPh.208..315M}.
{We have proposed a new and simple variant of the Constrained Transport (CT) method, termed as the Central Upwind CT method (Eq. (\ref{eq:63})), which is consistent with the base one-dimensional scheme, attains a designed order of accuracy in a multidimensional flow, and preserves the divergence-free condition of the magnetic field as designed high-order finite difference representation.
Furthermore, this method implements a two-dimensional directional biasing so as to avoid the numerical oscillation in a specific coordinate axis pointed out by \cite{2013JCoPh.243..269L}.}

The performance of the present scheme has been measured through linear and nonlinear problems.
The scheme achieves the designed order of accuracy, and it is more cost effective than lower-order schemes in multidimensional problems (Sec. \ref{sec:circ-polar-alfv}-\ref{sec:kelv-helmh-inst}).
MHD shocks and discontinuities, smooth flows, and their interactions are resolved without generating unphysical divergence errors (Sec. \ref{sec:orszag-tang-vortex}-\ref{sec:blast-wave-strongly}).
These problems clearly demonstrate that the CUCT method can improve the robustness of the MHD scheme for strong shocks.
The HLLD-CUCT scheme correctly captures the tangential discontinuity in the multidimensional flow irrespective of the plasma beta by virtue of the combination of the multi-state Riemann solver and the divergence-free method (Sec. \ref{sec:field-loop-advection}).
One has to be cautious to use a single-state Riemann solver for high beta plasmas, in which the transit time of the fastest mode may be much faster than a dynamical time scale that seriously overestimates the amount of numerical diffusion.
{The scheme successfully resolves the magnetic field amplification associated with the Richtmyer-Meshkov instability (Sec. \ref{sec:richtmy-meshk-inst}). 
The saturation level of the maximum field strength depends on a scheme design as well as a grid resolution: a high saturation level is obtained by a high resolution run and/or a high accurate scheme.}
{The evolution of the current sheet turns to be susceptible to the calculation method of the electric field (Sec. \ref{sec:magn-reconn}).
The HLLD-CUCT scheme tends to form an elongated current sheet and subsequent multiple plasmoids.
The evolution of the plasmoid considerably differs between the arithmetic average (Eq. (\ref{eq:62})) and the direction-biased average (Eq. (\ref{eq:63})).
On the other hand, the HLL-CUCT scheme tends to localize the current density and trigger the magnetic reconnection there.
}

\cite{1999JCoPh.150..561J} pointed out that ideal MHD simulations with different Godunov-type schemes may not lead to identical or convergent results in the presence of magnetic field rotation, because the amount of numerical diffusion for the velocity, magnetic field, and temperature vary according to a scheme design.
\cite{2015ApJ...808..54M} reported that the saturation level of the magnetorotational instability is susceptible to the numerical magnetic Prandtl number, which relies on the choice of the Riemann solver and the treatment for the divergence-free condition.
It would be preferable to involve explicit diffusion terms, viscosity, resistivity, and thermal conductivity for practical MHD simulations rather than the ideal approximation.
High accurate schemes allow us to use small diffusion coefficients for application to space and astrophysical plasma phenomena.

The CUCT method (Eq. (\ref{eq:63})) has upwind-correction terms twice as large as the baseline FluxCT method (Eq. (\ref{eq:54})), thus it would give a more robust solution as is demonstrated in Section \ref{sec:blast-wave-strongly}.
We also test the magnetic reconnection problem in Section \ref{sec:magn-reconn} with the W4I4D5-HLLD-FluxCT method, but numerical oscillation grows to an intolerable level at an early stage and prevents the system to evolve.
We then adopt the CUCT method rather than the FluxCT method to combine with the WCNS scheme and the HLLD Riemann solver.
In contrast, \citet{2000JCoPh.161..605T} conducted various benchmark tests and concluded that the FluxCT method is superior to other selected CT methods, one of which recovers the proper amount of numerical diffusion against the FluxCT method \citep{1998ApJ...509..244R}.
The difference between this paper and \cite{2000JCoPh.161..605T} might be partly owing to the order of accuracy in space: the reduction of the upwind property in the FluxCT method could be more problematic for higher-order schemes.
The CUCT method would be applicable to radiation and relativistic MHD equations because they solve the same Ohm's law.
The combination of the high-order WCNS scheme, the HLLD Riemann solver, and the CUCT method possesses good capabilities in accuracy, robustness, and computation cost.
{We expect that the present scheme is useful for practical MHD simulations of global space and astrophysical objects such as planetary magnetosphere, heliosphere, and accretion disks. 
They contain a wide variety of waves, shocks, and magnetic field inhomogeneities that would be faithfully resolved by the high-order, shock-capturing, and divergence-free scheme.}
\begin{acknowledgements}
We thank the anonymous referee for carefully reviewing our manuscript and giving insightful comments that greatly improved the manuscript.
This research is supported by JSPS KAKENHI Grant Numbers JP15K05369 (T. Minoshima) and JP15K04756 (T. Miyoshi).
Numerical simulations were in part carried out on Cray XC30 at Center for Computational Astrophysics, National Astronomical Observatory of Japan.
\end{acknowledgements}

\appendix
\section{Sixth and seventh-order finite difference schemes}\label{sec:sixth-order-finite}
The sixth-order linear central difference can be written as
\begin{eqnarray}
D_6(\tilde{U})_i &=& \frac{2250\left({\tilde{U}}_{i+1/2}-{\tilde{U}}_{i-1/2}\right)-125\left({\tilde{U}}_{i+3/2}-\tilde{{U}}_{i-3/2}\right)+9\left({\tilde{U}}_{i+5/2}-\tilde{{U}}_{i-5/2}\right)}{1920 \Delta x} \label{eq:34}\\ 
&=& \left.\pdif{{U}}{x}\right|_{i} + \frac{5}{7168}\left.\pdif[7]{{U}}{x}\right|_{i} \Delta x^6 + O(\Delta x^8) +O(\Delta x^p).\nonumber
\end{eqnarray}
If we apply the seventh-order left-side biased interpolation to $\tilde{U}$,
\begin{eqnarray}
 {\tilde{U}}^{L}_{i+1/2} &=& \frac{-5{U}_{i-3}+42{U}_{i-2}-175{U}_{i-1}+700{U}_{i}+525{U}_{i+1}-70{U}_{i+2}+7{U}_{i+3}}{1024}\label{eq:35}\\
&=&{U}_{i+1/2}+\frac{5}{2048} \left.\pdif[7]{{U}}{x}\right|_{i} {\Delta x^7} + O(\Delta x^8),\nonumber 
\end{eqnarray}
the resulting scheme has the sixth-order dispersion as a leading error,
\begin{eqnarray}
D_6(\tilde{U}^{L})_i = \left.\pdif{{U}}{x}\right|_{i} + \frac{5}{7168}\left.\pdif[7]{{U}}{x}\right|_{i} \Delta x^6 + \frac{5}{2048}\left.\pdif[8]{{U}}{x}\right|_{i} \Delta x^7 + O(\Delta x^8).\label{eq:39}
\end{eqnarray}
Similar to Eq. (\ref{eq:19}), we derive the following sixth-order left-side biased interpolation that is appropriate to Eq. (\ref{eq:34}),
\begin{eqnarray}
 {\tilde{U}^{L}}_{i+1/2} &=& \frac{-10{U}_{i-3}+81{U}_{i-2}-325{U}_{i-1}+1250{U}_{i}+900{U}_{i+1}-115{U}_{i+2}+11{U}_{i+3}}{1792}\label{eq:42}\\
&=&{U}_{i+1/2}-\frac{5}{7168} \left.\pdif[6]{{U}}{x}\right|_{i} {\Delta x^6} +\frac{5}{2048} \left.\pdif[7]{{U}}{x}\right|_{i} {\Delta x^7} + O(\Delta x^8).\nonumber 
\end{eqnarray}
Substituting Eq. (\ref{eq:42}) into Eq. (\ref{eq:34}) cancels the $O(\Delta x^6)$ error, and the resulting scheme has the seventh-order dissipation as a leading error,
\begin{eqnarray}
D_6(\tilde{U}^{L})_i = \left.\pdif{{U}}{x}\right|_{i} + \frac{5}{1792}\left.\pdif[8]{{U}}{x}\right|_{i} \Delta x^7 + O(\Delta x^8)\label{eq:52}.
\end{eqnarray}
Comparing Eqs. (\ref{eq:39}) and (\ref{eq:52}), one expects that the seventh-order scheme improves the dispersion error but the amount of numerical dissipation is increased by a factor of $2048/1792=1.14$ against the sixth-order scheme.
This is readily confirmed from the Fourier analysis performed in Section \ref{sec:nonl-interp}.

\section{Parameter dependence of the rotated shock tube problem with the hyperbolic divergence cleaning method}\label{sec:c_h-dependence-brio}
The hyperbolic divergence cleaning method (Eqs. (\ref{eq:66}) and (\ref{eq:67})) has two free parameters, $c_h$ and $c_p$, to diminish the \divb error through propagation and diffusion.
In Figure \ref{fig:chdependence_bwshock}, we examine the $c_h$ dependence of \divb and density in the Brio-Wu rotated shock tube problem (Sec. \ref{sec:oblique-shock}) solved by the W4I4D5-HLLD-GLM scheme.
The shock tube is tilted by an angle of $45^{\circ}$ relative to the $x$ axis.
The value of $c_h$ is set to the corresponding CFL numbers ${\rm CFL}(c_h)=c_h \Delta t/ \Delta x = 0.4,0.5,0.6,0.7,0.8$ (note that the CFL number for the MHD part is fixed to be 0.4).
The value of $c_p$ is fixed to be an optimal parameter proposed by \cite{2002JCoPh.175..645D}, $c_p^2 = 0.18 c_h$.
A \divb error of $\sim 2$ is generated at the slow compound wave and the slow shock irrespective of the value of $c_h$.
Subsequently, the error propagates and diffuses toward the upstream and downstream directions.
Grid-scale oscillations seen in ${\rm CFL}(c_h) \le 0.5$ is largely reduced for ${\rm CFL}(c_h) \ge 0.6$.
{Although the density profile in panel (b) is in good agreement with the W4I4D5-HLLD-CUCT scheme for ${\rm CFL}(c_h) \le 0.5$, it gets worse for ${\rm CFL}(c_h) \ge 0.6$ and this makes it hard to conclude that a solution with higher $c_{h}$ is always better than with lower $c_{h}$}.
The solution becomes more diffusive around the slow compound wave, the contact discontinuity, and the slow shock for higher $c_h$ value.
The thickness of the slow shock for ${\rm CFL}(c_h) = 0.8$ is 2.5 times broader toward upstream than for ${\rm CFL}(c_h) = 0.4$.
The solution is almost equivalent to one using a half CFL number for the MHD part, and the error is also observed with a tilt angle of $26.6^{\circ}$, indicating the acceptable upper limit of $c_h$.

\bibliographystyle{aasjournal}

\clearpage
\gdef\thetable{\arabic{table}}

\begin{table}[h!]
{\scriptsize
\caption{$L_1$ errors (orders) of the {\Alfven} wave propagation problem.}
      \begin{tabular}{ccccccc}
        \hline
       $B_{\perp}$ & $N=16$  &32   & 64 & 128 & 256 & 512\\ \hline
       First &4.80e-03 (-) &4.44e-03 ( 0.11) &3.27e-03 ( 0.44) &2.06e-03 ( 0.67) &1.17e-03 ( 0.82) &6.27e-04 ( 0.90)\\
       MUSCL &4.76e-03 (-) &3.95e-04 ( 3.59) &1.07e-04 ( 1.88) &2.85e-05 ( 1.91) &7.27e-06 ( 1.97) &1.82e-06 ( 2.00)\\
       W3I4D3 &1.57e-03 (-) &5.55e-04 ( 1.50) &1.26e-04 ( 2.14) &2.67e-05 ( 2.23) &5.32e-06 ( 2.33) &1.05e-06 ( 2.34)\\
       W4I4D5 &1.84e-04 (-) &4.85e-06 ( 5.24) &1.39e-07 ( 5.12) &5.51e-09 ( 4.66) &2.90e-10 ( 4.25) &1.74e-11 ( 4.06)\\
       W5I4D4 &1.79e-04 (-) &5.01e-06 ( 5.16) &1.69e-07 ( 4.89) &8.30e-09 ( 4.35) &4.86e-10 ( 4.09) &3.00e-11 ( 4.02)\\ \hline
       $B_{z}$ & $N=16$  &32   & 64 & 128 & 256 & 512\\ \hline
       First &2.48e-03 (-) &3.32e-03 (-0.42) &2.48e-03 ( 0.42) &1.54e-03 ( 0.68) &8.68e-04 ( 0.83) &4.61e-04 ( 0.91)\\
       MUSCL &1.78e-03 (-) &4.05e-04 ( 2.14) &1.05e-04 ( 1.94) &2.46e-05 ( 2.10) &5.30e-06 ( 2.21) &1.22e-06 ( 2.12)\\
       W3I4D3 &1.20e-03 (-) &4.32e-04 ( 1.48) &9.15e-05 ( 2.24) &1.76e-05 ( 2.38) &3.23e-06 ( 2.45) &5.83e-07 ( 2.47)\\
       W4I4D5 &1.59e-04 (-) &3.83e-06 ( 5.37) &9.48e-08 ( 5.34) &2.57e-09 ( 5.21) &7.41e-11 ( 5.12) &2.23e-12 ( 5.05)\\
       W5I4D4 &1.55e-04 (-) &3.76e-06 ( 5.36) &9.79e-08 ( 5.26) &3.29e-09 ( 4.90) &1.54e-10 ( 4.42) &8.82e-12 ( 4.12)\\ \hline
      \end{tabular}
\label{tab:alf_errors}
}
\end{table}

\begin{table}[h!]
\caption{Relative differences (convergence rates) of the Orszag-Tang vortex problem from the reference solution. The solution is smooth at $t=0.2\pi$, but becomes discontinuous at $t=\pi$.}
      \begin{tabular}{ccccc}
        \hline
       $t=0.2 \pi$ & $N=100$ & 200 & 400 & 800 \\ \hline
       MUSCL & 1.17e-03 (-) & 2.65e-04 ( 2.15) & 6.05e-05 ( 2.13) & 1.43e-05 ( 2.08) \\
       HYBRID & 7.74e-04 (-) & 1.52e-04 ( 2.34) & 2.87e-05 ( 2.41) & 5.28e-06 ( 2.44) \\
       W4I4D5 & 1.25e-04 (-) &8.14e-06 ( 3.94) &5.35e-07 ( 3.93) &3.71e-08 ( 3.85) \\ \hline
       $t= \pi$ & $N=100$ & 200 & 400 & 800 \\ \hline
       MUSCL & 3.42e-02 (-) & 1.65e-02 ( 1.05) & 7.80e-03 ( 1.08) & 3.23e-03 ( 1.27) \\
       HYBRID & 3.17e-02 (-) & 1.49e-02 ( 1.09) & 6.68e-03 ( 1.16) & 2.52e-03 ( 1.41) \\
       W4I4D5 &3.01e-02 (-) &1.38e-02 ( 1.12) &6.19e-03 ( 1.16) &2.25e-03 ( 1.46) \\ \hline
\end{tabular}
\label{tab:ot_table}
\end{table}

\clearpage
\gdef\thefigure{\arabic{figure}}

\begin{figure}[htbp]
\centering
\includegraphics[clip,angle=0,scale=0.5]{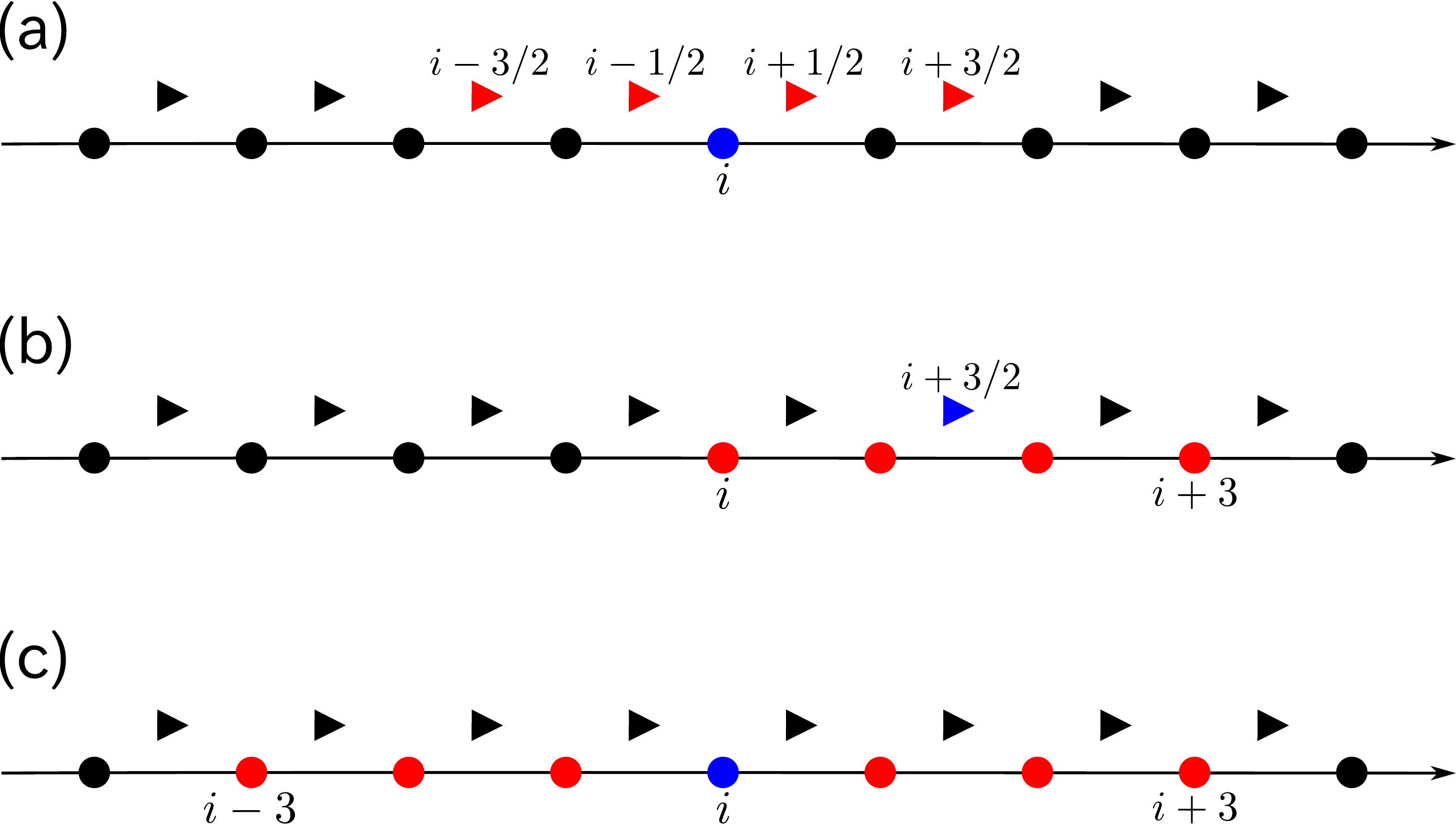}
\caption{One-dimensional grid spacing for the explicit WCNS scheme. Computational nodes and midpoints located at integer and half-integer grid points are represented by circle and triangle symbols, respectively.
(a) Location of the midpoints used for the fourth-order difference at the node $i$ (blue circle) is highlighted by red triangles.
(b) Stencil for the third-order interpolation to the midpoint $i+3/2$ (blue triangle) is highlighted by red circles.
(c) Stencil for the update at the node $i$ (blue circle) is highlighted by red circles.}
\label{fig:grid1d} 
\end{figure}

\begin{figure}[htbp]
\centering
\includegraphics[clip,angle=0,scale=0.7]{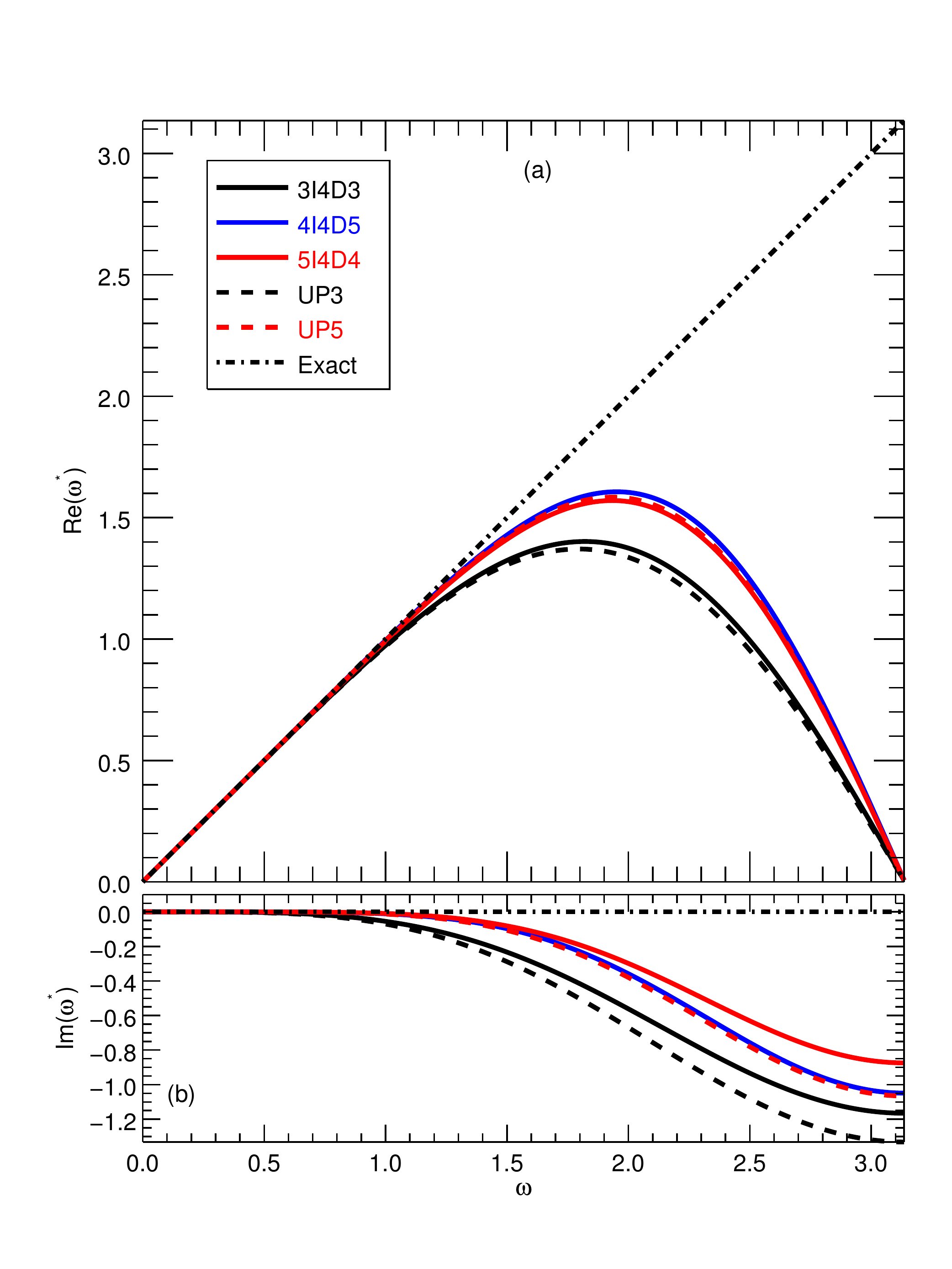}
 \caption{Fourier analysis of high-order finite difference schemes. (a) Real and (b) imaginary parts of the modified wave number are shown.}
\label{fig:wcns_fourier}
\end{figure}

\begin{figure}[htbp]
\centering
\includegraphics[clip,angle=0,scale=0.55]{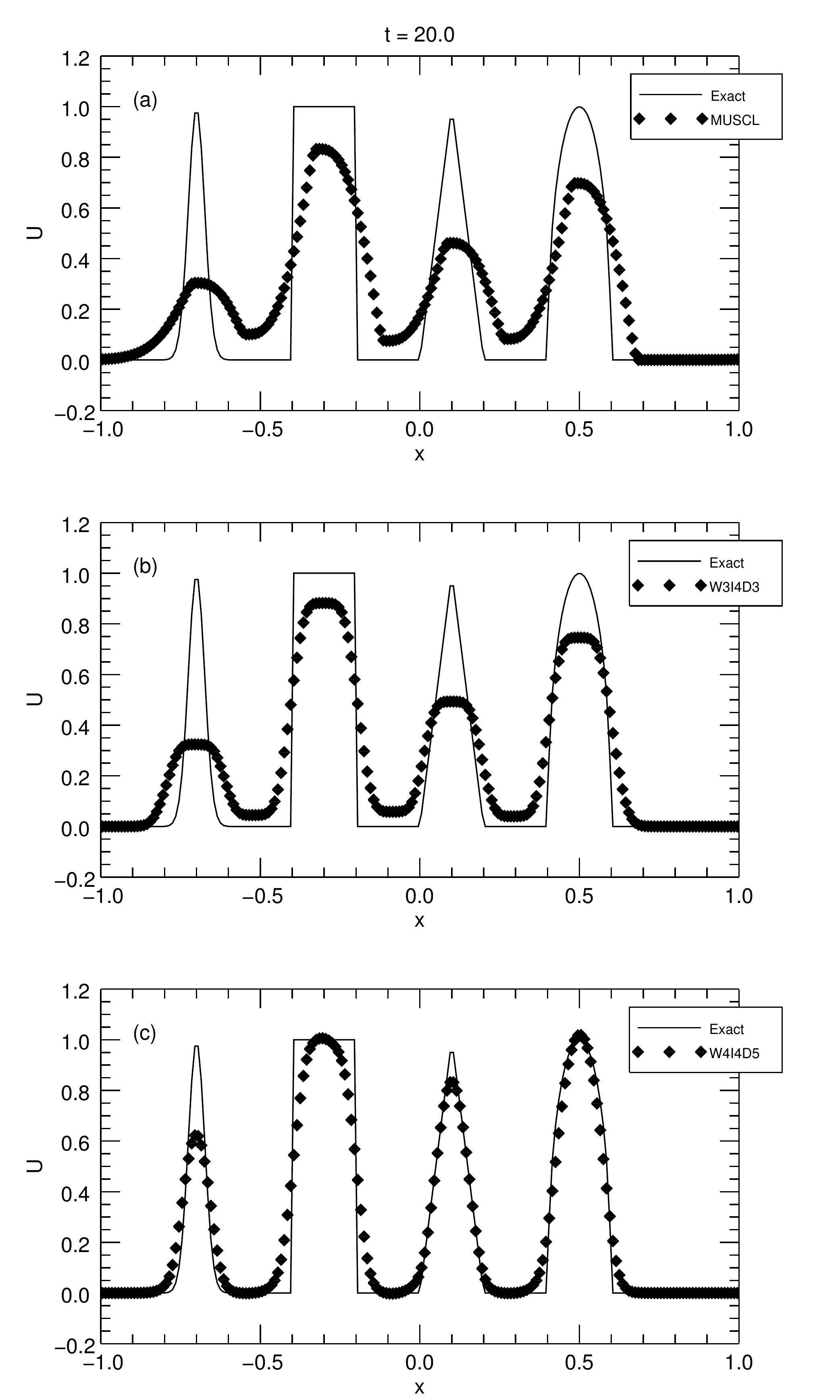}
 \caption{One-dimensional linear advection over ten periods solved by (a) the MUSCL-MC scheme, (b) the W3I4D3 scheme, and (c) the W4I4D5 scheme. The number of grid points is 200 and the CFL number is 0.4.}
\label{fig:adv1d_scalar}
\end{figure}

\begin{figure}[htbp]
\centering
 \includegraphics[clip,angle=0,scale=0.6]{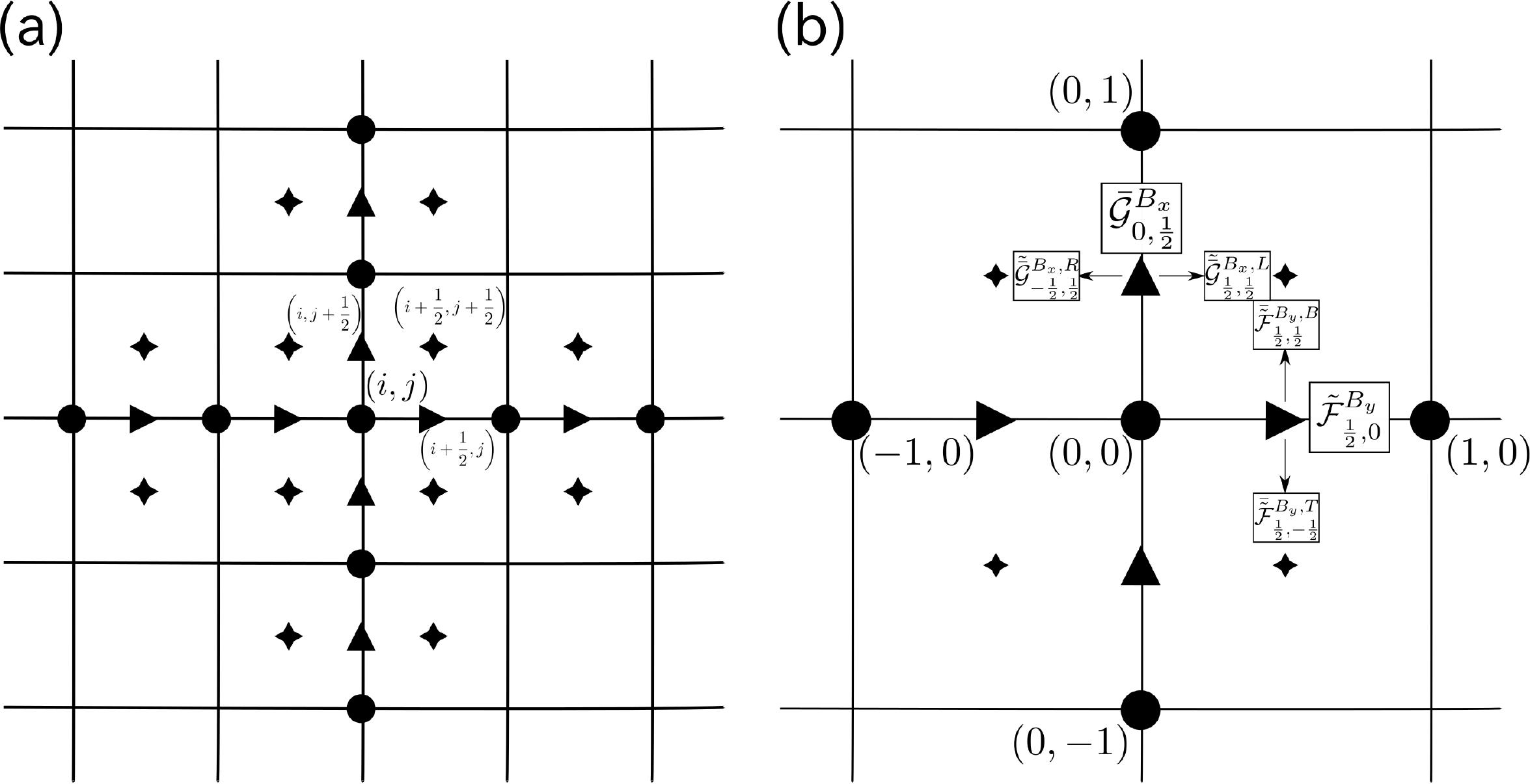}
\caption{(a) Two-dimensional grid spacing for the present finite difference scheme. Fluid variables are defined at integer circle points $(i,j)$, numerical fluxes and the in-plane magnetic fields at half-integer triangle points $(i+1/2,j)$ and $(i,j+1/2)$, and the out-of-plane electric field at staggered star points $(i+1/2,j+1/2)$, respectively. (b) Position of the numerical flux to calculate the electric field at staggered grid points. The numerical fluxes at half-integer points, $\tilde{\cal{F}}^{B_y}_{1/2,0}$ and $\bar{\cal{G}}^{B_x}_{0,1/2}$, are interpolated in the orthogonal direction (indicated by arrows).}
\label{fig:grid2d}
\end{figure}

\begin{figure}[htbp]
\centering
\includegraphics[clip,angle=0,scale=0.7]{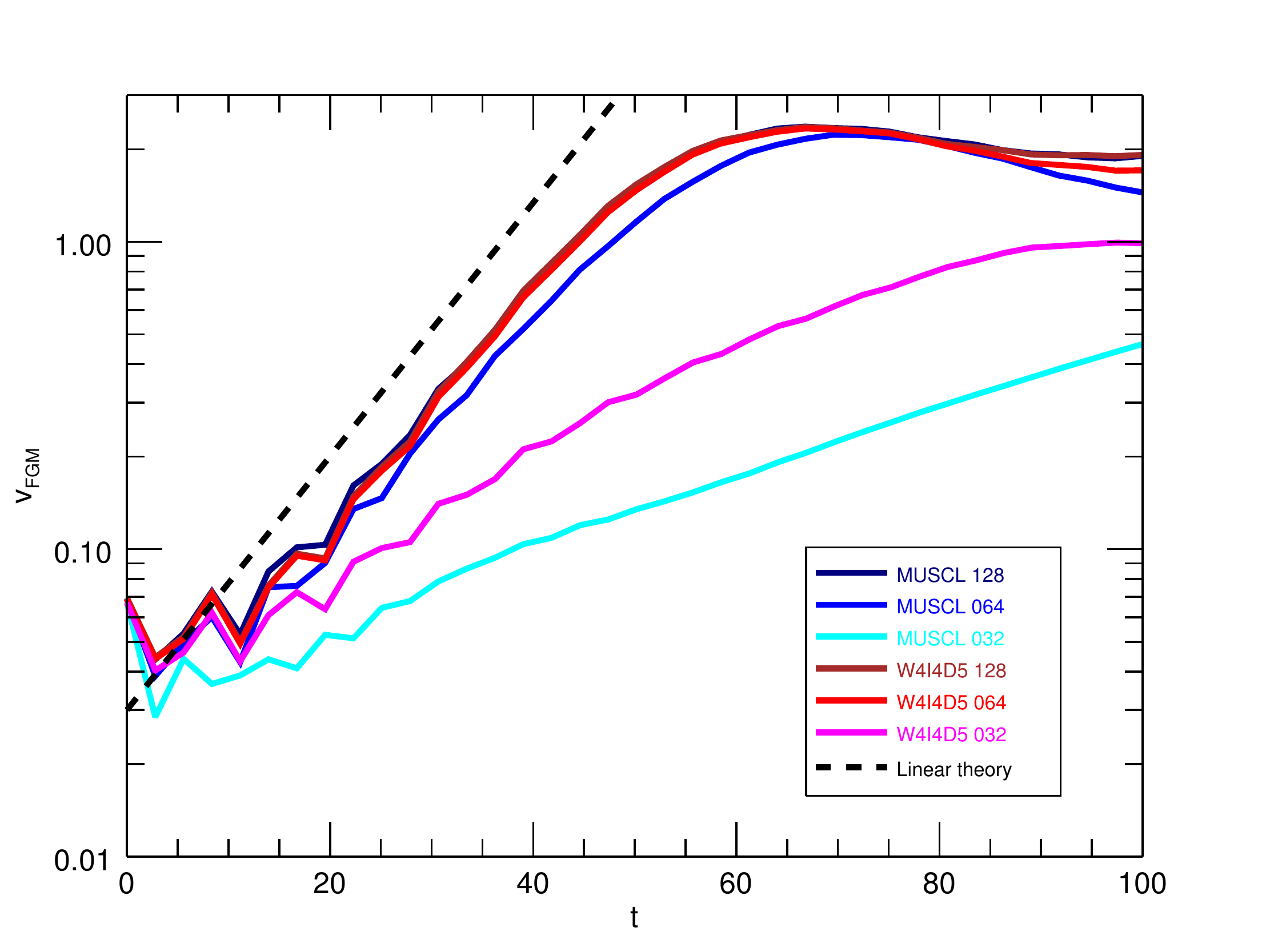}
 \caption{Time profile of the fastest growing mode in the Kelvin-Helmholtz instability solved by the MUSCL-MC and the W4I4D5 schemes with $N=32,64,128$. The dashed line indicates the solution obtained from the linear theory.}
\label{fig:kh_lgrowth}
\end{figure}

\begin{figure}[htbp]
 \centering
\includegraphics[clip,angle=0,scale=0.25]{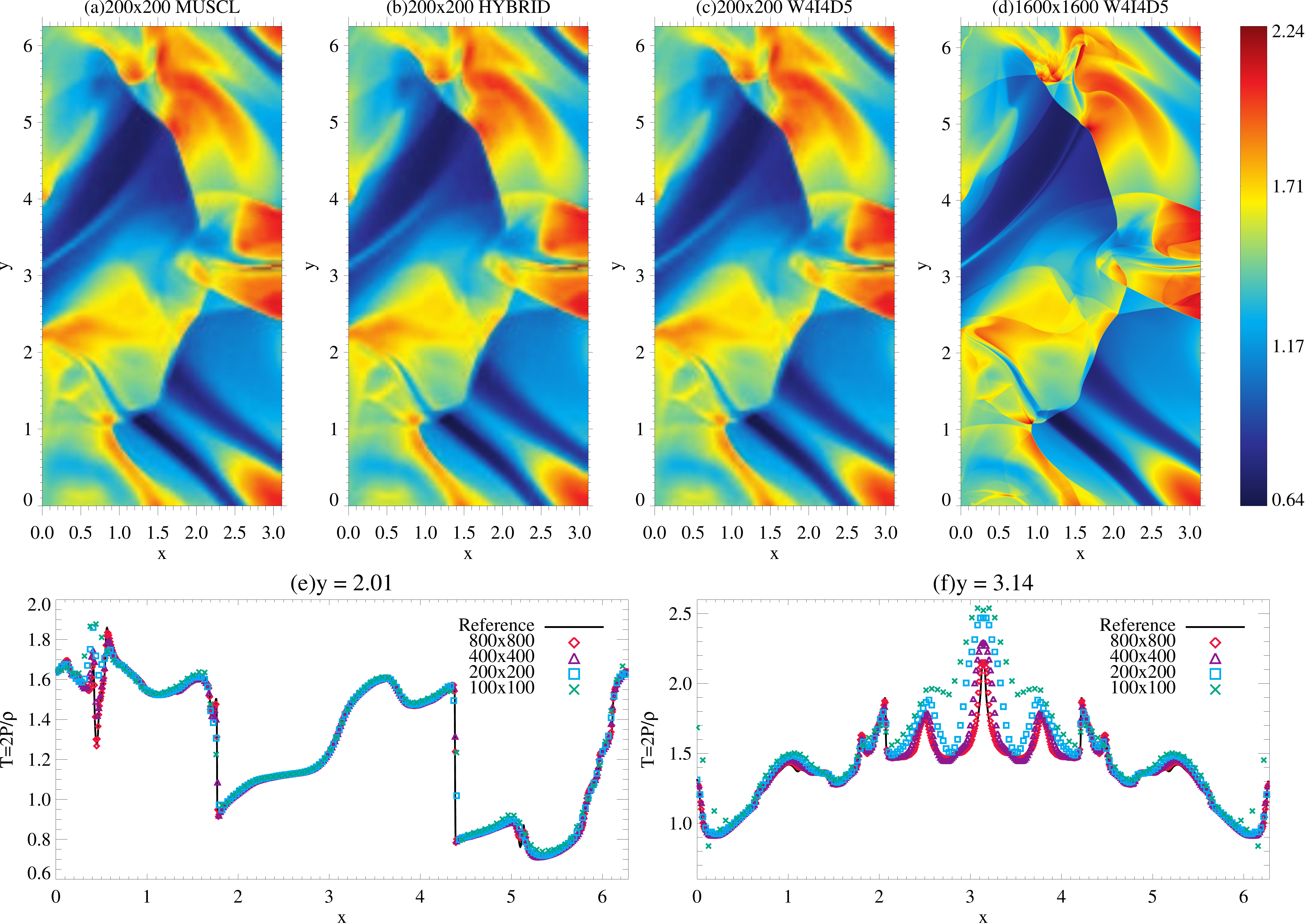}
\caption{Temperature profile in the Orszag-Tang vortex problem. (a)-(d) Two-dimensional profile solved by the MUSCL-HLLD-CUCT scheme, the HYBRID-HLLD-CUCT scheme, and the W4I4D5-HLLD-CUCT scheme with $N = 200$, and the reference with $N=1600$. The left half of the domain is shown. (e)-(f) One-dimensional profile at $y=0.64\pi$ and $y=\pi$ solved by the W4I4D5-HLLD-CUCT scheme with $N=100$ (green), $200$ (cyan), $400$ (purple), $800$ (red), and the reference (solid line).}
\label{fig:ot_vortex}
\end{figure}

\begin{figure}[htbp]
\centering
\includegraphics[clip,angle=0,scale=0.45]{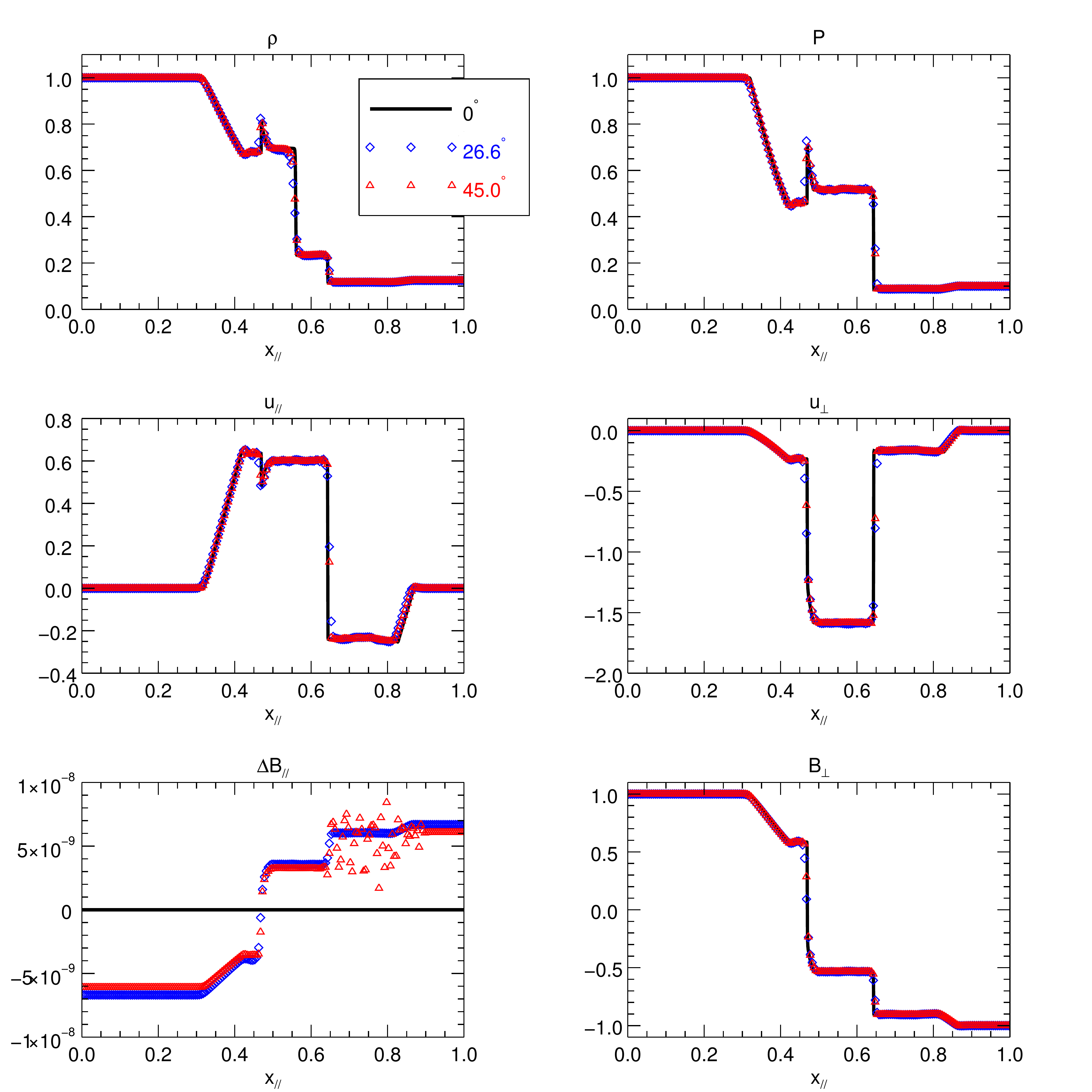}
 \caption{Brio-Wu rotated shock tube problem at $t=0.1$ solved by the W4I4D5-HLLD-CUCT scheme with $N=200$. The blue and red symbols correspond to the cases with propagation angles of $\alpha=26.6^{\circ}$ and $45^{\circ}$, respectively. The black line is the one-dimensional reference solution with 2000 grid points.}
\label{fig:bwrshock_uct}
\end{figure}

\begin{figure}[htbp]
\centering
\includegraphics[clip,angle=0,scale=0.45]{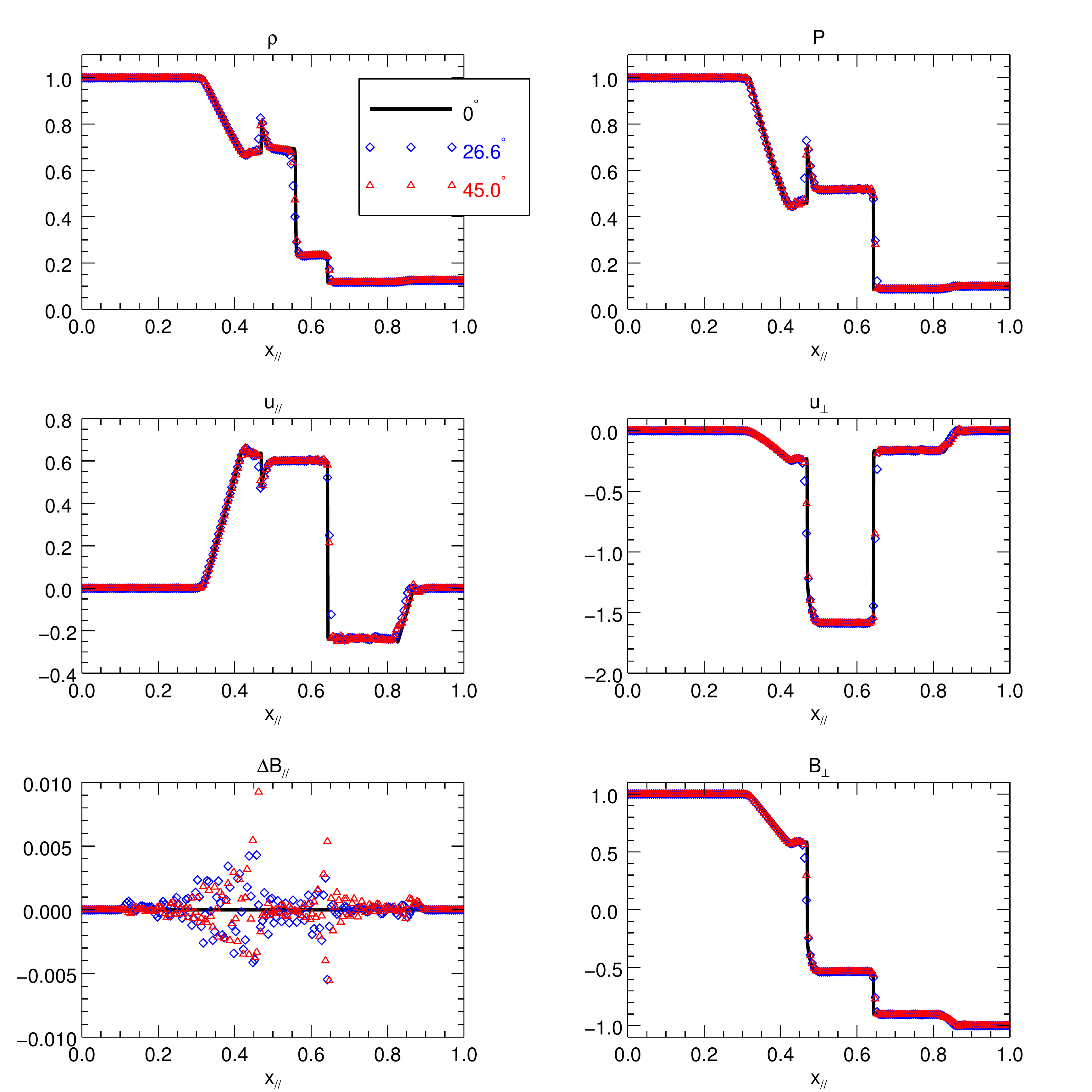}
\caption{Same as Figure \ref{fig:bwrshock_uct}, but solved by the W4I4D5-HLLD-GLM scheme.} 
\label{fig:bwrshock_glm}
\end{figure}

\clearpage

\begin{figure}[htbp]
\centering
\includegraphics[clip,angle=0,scale=0.35]{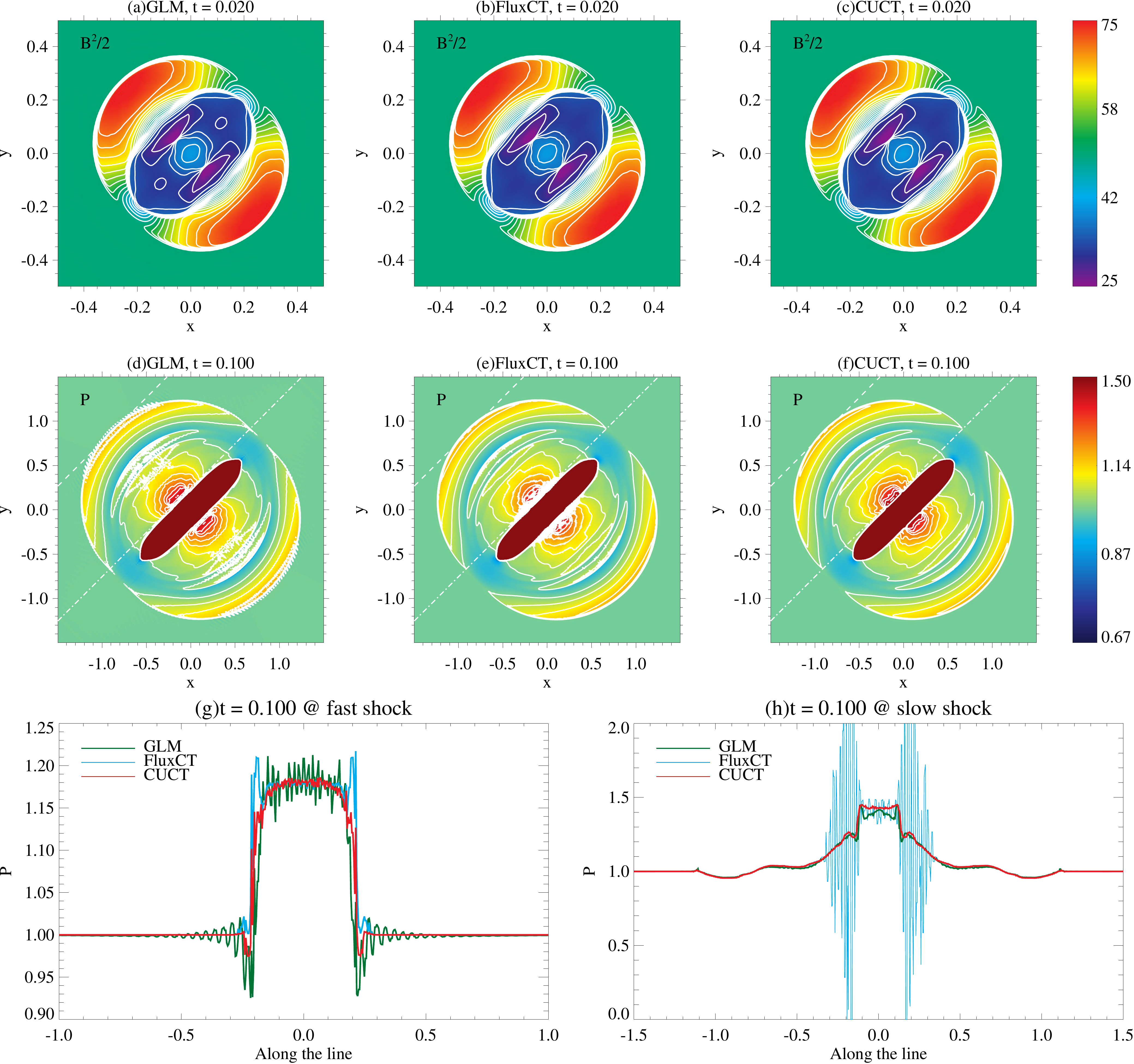}
\caption{The blast wave problem. (a)-(c) Two-dimensional magnetic energy profile at $t=0.02$ solved by the HLLD-GLM, the HLLD-FluxCT, and the HLLD-CUCT schemes. (d)-(f) Two-dimensional gas pressure profile at $t=0.1$ solved by the three schemes. (g) One-dimensional profile along the fast shock front denoted by dashed lines in panels (d)-(f). The green, light blue, and red lines correspond to the GLM, the FluxCT, and the CUCT schemes. (h) One-dimensional profile along the slow shock front denoted by dot-dashed lines in panels (d)-(f).}
\label{fig:blast_summary} 
\end{figure}

\begin{figure}[htbp]
\centering
\includegraphics[clip,angle=0,scale=0.35]{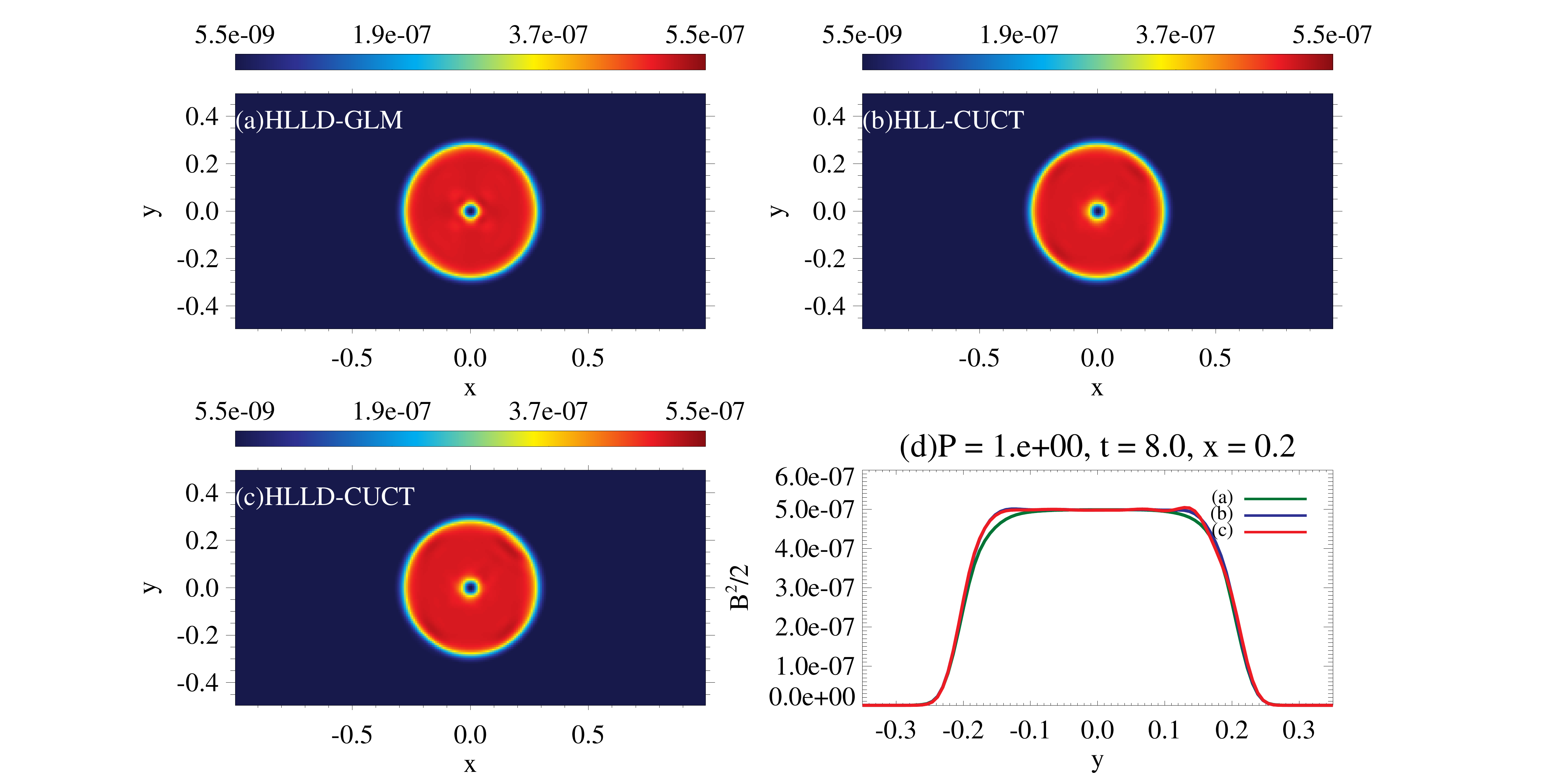}
 \caption{(a)-(c) Magnetic pressure (linear scale) in the field loop advection problem over eight periods solved by the W4I4D5-HLLD-GLM, the W4I4D5-HLL-CUCT, and the W4I4D5-HLLD-CUCT schemes. The ambient gas pressure is $1$. (d) One-dimensional profile of the magnetic pressure at $x=0.2$.}
\label{fig:floop_p=1e0}
\end{figure}

\begin{figure}[htbp]
\centering
\includegraphics[clip,angle=0,scale=0.35]{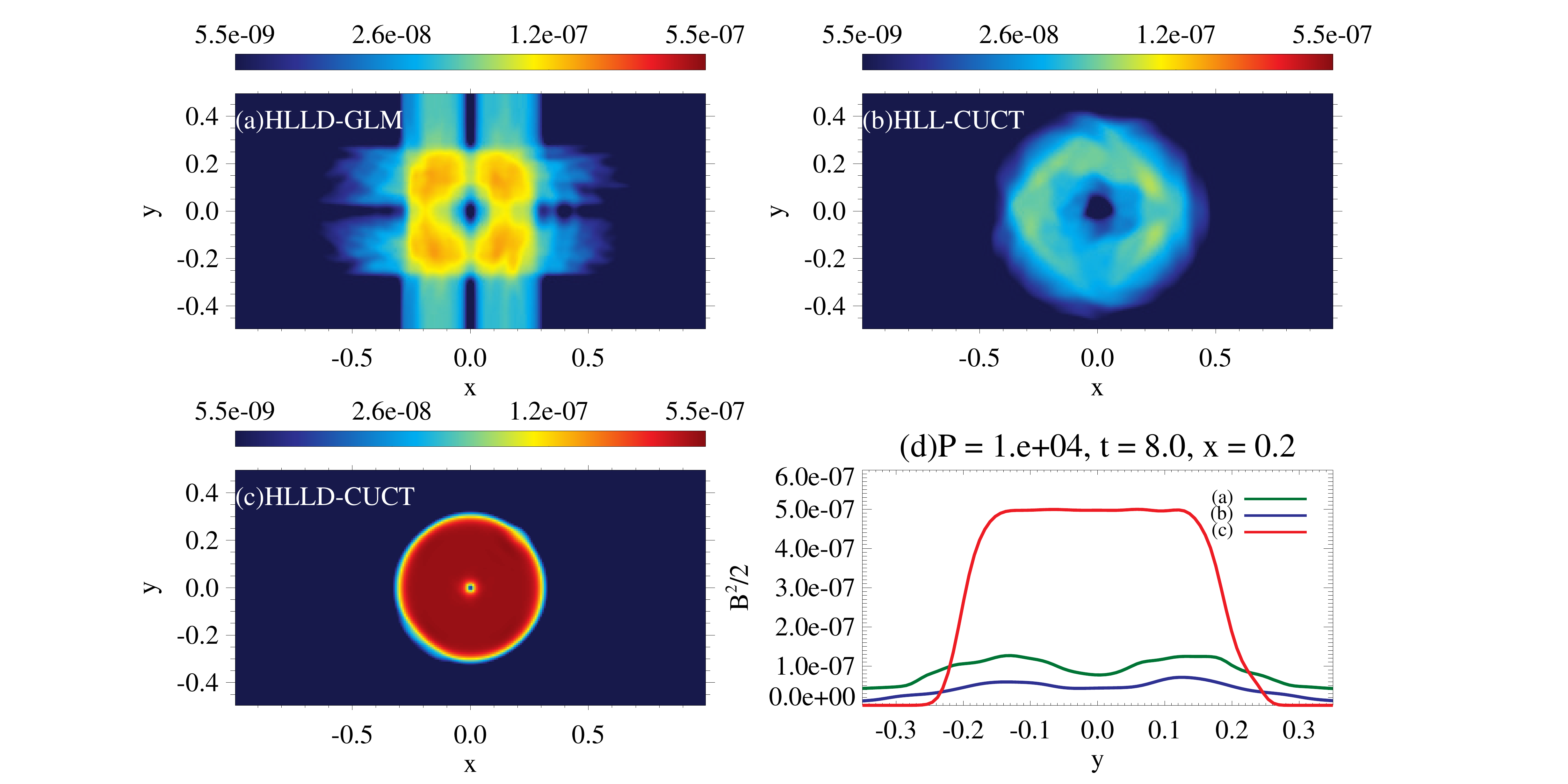}
\caption{Same as Figure \ref{fig:floop_p=1e0}, but with an ambient gas pressure of $10^4$. The color images are shown in the logarithmic scale.}
\label{fig:floop_p=1e4}
\end{figure}

\begin{figure}[htbp]
\centering
\includegraphics[clip,angle=0,scale=0.35]{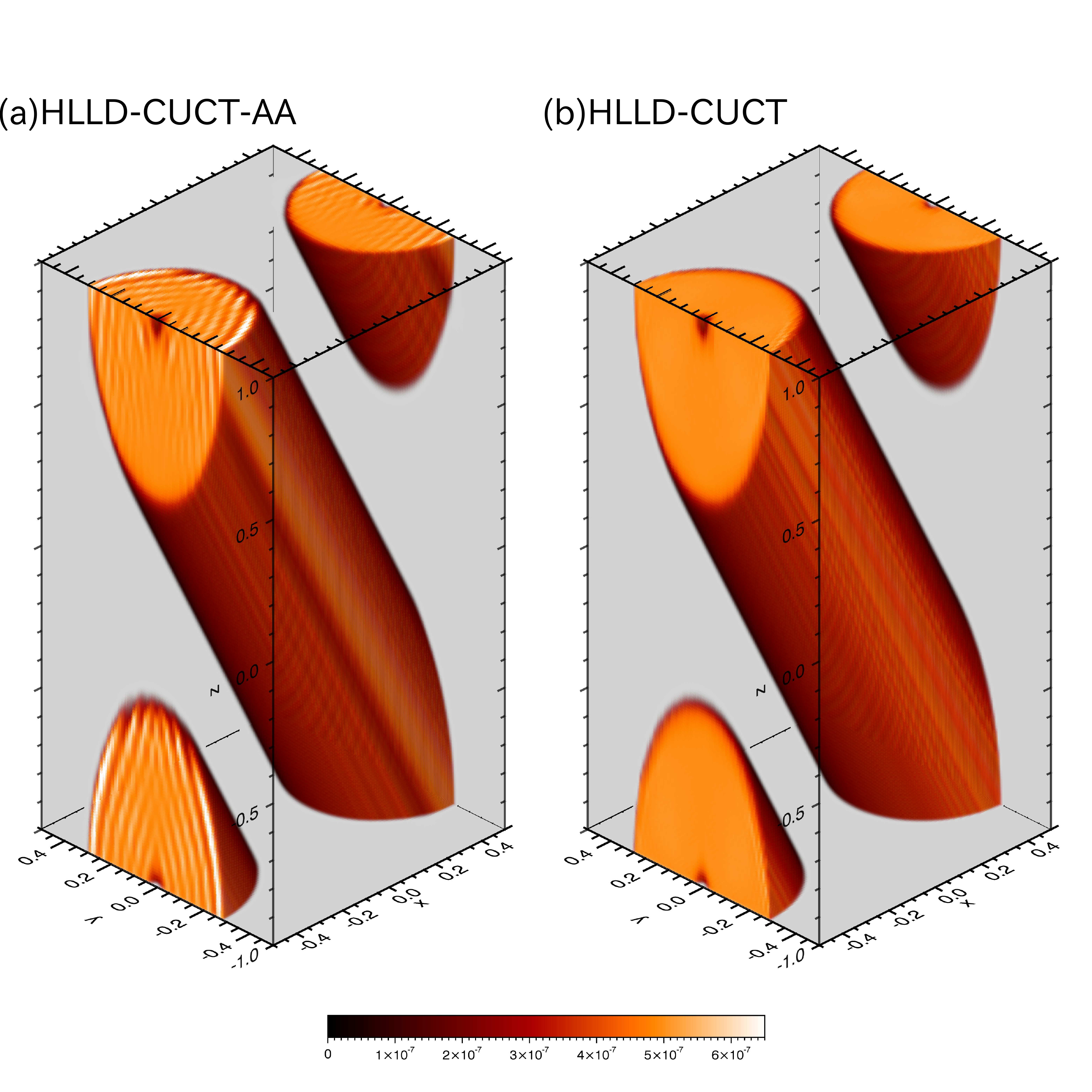}
 \caption{Magnetic pressure (linear scale) in the three-dimensional small angle advection problem over one period using the CUCT method with (a) the arithmetic average (Eq. (\ref{eq:62})) and (b) the direction-biased average (Eq. (\ref{eq:63})).}
\label{fig:3dloop_p=1e0_smallangle}
\end{figure}

\begin{figure}[htbp]
\centering
\includegraphics[clip,angle=0,scale=0.25]{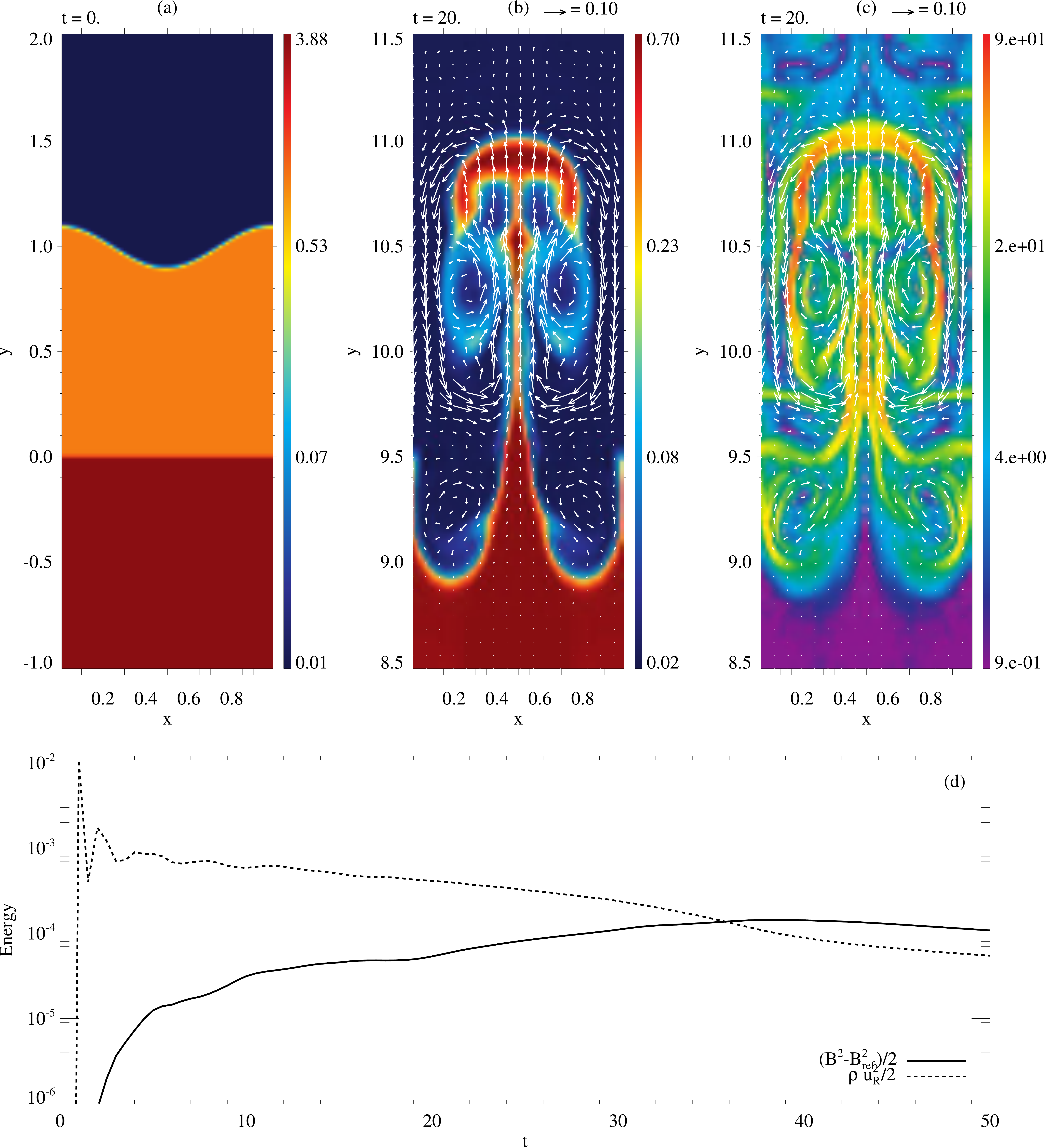}
\caption{The Richtmyer-Meshkov instability solved by the W4I4D5-HLLD-CUCT scheme with $N=64$. (a)-(b) Density at $t=0,20$, and (c) magnetic field strength at $t=20$. Arrows represent the direction of the rotational velocity $\vect{u}_{R}$ in Eq. (\ref{eq:70}). (d) Time profile of the magnetic energy increased by the instability (solid line) and the rotational energy (dashed line).}
\label{fig:rmi_summary}
\end{figure}

\begin{figure}[htbp]
\centering
\includegraphics[clip,angle=0,scale=0.4]{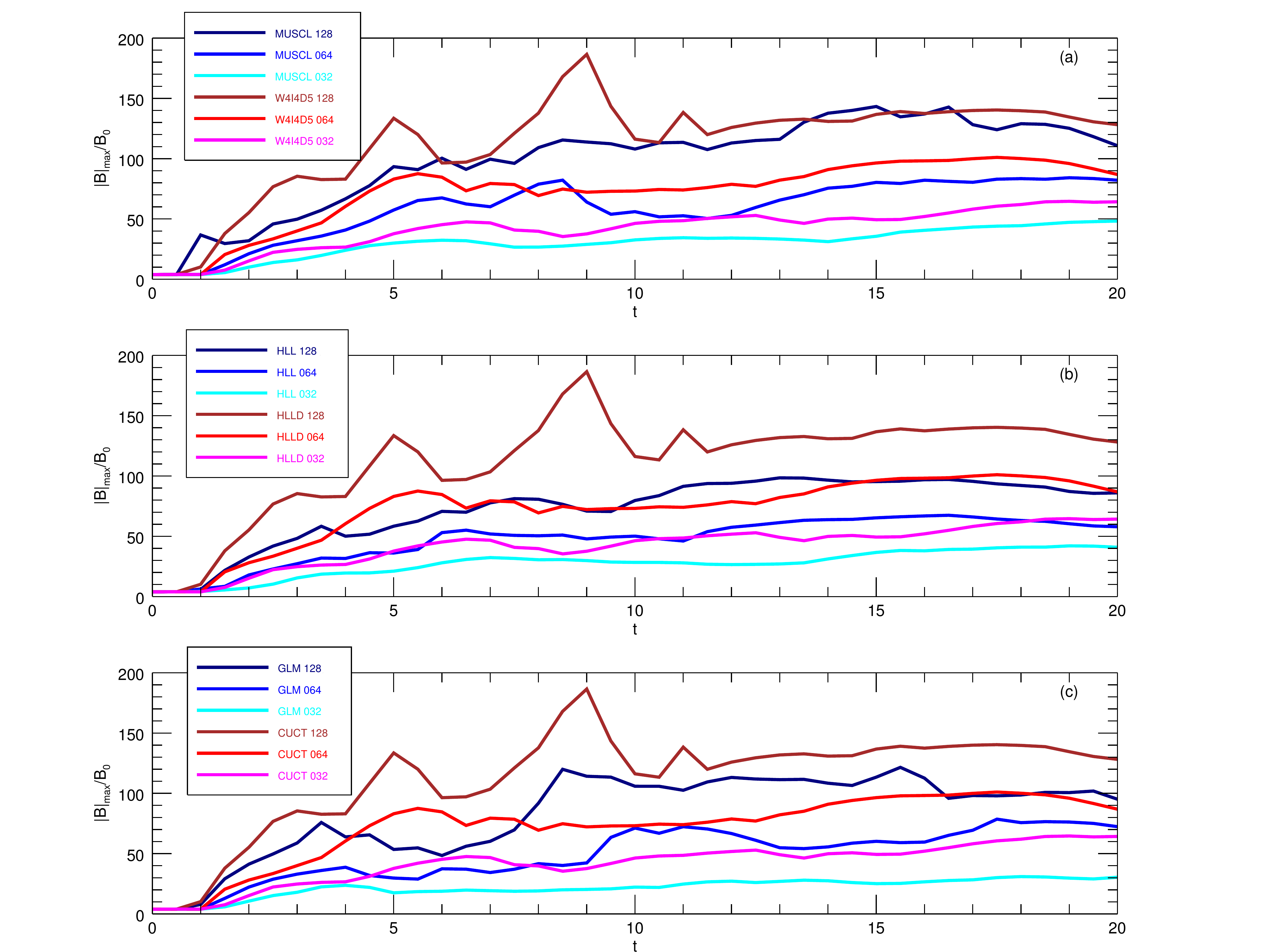}
\caption{Time profile of the maximum of the magnetic field strength $|\vect{B}|_{\rm max}$ in the Richtmyer-Meshkov instability with $N=32,64,128$. The W4I4D5-HLLD-CUCT scheme is compared with (a) the MUSCL-HLLD-CUCT scheme, (b) the W4I4D5-HLL-CUCT scheme, and (c) the W4I4D5-HLLD-GLM scheme. Since the value is normalized by the upstream magnetic field strength, the initial $|\vect{B}|_{\rm max}$ is equal to the compression ratio for the perpendicular MHD shock. }
\label{fig:rmi_bmaxprof}
\end{figure}

\begin{figure}[htbp]
\centering
\includegraphics[clip,angle=0,scale=0.5]{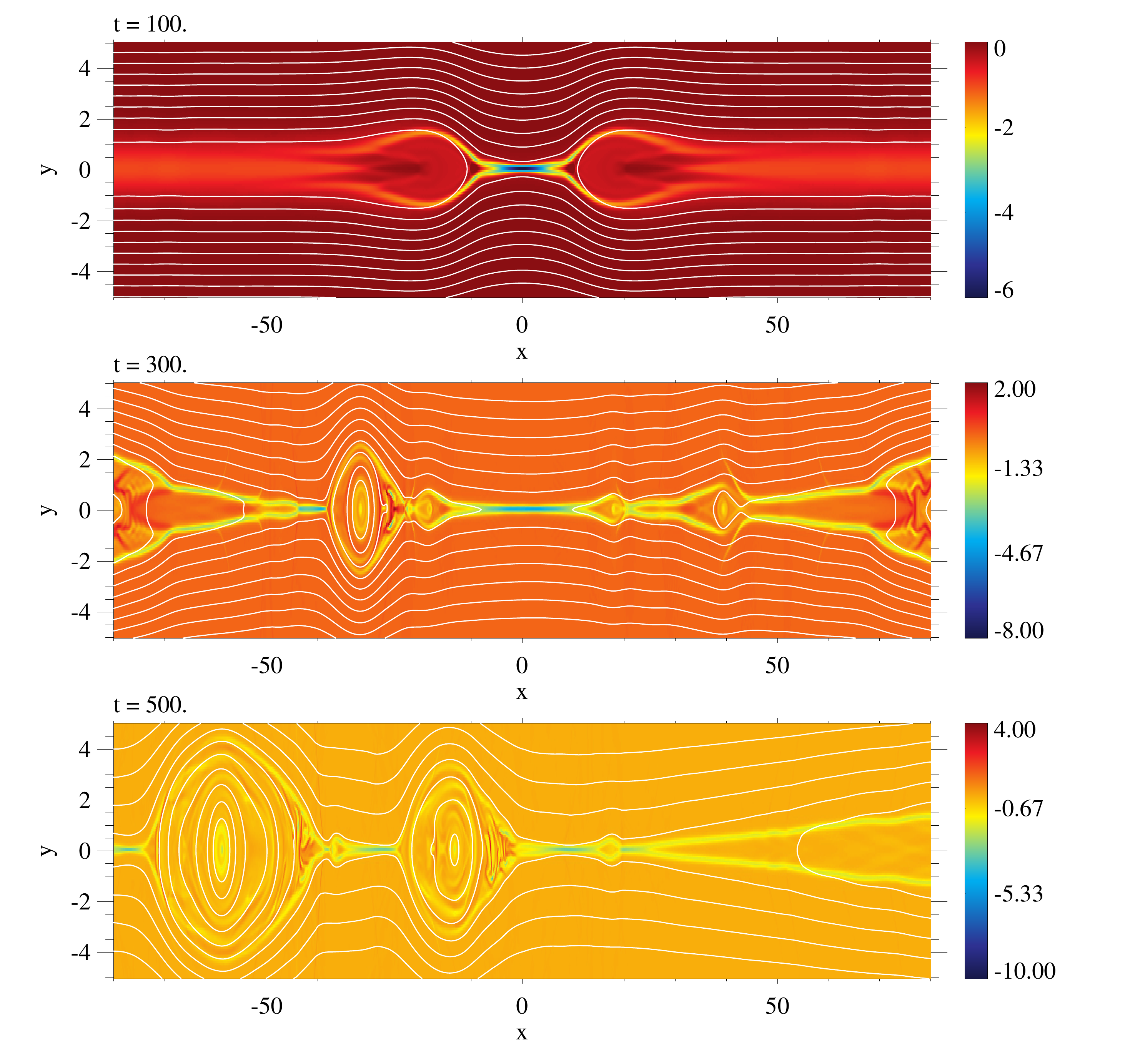}
 \caption{Out-of-plane current density in the magnetic reconnection at $t=100,300,500$ solved by the W4I4D5-HLL-CUCT scheme with $N=8192$. The contour lines represent the magnetic field lines. The figures are enlarged by a factor of 5 in the $y$ direction for better illustration.}
\label{fig:mrxjz_hlluct}
\end{figure}

\begin{figure}[htbp]
\centering
\includegraphics[clip,angle=0,scale=0.5]{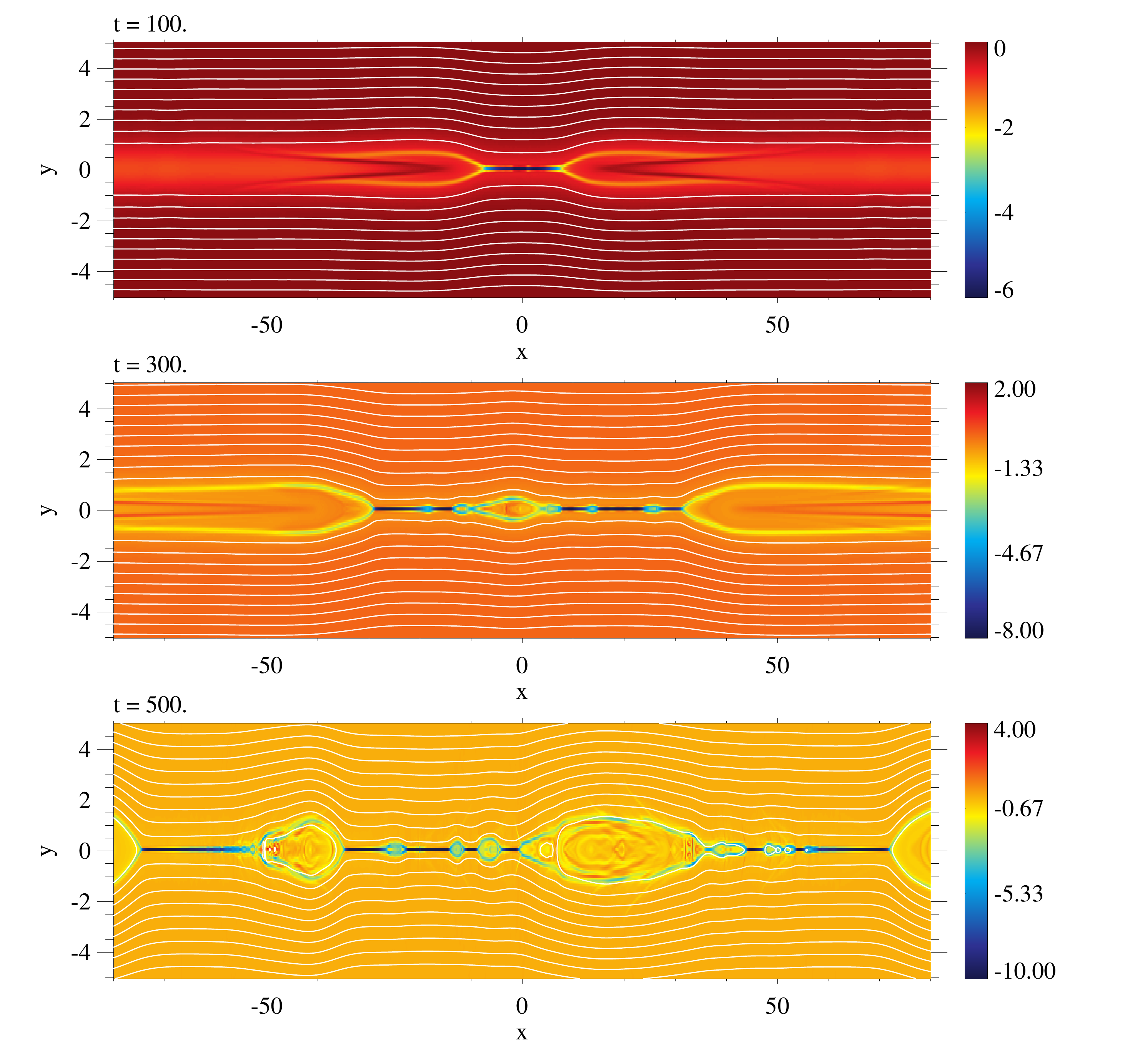}
\caption{Same as Figure \ref{fig:mrxjz_hlluct}, but solved by the W4I4D5-HLLD-CUCT-AA scheme.}
\label{fig:mrxjz_hlltcuctctw0}
\end{figure}

\begin{figure}[htbp]
\centering
\includegraphics[clip,angle=0,scale=0.5]{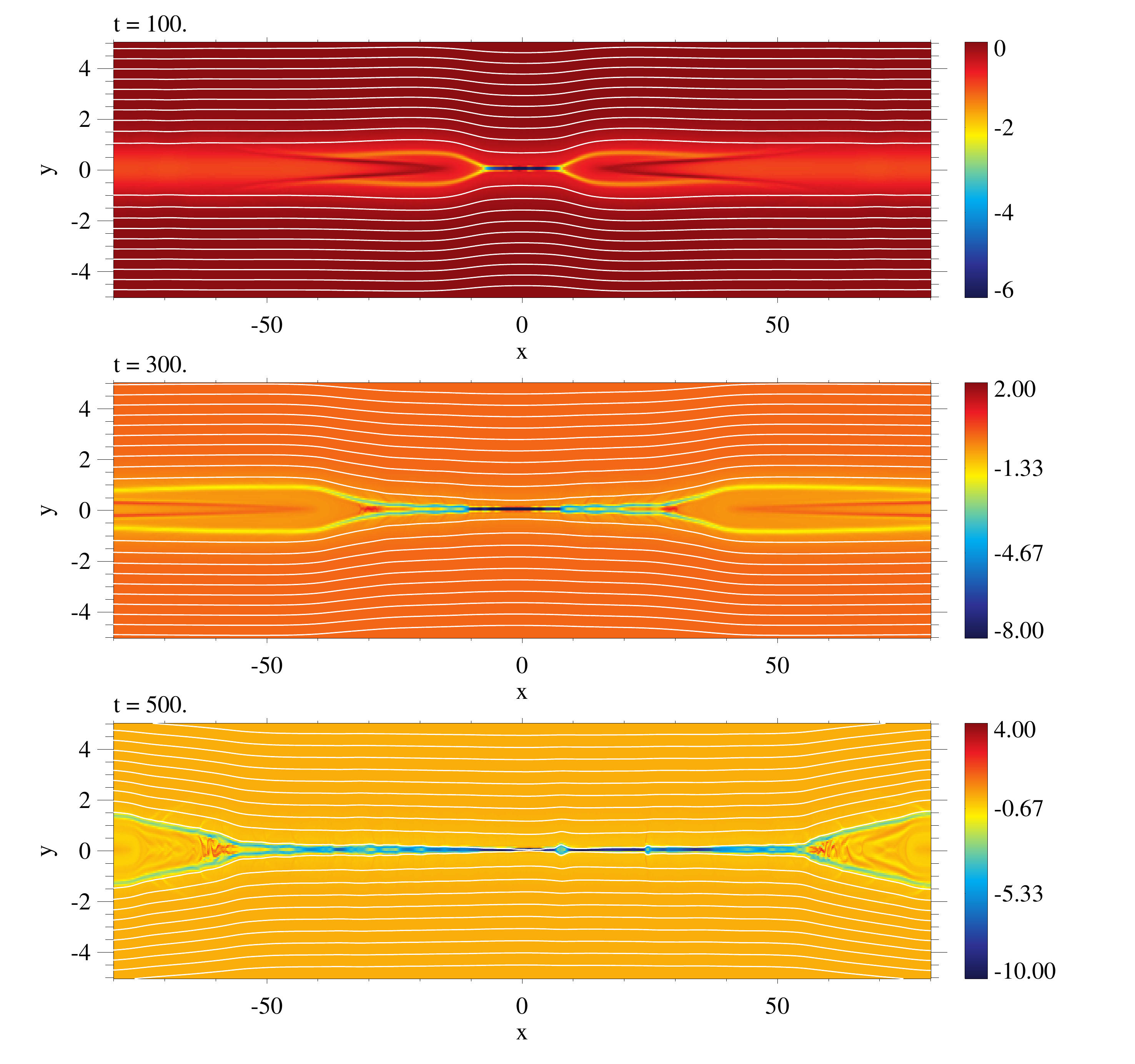}
\caption{Same as Figure \ref{fig:mrxjz_hlluct}, but solved by the W4I4D5-HLLD-CUCT scheme.}
\label{fig:mrxjz_hlldcuct}
\end{figure}

\begin{figure}[htbp]
\centering
\includegraphics[clip,angle=0,scale=0.7]{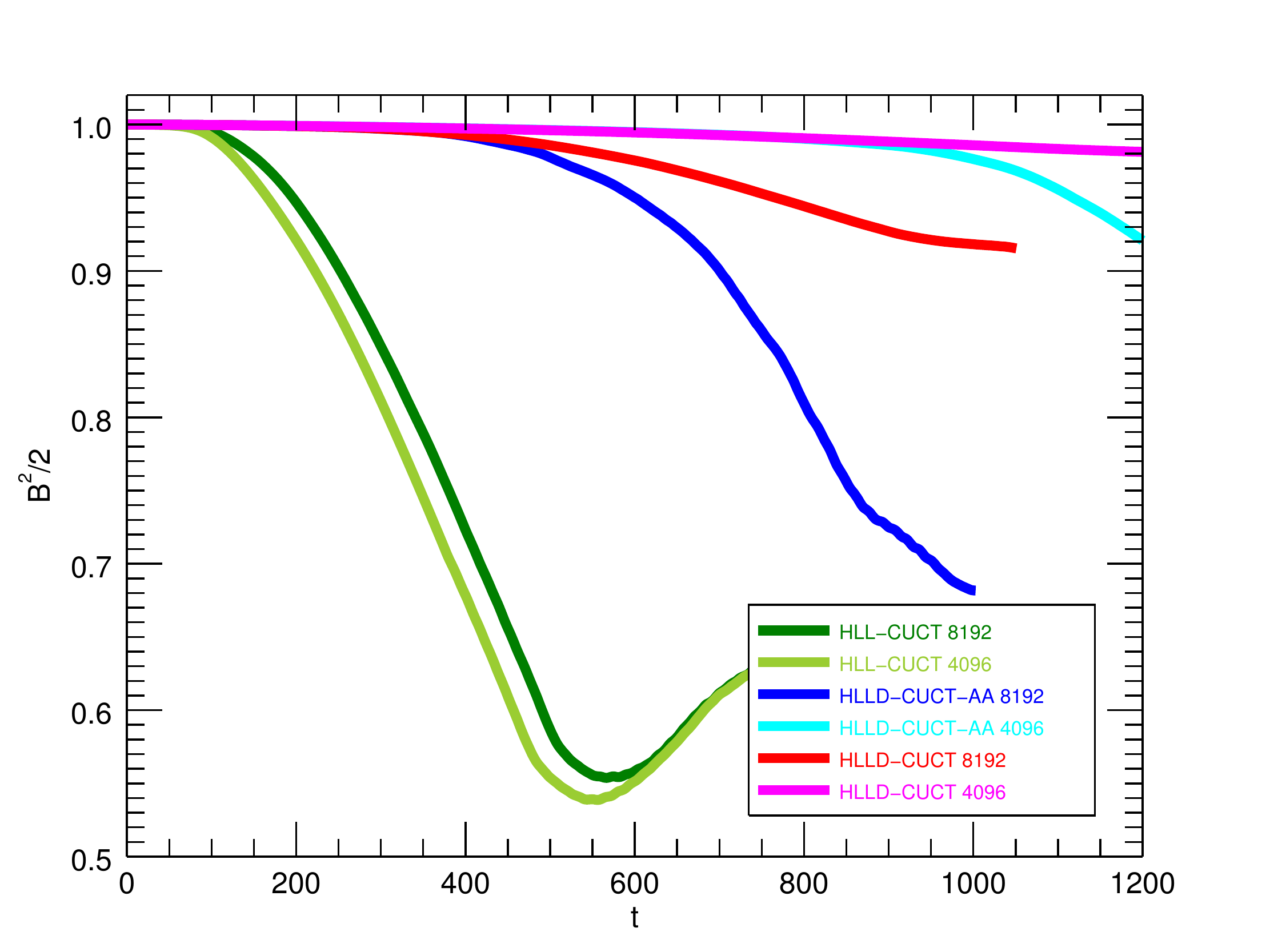}
 \caption{Time profile of the spatially-integrated magnetic energy in the magnetic reconnection solved by the W4I4D5-HLL-CUCT, the W4I4D5-HLLD-CUCT-AA, and the W4I4D5-HLLD-CUCT schemes with $N=4096,8192$.}
\label{fig:mrx_flux}
\end{figure}

\begin{figure}[htbp]
 \centering
\includegraphics[clip,angle=0,scale=0.5]{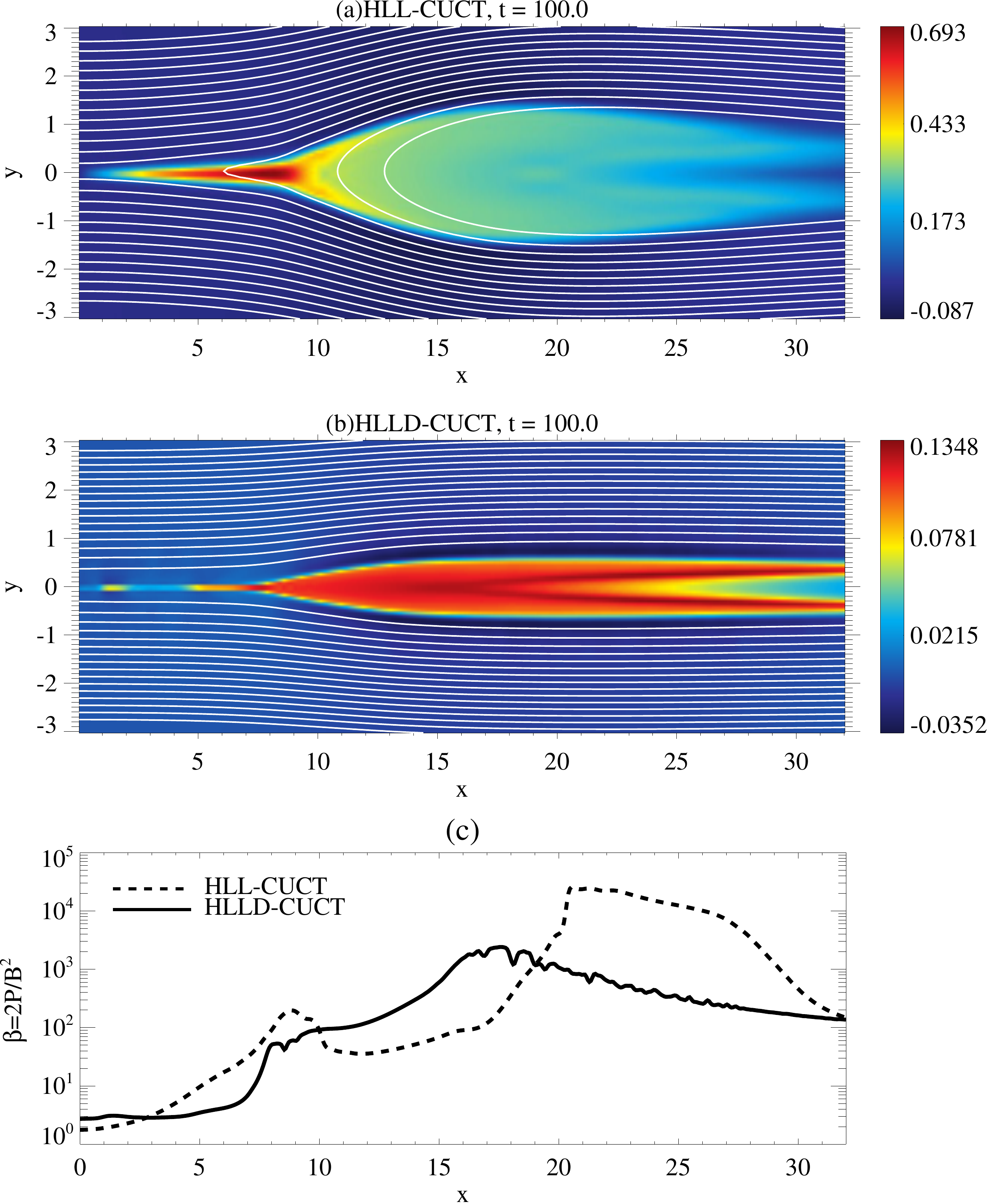}
 \caption{(a)-(b) The $x$-component of the velocity in the magnetic reconnection at the early period $t=100$ solved by the W4I4D5-HLL-CUCT and the W4I4D5-HLLD-CUCT schemes. The contour lines represent the magnetic field lines. The figures are enlarged by a factor of 2 in the $y$ direction for better illustration. (c) Plasma beta value in the current sheet at $t=100$ solved by the HLL-CUCT (dashed line) and the HLLD-CUCT (solid line) schemes.}
\label{fig:mrx_plt2d_ux_beta1d}
\end{figure}

\begin{figure}[htbp]
 \centering
\includegraphics[clip,angle=0,scale=0.5]{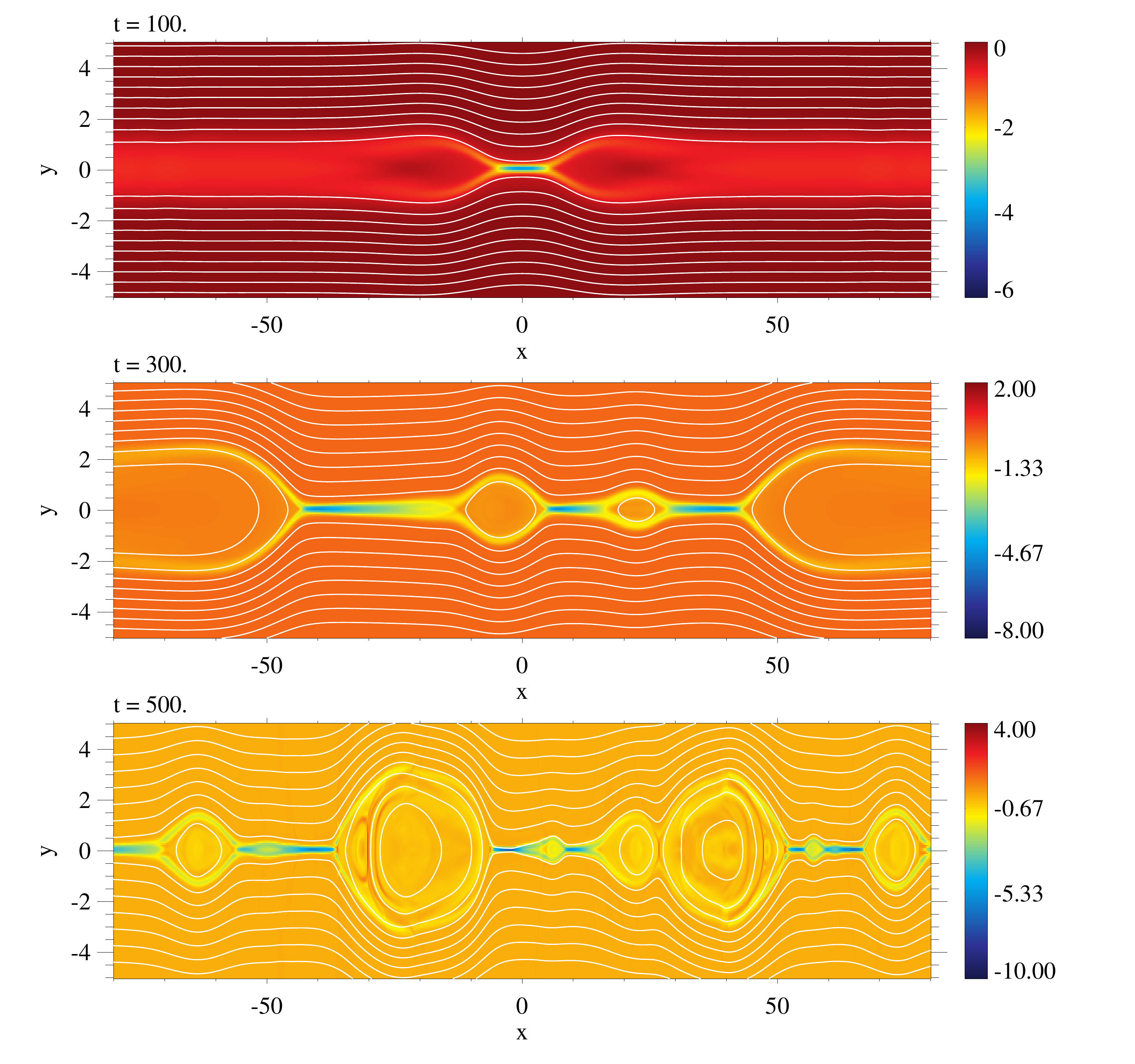}
\caption{Same as Figure \ref{fig:mrxjz_hlluct}, but obtained from the visco-resistive MHD simulation with the W4I4D5-HLLD-CUCT scheme.}
\label{fig:vrmrx}
\end{figure}

\begin{figure}[htbp]
\centering
\includegraphics[clip,angle=0,scale=0.7]{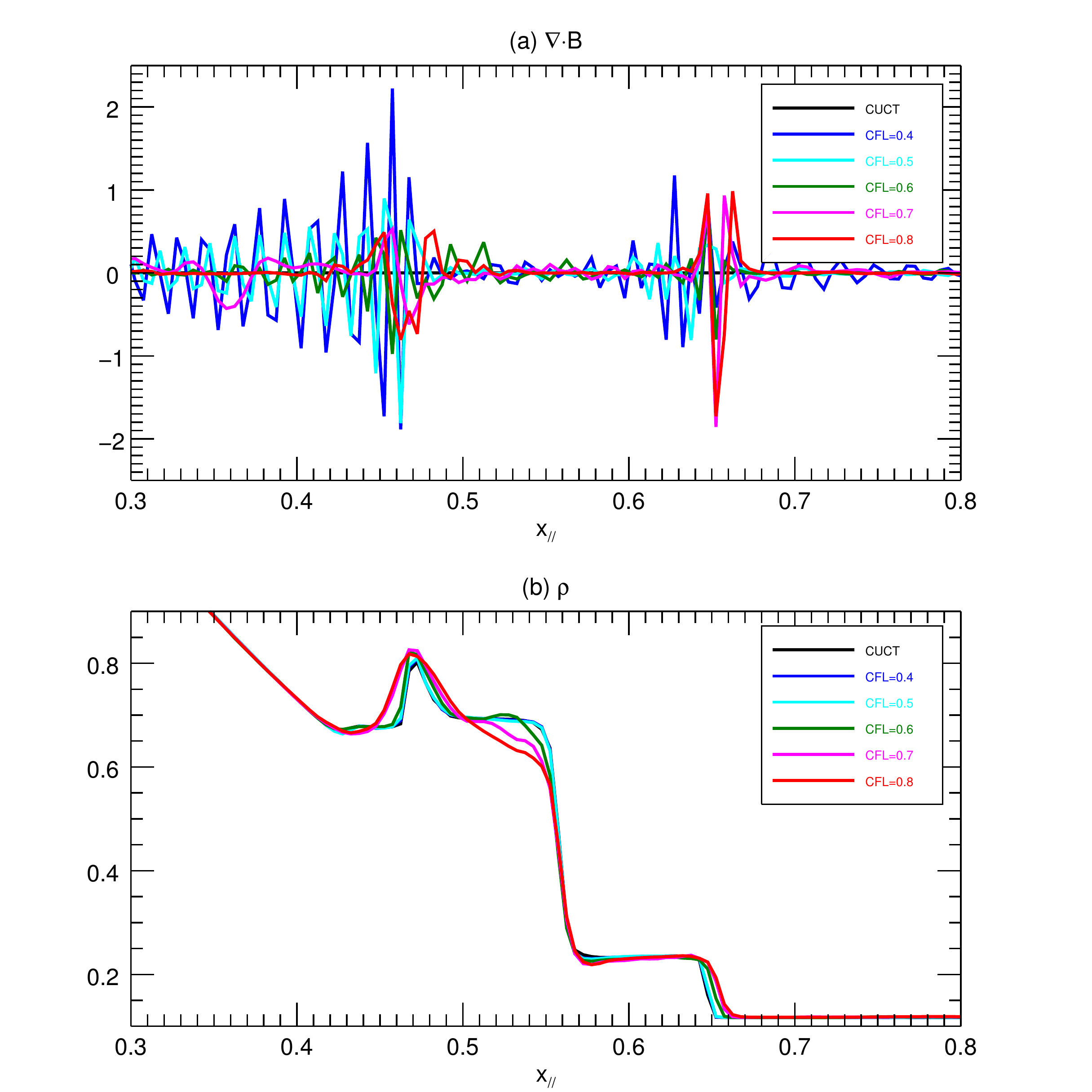}
 \caption{The $c_h$ dependence of (a) $\nabla \cdot \vect{B}$ and (b) density in the Brio-Wu rotated shock tube problem solved by the W4I4D5-HLLD-GLM scheme. The propagation angle is $45^{\circ}$.}
\label{fig:chdependence_bwshock}
\end{figure}

\end{document}